\documentclass[12pt,3p,number,sort&compress,longtitle,letterpaper]{elsarticle}
\pdfoutput=1
\usepackage{natbib}
\usepackage{amsmath,amssymb,amsthm}
\usepackage{url}
\usepackage{epsfig,lineno,textcomp,subfigure,units}
\usepackage{wrapfig}
\usepackage{caption}

\usepackage{lineno}
\usepackage{cuted,titlesec}
\usepackage{soul}
\titleformat{\paragraph}
{\normalfont\normalsize\itshape}{\theparagraph}{1em}{}
\titlespacing*{\paragraph}
{0pt}{3.25ex plus 1ex minus .2ex}{1.5ex plus .2ex}

\usepackage[breaklinks=true,colorlinks=true,linkcolor=blue]{hyperref}
\usepackage[normalem]{ulem} 
\usepackage{amsmath}
\usepackage{graphicx}
\usepackage{dcolumn}
\usepackage{bm}
\usepackage{hyperref}
\usepackage{xcolor}
\usepackage{booktabs}
\hypersetup{colorlinks=true,            allcolors=blue,            bookmarksopen=false,            bookmarksnumbered}   
\usepackage{acronym} 
\usepackage{epstopdf}
\usepackage{color,multirow}
\usepackage{latexsym}
\usepackage{graphicx}

\newcommand{\Qweak}{$Q{\rm{_{weak}}}$ } 

\newcommand{\Qwp}{$Q{\rm{_w^p}}$}
\newcommand{\sigb}{$\Delta { {A_{\, \rm{tgt}}}}$ }
\newcommand{\sigqrt}{$\Delta { {A_{\, \rm{qrt}}}}$ }
\newcommand{\tgtnoise}{52 ppm}
\newcommand{\LH}{LH$_2$}
\newcommand{\Nqrt}{$ { {N_{\, \rm{qrt}}}}$ }
\newcommand{\Apv}{$ { {A_{\, \rm{PV}}}}$ }
\newcommand{\dApv}{$\Delta { {A_{\, \rm{PV}}}}$ }
\newcommand{\dAstat}{$\Delta { {A_{\, \rm{stat}}}}$ }

\newcommand{\dAqrt}{$\Delta { {A_{\, \rm{qrt}}}}$ }
\newcommand{\dAtgt}{$\Delta { {A_{\, \rm{tgt}}}}$ }
\newcommand{\dAbcm}{$\Delta { {A_{\, \rm{BCM}}}}$ }

\begin{document} 

\begin{frontmatter}
\date{\today}
\title{The Q$_{\rm weak}$ High Performance LH$_2$ Target}
\date{\today}

\author[JLab]{J.~Brock}
\author[JLab]{S.~Covrig Dusa}
\author[MSU]{J.~Dunne} 
\author[JLab]{C.~Keith}
\author[JLab]{D.~Meekins}
\author[JLab]{J.~Pierce} 
\author[JLab]{G.R.~Smith\fnref{fn1}}
\ead{grsmith02@wm.edu}
\author[MSU]{A.~Subedi} 

\fntext[fn1]{Corresponding author}

\address[JLab]{Thomas Jefferson National Accelerator Facility, Newport News, VA 23606 USA}
\address[MSU]{Mississippi State University, Mississippi State, MS 39762 USA}

\begin{abstract}
A high-power liquid hydrogen target was built for the Jefferson Lab \Qweak experiment, which 
measured the tiny parity-violating asymmetry in $\vec{e}$p scattering at an incident energy of 1.16 GeV, and a Q$^2 = 0.025$ GeV$^{2}$. To achieve the luminosity of  $1.7 \times 10^{39}$ cm$^{-2}$ s$^{-1}$, a 34.5 cm-long target was used with a beam current of 180 $\mu$A. The ionization energy-loss deposited by the beam in the target was 2.1 kW. The target temperature was controlled to within $\pm$0.02~K and the target noise (density fluctuations) near the experiment's beam helicity-reversal rate of 960 Hz was only 53 ppm. The 58 liquid liter target achieved a head of 11.4 m (7.6 kPa) and a mass flow of 1.2 $\pm$ 0.3 kg/s (corresponding to a volume flow of 17.4 $\pm$ 3.8 l/s) at the nominal 29 Hz rotation frequency  of the recirculating centrifugal pump. We describe aspects of the design, operation, and performance  of this target, the highest power \LH ~target ever used in an electron scattering experiment to date.
\end{abstract}

\end{frontmatter}
\clearpage

\tableofcontents
\listoffigures
\listoftables
\clearpage

\section{Introduction}
\label{sec:Intro}

The \Qweak experiment~\cite{QweakNIM} provided the first determination of the proton's weak charge \Qwp, and used it to  probe for physics beyond the standard model (SM) of particle physics. To reach for new physics at TeV-scales, the experiment sat at the precision/intensity frontier, where precise measurements can be compared to precise predictions of SM observables like \Qwp.   

The weak charge is the electroweak analog of the familiar electromagnetic charge. The weak interactions in  electron-proton scattering that occur as a result of  neutral $Z^0$ exchange have to be separated from among the much more copious electromagnetic interactions that occur when a photon is exchanged between the electron and proton. 
This was accomplished using parity violation: although parity is conserved in the electromagnetic interaction, it is violated in the weak interaction~\cite{LeeYang,Wu}. 

The \Qweak experiment exploited this distinguishing feature by measuring the spin-asymmetry in the elastic scattering of longitudinally-polarized electrons from protons
\begin{equation}\label{eq:pv_asymmetry}
    A_{\text{PV}} = \frac{\sigma_{+}(\theta)-\sigma_{-}(\theta)}{\sigma_{+}(\theta)+\sigma_{-}(\theta)} 
\end{equation}
\noindent  where the beam helicity subscript $\pm$ denotes whether the incoming electron is polarized parallel or anti-parallel to its momentum of about 1.16 GeV. As described in \cite{QweakNature}, it was crucial to perform the experiment at small angles ($\left< \theta \right>=7.9^\circ$) and small four-momentum-transfer squared ($\left< Q^2 \right> =0.0248 $ GeV$^2$) to minimize the contributions of hadronic (internal proton structure) corrections relative to the weak charge. The final results of the experiment and corresponding physics insights were published in \cite{QweakNature}.

\subsection[Performance Requirements]{Performance Requirements}
\label{sec:Performance_Requirements}
Because the parity violating asymmetry was expected to be small ($\mathrm{A_{\, PV}} \approx -230$ ppb) and had to be measured with precision ($\lesssim 10$ ppb), the beam had to be intense and the target had to be thick. To reach the desired precision goal in roughly a year's worth of 
beam delivery, the electron beam current used in the JLab \Qweak experiment was 180 $\mu$A, and the liquid hydrogen (LH$_2$) target was 34.5 cm thick. This resulted in the highest luminosity ($1.7 \times 10^{39}$ cm$^{-2}$ s$^{-1}$) ever employed with a LH$_2$ target in an $ep$ scattering experiment at Jefferson Lab, or any other laboratory we're aware of. 

However, the cost of high luminosity is more beam heating. Over 2 kW of heat deposited by the beam in the LH$_2$ had to be removed to maintain the target temperature within about 10 mK of its  nominal value of 20.00 K. This exceeded the nominal cooling power available from the JLab End Station Refrigerator (ESR) and led to the development of a novel hybrid heat exchanger (see Sec.~\ref{sec:HX}) for the target which simultaneously made use of 15~K high-pressure helium gas coolant normally used for cryotargets as well as low-pressure 4~K helium liquid normally used for superconducting magnets.

Moreover, high luminosity also leads to more boiling in the LH$_2$. 
Density fluctuations from target boiling \sigb\hspace{-0.5em}  near the helicity-reversal frequency of the beam contribute in quadrature to the total asymmetry width \dAqrt measured over beam-helicity quartets. Thus boiling  increases the time required to achieve a given precision goal, and must be minimized. Typically, \sigqrt\ was 225-230 ppm, and consisted of the quadrature sum of detector statistics (\dAstat $\approx 215$ ppm), beam current monitor (BCM) resolution (\dAbcm $ \approx 43$ ppm), and  a target boiling component \sigb\ of about \tgtnoise ~(see Sec.~\ref{sec:Boiling}).  The statistical width (uncertainty) \dApv of the measured parity-violating asymmetry \Apv depends on \sigqrt:
\begin{equation} \label{eq:QuartetdA}
\Delta A_{\rm{PV}}=\Delta A_{\rm{qrt}}/(P \sqrt{N_{qrt}}),
\end{equation}
where $P$ is the beam polarization, and \Nqrt is the total number of beam-helicity quartets. At 100\% efficiency, \Nqrt $ = 10^7$ per day with the 240 Hz quartet helicity patterns ($+--+$ or $-++-$) used in the experiment. The time penalty for the experiment from target boiling (also referred to as target noise) is thus the square of the ratio of the asymmetry width \dAqrt with and without the boiling contribution \sigb \hspace{-0.25em}. The design goal was to limit the time penalty from target boiling to less than 10\%. Despite the record luminosity of the \Qweak target, the penalty achieved was even smaller: only 5\%.

The successful development of a \LH ~target that could meet all these conflicting requirements is the subject of this article.

\subsection[Performance Scaling]{Performance Scaling}
\label{sec:Performance_Scaling}
In the \Qweak experiment's proposal stage, the target noise \dAtgt that might be achievable was estimated by scaling the well-studied low-noise target used for the G0 experiment~\cite{G0tgt}. The scaling was estimated as follows:

\begin{align} \label{eq:scaling}
    \Delta {{A_{\, \rm{tgt}}}}(Q_{\rm{weak}}) \sim & \Delta {{A_{\, \rm{tgt}}}}(G0) 
    \times L_{\rm{tgt}} \left( \frac{Q_{\rm{weak}}}{G0}
    \right)
    \times R_{\rm{width}} \left( \frac{G0}{Q_{\rm{weak}}}
    \right)^2 \nonumber \\
    & \times I_{\rm{beam}} \left( \frac{Q_{\rm{weak}}}{G0}
    \right)
    \times \nu_{\rm{beam}} \left( \frac{Q_{\rm{weak}}}{G0}
    \right)^{-0.4}
    \times \dot{m}_{\rm{LH2}} \left( \frac{G0}{Q_{\rm{weak}}}
    \right),
\end{align}

\noindent
where $L_{\rm{tgt}}$ refers to the target length, $R_{\rm{width}}$  to the square raster dimension, $I_{\rm{beam}}$ the incident beam current, 
$\nu_{\rm{beam}}$ the beam helicity-reversal rate, 
and $\dot{m}_{\rm{LH2}}$ the \LH ~mass or volume flow rate across the beam axis. The values used in this scaling Eq.~\ref{eq:scaling} are tabulated in Table~\ref{tab:PerformanceScaling}. The G0 values come from Ref.~\cite{G0tgt}, and the values for the \Qweak target described here are what were initially proposed and actually used. 

\begin{table}[!hhbt]
\centering
\begin{tabular}{ c  c  c c c c c }
\toprule
  &        & Beam & Raster & Helicity & Volume & Noise \\
  & Length & Current & Area & Reversal & Flow & \dAtgt \\
 Target & (cm) & ($\mu$A) & (mm$^2$) & (Hz) & (l/s) & (ppm)  \\
 \midrule
G0 \cite{G0tgt} & 20 & 40 & 4 & 30 & 4 & 238  \\
\Qweak & 35 & 180 & 16 & 960 & 15 & 31 \\
G0 factor & 1.75 & 4.5 &  0.25 & 0.25 & 0.27 & 0.13\\
\bottomrule
\end{tabular}
\caption[Performance Scaling]{
Parameters used in Eq.~\ref{eq:scaling} to provide an initial estimate of the helicity-quartet target noise that might be achievable in the \Qweak target, based on the performance of the G0 target reported in \cite{G0tgt}. The last row lists the multiplicitive factors that scale the G0 target noise \dAtgt to the target noise expected for the \Qweak target, using the assumptions noted in the text.  
}
\label{tab:PerformanceScaling}
\end{table}

The assumption that the target noise is the same for transverse and longitudinal flow was  untested, so the mass flow was scaled linearly instead of quadratically or even cubically as inferred in~\cite{G0tgt}. The power scaling  used for the faster helicity reversal was  
based on results obtained~\cite{Leacock} for just three simulated helicity reversal frequencies on the standard Hall C cryogenic target which did not have and was not designed for small target noise.  
It is however clear a priori that faster helicity reversal results in better performance because the statistical width of faster (shorter) helicity patterns at a fixed beam current must be larger, and thus a given target noise makes a smaller relative contribution to the total asymmetry width, as shown in Sec.~\ref{sec:helicity}. Moreover, Fourier transforms of the noise spectrum showed that there is more noise at lower frequencies, especially below the 60 Hz line frequency and due to mechanical vibrations from the 30 Hz LH$_2$ recirculation pump, for example.

To summarize, this simple scaling provided early reassurance that the target noise goals of the experiment might be met with reasonable improvements to existing technology on several fronts. 

\section{The Target Components}
\subsection{Overview}

As noted above, in a parity experiment it is important to design a \LH ~target capable of handling an intense beam with correspondingly large  beam-related heat deposition, as well as to minimize density fluctuations near the helicity reversal frequency which cause noise that degrades the uncertainty \dApv of the asymmetry measurement associated with the experiment. The density fluctuations can arise from boiling associated with beam heating in the \LH ~fluid and the target cell windows where the beam enters and exits the cell containing the \LH. These target cell windows also present a background which must be measured and corrected for in order to isolate the results that arise from the hydrogen.

\begin{figure}
\centering
\includegraphics[width=0.75\textwidth]{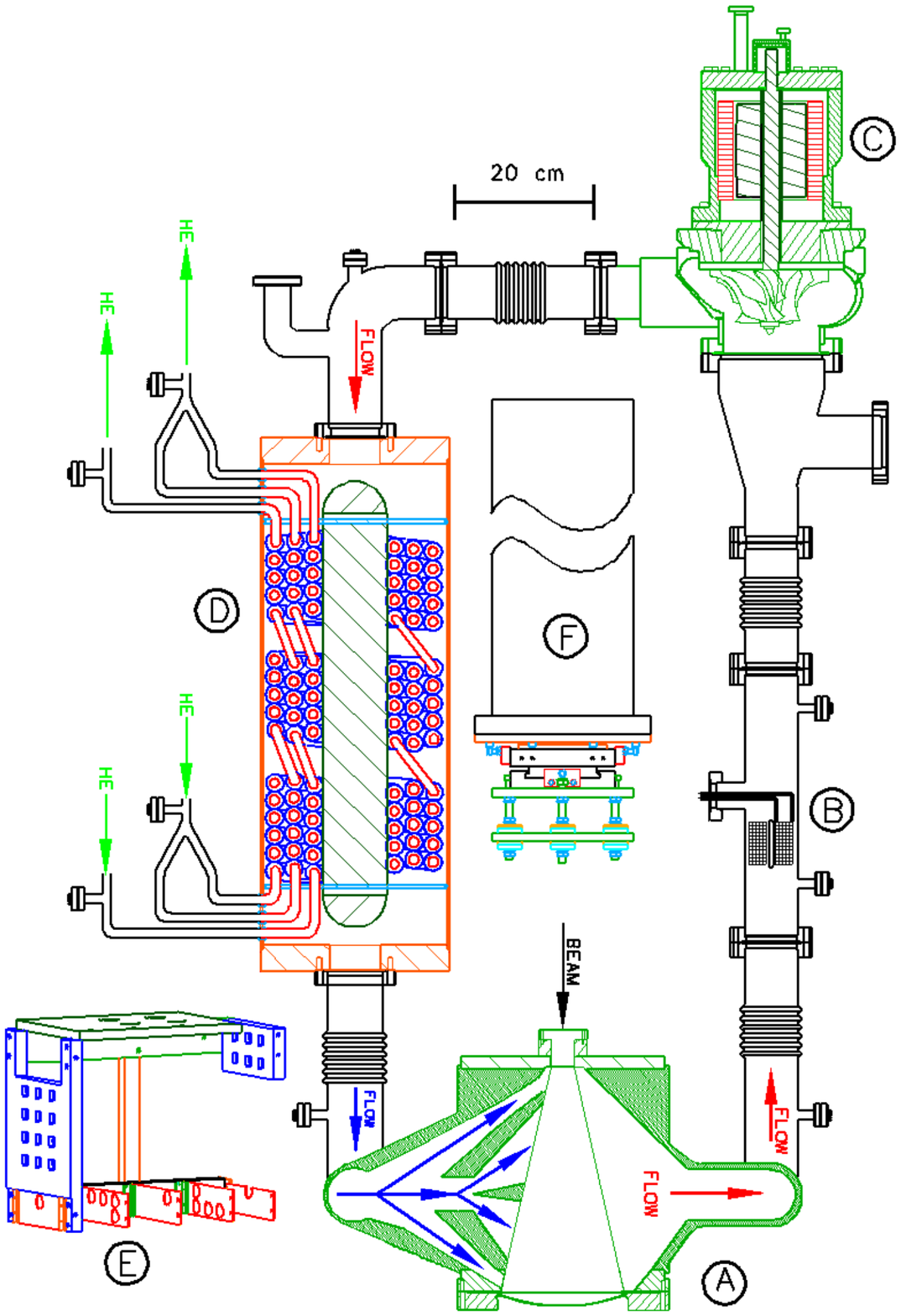}
\caption[Target layout]{\label{fig:tgt-layout} A schematic showing the components of the \Qweak target. A: The beam interaction cell (pitched 90$^\circ$ in this figure in order to illustrate the flow pattern), B: the resistive heater, C: the centrifugal \LH ~re-circulation pump, D: the hybrid heat exchanger, E: the solid target ladder, which was mounted directly below the cell, and F: the long thin stainless-steel pipe which thermally isolated and mechanically supported the target loop, as well as the manual cell adjustment mechanism at its lower end. }
\end{figure}

The basic design of the \Qweak \LH ~target is shown in Fig.~\ref{fig:tgt-layout}. Like most cryogenic targets it is based on a loop of recirculating \LH ~in an insulating vacuum provided by a scattering chamber. The \LH ~circulation is provided by a pump. The beam interactions with the \LH ~take place in a target cell which separates the \LH ~volume from the beamline vacuum with thin windows where the beam can enter and exit the cell. The heat associated with the ionization energy loss of the beam passing through the \LH ~and associated cell windows is removed with a cold helium heat exchanger, which is also used to condense the hydrogen in the system. The temperature of the \LH ~is regulated using a resistive heater immersed in the \LH ~flow which is continuously adjusted by means of a Proportional-Integral-Differential (PID) feedback loop.

The \Qweak target was built to code~\cite{ASME}. Target operators were trained in the physics principles and  operational procedures of the target, and given practical training specific to this target by a subset of the authors of this article. A dedicated target operator staffed the target 24/7 whenever hydrogen was condensed in the target, and the same people who provided the training were available on-call for any problems the target operators couldn't solve on their own.

\subsection{The Target Cell}
\label{sec:Cell_Design}

The target cell defines the volume where the \LH ~flows across the beam axis and electron-proton interactions occur. It separates the pressurized flow space of the \LH ~from the vacuum of the beamline and the scattering chamber.  

Although the requirements that the \Qweak experiment placed on the target were demanding, they also presented some design opportunities because the target needed to serve the needs of only one experiment. In particular, the experiment's acceptance was limited to forward angles  $5.8^\circ < \theta < 11.6^\circ$ by means of a collimator system downstream of the target.  This suggested a conical \LH ~volume whose axis coincided with the beam  such that all electrons scattered less than 
$\approx 14^\circ$ could pass 
out of a large thin exit window on the downstream end of the target cell.  

The precise geometry of the cell with its carefully tailored input and output \LH ~manifolds was arrived at iteratively using Computational Fluid Dynamics (CFD) simulations \cite{CFD}. 
The finite-element analysis software  used for the CFD simulations was developed by Fluent, Inc.\ now part of ANSYS, Inc. The CFD simulations were benchmarked to the G0 target~\cite{G0tgt} cell. Designing a high-power target before CFD became feasible was mostly based on experience and conjecture. With the proper use of CFD design, heating of \LH ~in the beam-illuminated volume of the cell can be mitigated by adjusting flow geometry and flow parameters to satisfy the physics requirements for target noise. 

The ANSYS-Fluent CFD engine \cite{CFD} solves  conservation and transport equations iteratively through gas, liquids, solids or even plasma. Of the first three, the evolution equations are most difficult to solve in fluids (gases and liquids), hence the developers kept the word ``fluid" in the name of the software. But the software is capable of solving the conservation and transport equations in solids too and to deal with fluid-structure interactions. The heat deposited by the electron beam into any medium/material it traversed was calculated using the collisional heat deposition formula described in Eq.~\ref{beampower} below. The CFD software can deal with fluids from subsonic to hypersonic regimes, and it can incorporate chemical reactions. We even used  CFD to simulate  H2 release and fire  in various accident scenarios in the experimental hall  to better define keep-out zones for ignition sources, etc.

The CFD process for the \Qweak target started by creating a geometry in a Computer Aided Design (CAD) program  with an appropriate mesh size  
to capture the flow details of interest. The meshed geometry was imported to Fluent and a case was set up. The case included boundary and bulk conditions, turbulence modelling, 
fluid-structure interactions, etc. Material properties were corrected for temperature dependence over the range 15-300~K and a 2-phase flow model for hydrogen was used to capture the liquid-vapor phase transition wherever it may occur in the geometry. The flow was calculated iteratively to convergence either in steady-state or transient mode. If the model converged, the next step was post-processing: comparison of the model predictions to the experiment's goals for the target, and using that comparison to inform parameter and geometry changes that might lead to better results in the next iteration. The design phase was completed once a geometric model satisfied the physics requirements in a robust way. 

CFD steady-state simulations are very reliable at predicting the equilibrium density loss in a target cell caused by beam heating.  We tried to develop CFD technologies to also predict  the \LH~density fluctuations at the electron beam-helicity reversal frequency of 960~Hz, but were limited by the available computational power at the time. We estimated that to acquire 1~sec of \LH~flow time with a top-of-the-line workstation ($\approx 2007\mbox{-}2008$) would require 5 years of continuous computer time, which was not feasible for our design purposes. 

\begin{figure}[!hhtb] \centering
\includegraphics[width=0.5\textwidth]{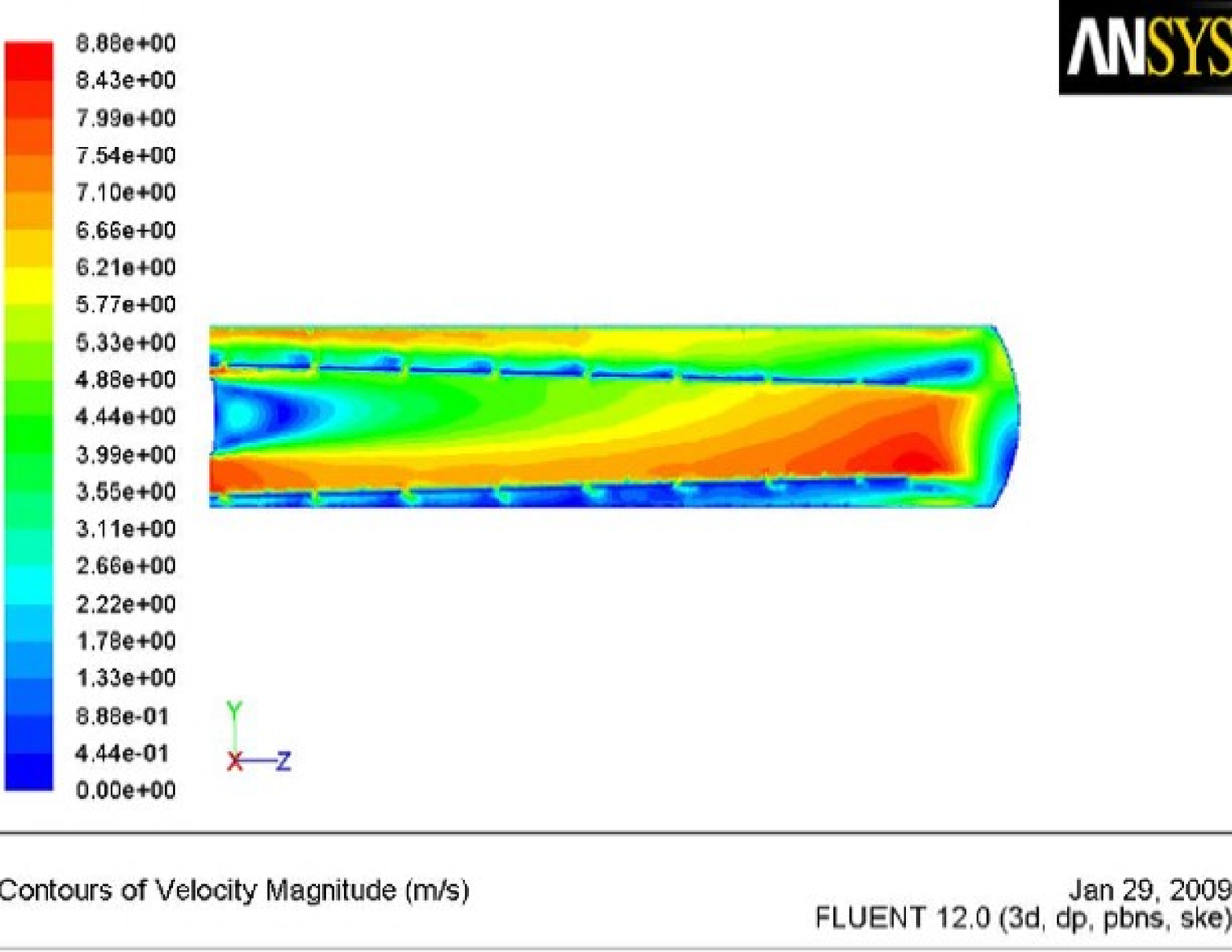}%
\includegraphics[width=0.5\textwidth]{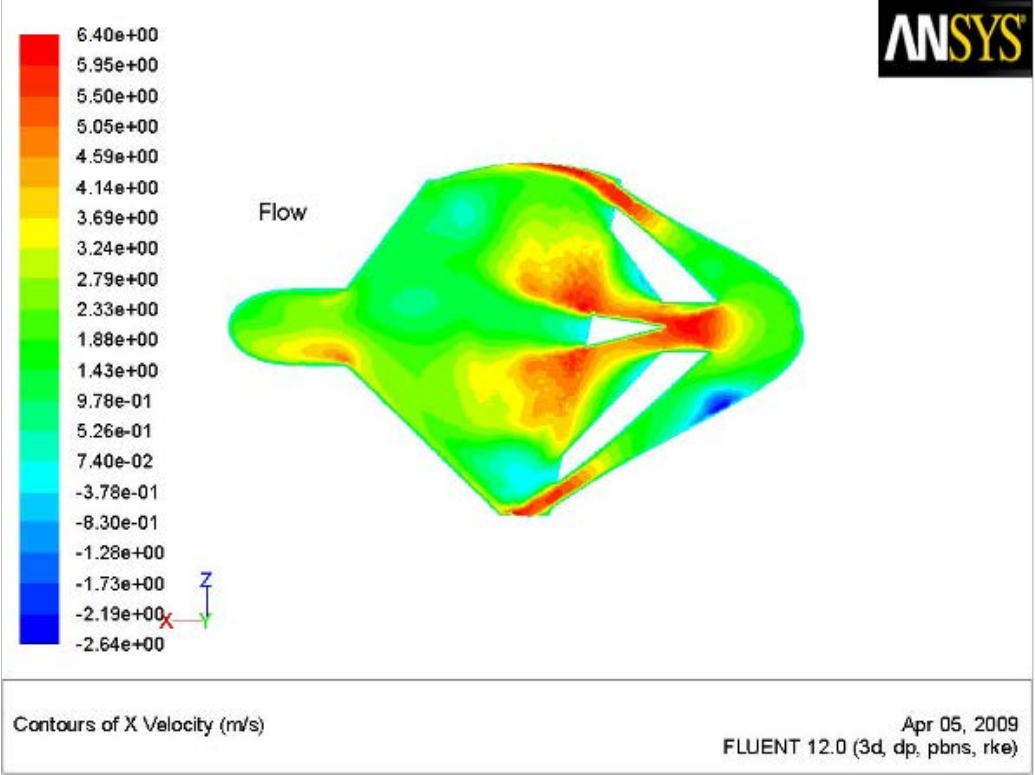}%
\caption[Target Cell CFD]{\label{fig:Cell-CFD} Flow velocity predictions from CFD models for a longitudinal, G0-like cell design with an offset flow diverter (left) and a transverse, conical cell design (right). The beam is incident from the left for the G0-like cell. The \LH ~flow is coaxial, entering from the left inside the perforated flow diverter and exiting the cell at larger diameters outside the flow diverter also on the left. For the transverse cell, the beam is incident from the bottom and the \LH ~flow enters the cell from the right and exits the cell on the left. The input manifold directs part of the \LH ~flow at the entrance and exit windows. The remainder is directed across the beam axis.  }
\end{figure}

\begin{figure}[!hhtb] \centering
\includegraphics[width=0.6\textwidth]{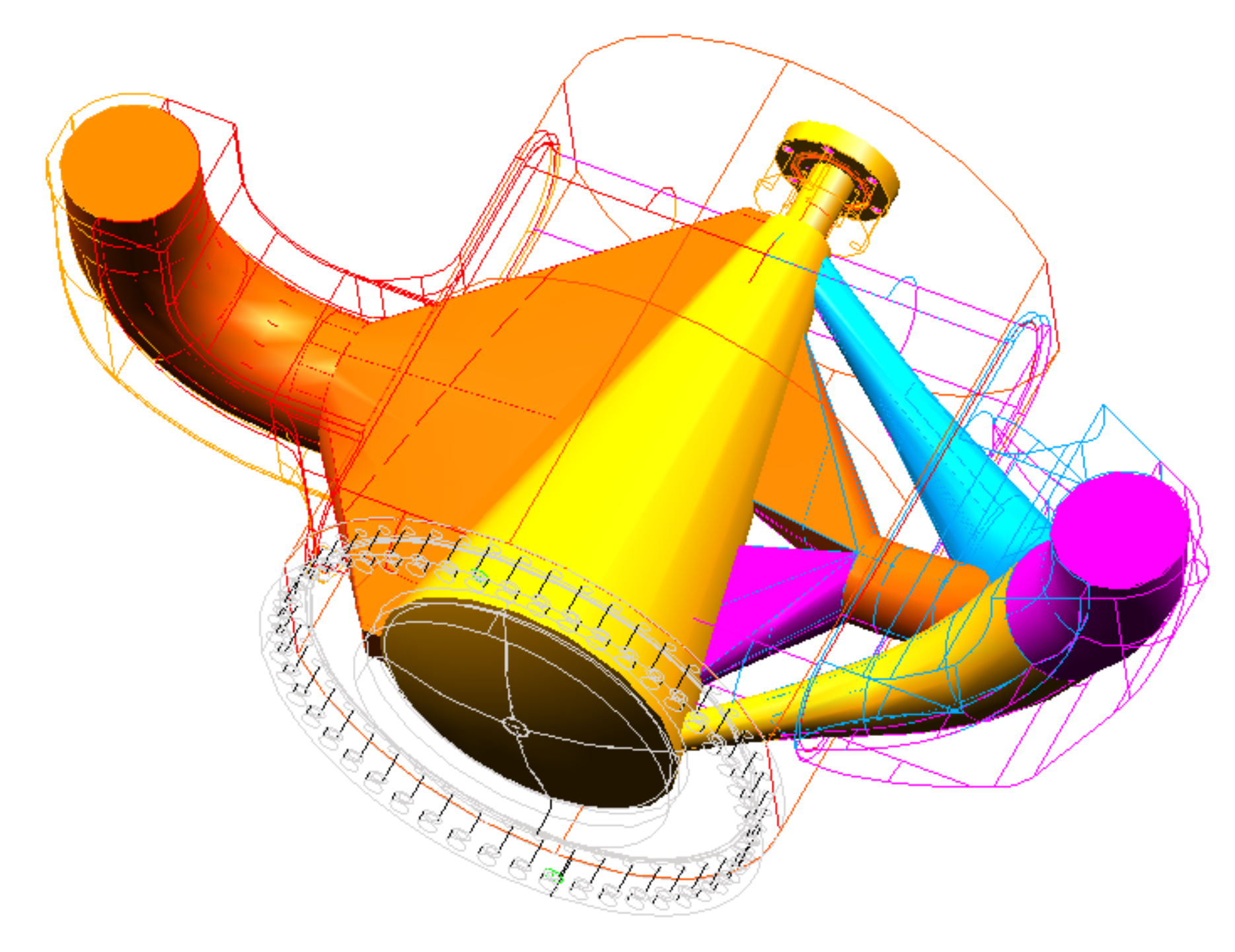}%
\includegraphics[width=0.4\textwidth]{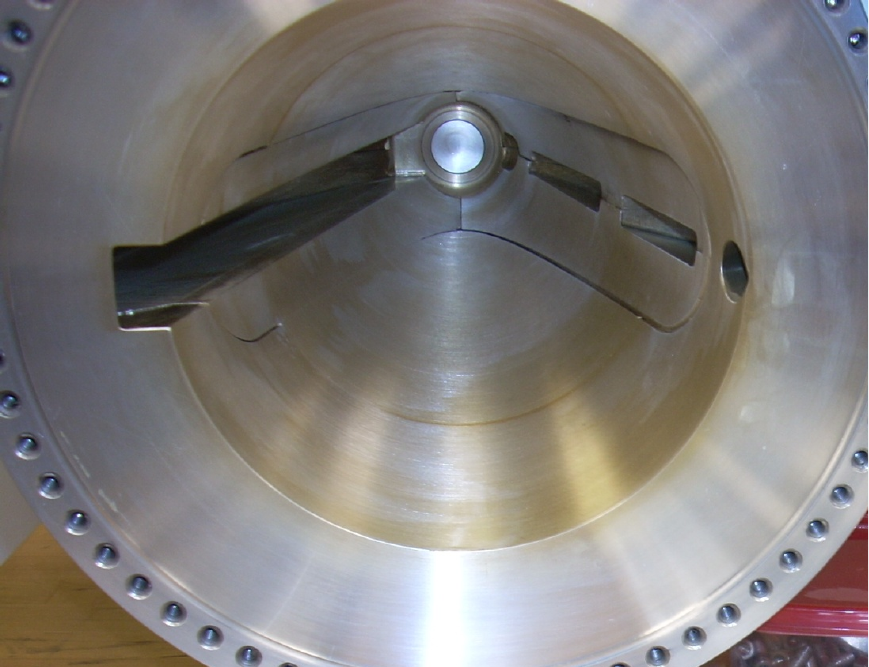}%
\caption[Target Cell Design]{\label{fig:Cell-Design} Left: CAD depiction of the \LH ~cell, showing the  beam and scattered electron \LH ~volume (solid yellow) inside a wire frame of the cylindrical aluminum alloy cell. The \LH ~exit manifold is denoted in orange on the left of the figure. The \LH ~flow is directed across the beam axis by the four sections of the \LH ~input manifold on the right. Right: The inside of the conical cell looking upstream is shown in the inset photo in the lower left. The conflat knife-edge is visible just inside the outer bolt pattern. The \LH ~flow is from right to left in both depictions. The incident electron beam is from the upper right to the lower left along the central axis of the yellow conical \LH ~volume in the CAD diagram.}
\end{figure}

After the baseline was  established by simulating the G0 target \cite{G0tgt} geometry, a stretched G0-like longitudinal flow design was studied which adopted off-center flow diverters~\cite{Mainz} to mitigate beam heating 
at the cell windows. Those results were then compared to a transverse flow design with a conical \LH ~target volume. Local heating at the entrance and exit windows was reduced by diverting some of the $\approx 3$ m/s transverse flow diagonally across the beam axis to the central region of each window at $\approx 7$ m/s, as shown in Fig.~\ref{fig:Cell-CFD}. Table~\ref{tab:CFD} compares the results obtained for both designs. Although both designs had local hot spots, and both were predicted to have maximum temperature increases $\Delta T_{\rm max}$ below the 3.7~K required for fluid boiling, the transverse design $\Delta T_{\rm max}$ was about half that of the longitudinal design. 
The transverse flow conical cell design 
was chosen for the \Qweak experiment, machined out of a cylindrical block of cast 2291 aluminum, as shown in Fig.~\ref{fig:Cell-Design}. The head associated with this cell and its inlet and outlet manifolds was determined from CFD calculations to be 2.5~m.

The cell main body and its inlet and outlet manifolds were machined from B209 aluminum 6061-T651 plate and welded together. Sections of B209 2219-T851 plate  were then welded to the upstream and downstream faces of the cell as well as to the outer ends of the inlet and outlet manifolds. Custom conflat flange knife-edges were machined into the 2219 surfaces as a last step (see Fig.~\ref{fig:Cell-Design}). The two alloys were used because welded 6061 is too soft to hold the conflat knife-edge, and we couldn't get the harder 2219 plate thick enough to build the whole cell.  

\begin{table}[htb]
\centering
\begin{tabular}{ r  c  c c c c c }
\toprule
 & P & $<v>$ & $\Delta \rho / \rho$ & $<q>$ & $<\Delta T>$ & $\Delta { T_{\rm{max}}}$         \\ 
 & W/cm$^3$ & m/s & \% & W/cm$^2$ & K & K \\ \midrule
Windows & 3950 & 7 & -& 22.3 & 15.2 & 22.7 \\ 
Transverse & 245 & 2.8 & 0.8 & -& 0.476 & 1.73 \\ 
Longitudinal & 245 & $0.28 - 3.8$ & 1.8 & -& 1.1 & 2.97
\\ \bottomrule
\end{tabular}
\caption[CFD predictions]{Predictions from CFD simulations for various properties of two different target designs, assuming 180 $\mu$A $e^-$ beam rastered $5\times 5$ mm$^2$ on a 35-cm-long \LH ~target held at 20~K and 35 psia (3.7~K sub-cooled) with a 1 kg/s mass flow (15 liters/s). The beam power in the \LH ~is 2120 W and 25 W in the two 0.125 mm thick Al windows. The columns represent the volume power density P, the average \LH ~flow velocity $v$, the relative change in density $\Delta \rho / \rho$, the areal power density $q$, the average overall temperature increase $\Delta T$, and the maximum temperature increase $\Delta T_{\rm max}$.}
\label{tab:CFD}
\end{table}

\begin{figure}[!hhhhtb] \centering
\includegraphics[width=0.48\textwidth,angle=0]{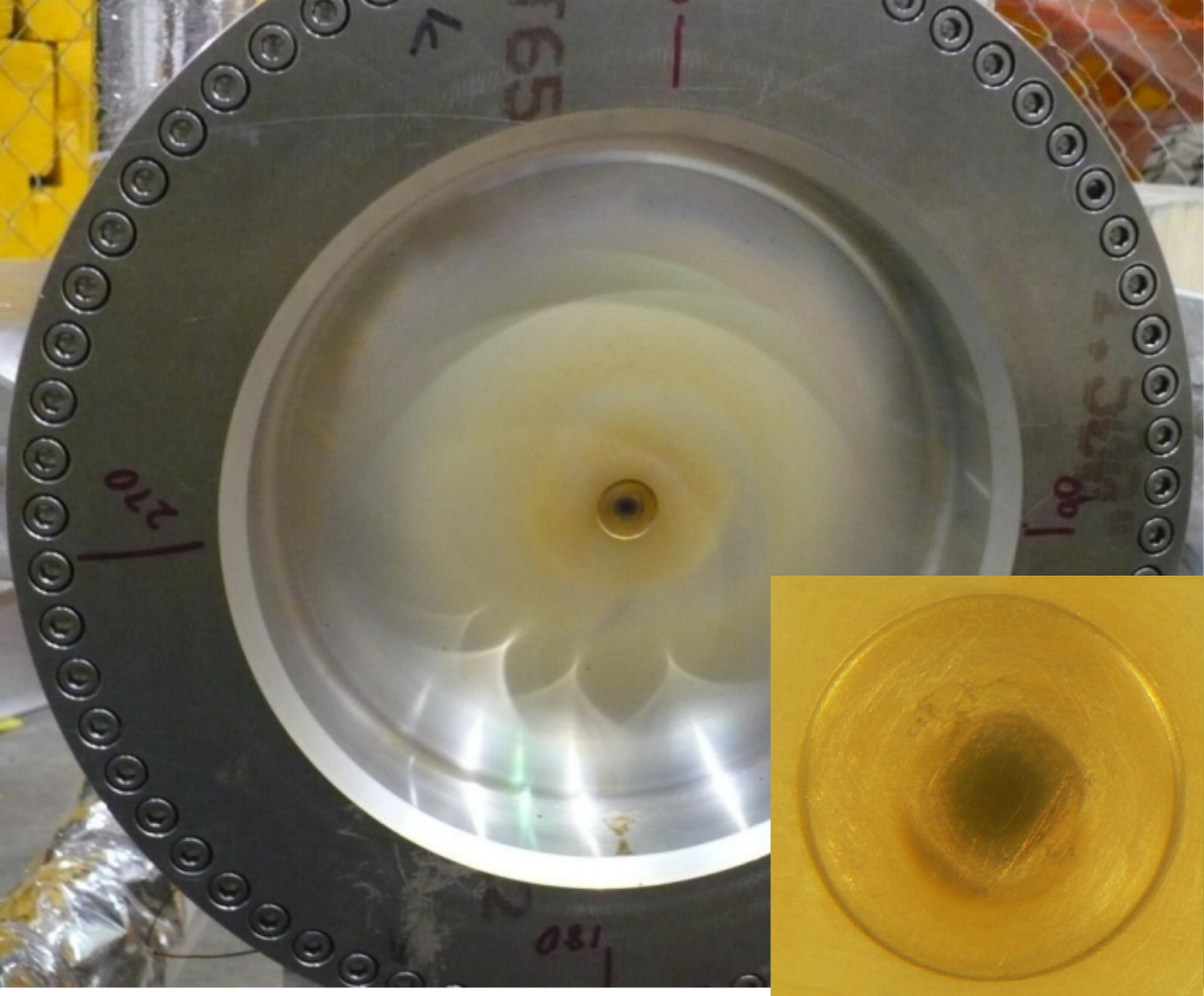}
\caption[Exit Window Spot]{\label{fig:Exit Window Spot} A photo of the downstream face of the LH$_2$ target cell window after about 6 months exposure to 140 $\mu$A beam, looking upstream from downstream at the vacuum side of the window. The discoloration pattern left by the $4\times 4$ mm$^2$ rastered beam spot is clearly visible in this photo, indicating that the beam was well centered on the thin nipple of the 190.5 mm diameter convex exit window machined from a 305 mm diameter flange. The inset in the lower right shows a closeup of the central 0.127 mm thick, 15 mm  diameter nipple with the $4\times 4$ mm$^2$ spot left by the beam clearly visible and well centered. No corresponding spot was made on the opposite (LH2) face of the window.}
\end{figure}

\subsubsection{Cell Windows}
The collaboration advocated for beryllium windows where the beam entered and exited the cell containing the \LH. With an atomic number of only 4, a very high melting point, good strength and thermal conductivity, beryllium seems like an ideal window material to use to minimize background and beam heating in the windows. 
However it can be brittle at low temperatures and the consequences of that proved fatal in the past~\cite{CEA} at another laboratory. 

With that tragic accident in mind,  the aluminum alloy 7075-T651 was chosen for the target windows instead of Be. Aluminum is more ductile then Be at low temperatures. This alloy was chosen for its superior strength,  and consisted of Al (89.2 wt\%), 
Zn (5.87 wt\%), Mg (2.63 wt\%), Cu (1.81 wt\%),
and other (0.47 wt\%), determined by chemical assay of the aluminum actually used for the target. The windows were machined from single billets of ASTM B209 7075\:-T651 aluminum plate that were extruded and hot rolled to minimize voids (relative to cast aluminum). Although target window background was the largest correction that had to be accounted for in the \Qweak experiment (about 17\%), no other problems related to the target windows were encountered.

The cell entrance window~\cite{entrance_window} design was similar to windows in use at JLab for many years.  Past applications include 1.55 MPa (225 psia) helium gas targets and (more typically) 170 kPa liquid hydrogen targets. The entrance window was tested to 3.45 MPa (500 psi). It consisted of a 12.7 mm-thick, 69.3 mm-diameter machined conflat flange with a 22.2 mm bore. The downstream end of the flange supported a 22.2 mm-i.d., 25.4 mm-o.d.\ cylindrical re-entrant tube 41.4 mm long which penetrated the target cell block into the \LH ~volume. The flange and the tube were machined as one piece from a single block.  The 0.097 mm-thick entrance window at the end of the tube separated the beamline/scattering chamber vacuum from the nominally 220 kPa \LH. The deflection of the center of the 22.2 mm-diameter entrance window measured at 300 K and the 221 kPa operating pressure of the target was only 0.18 mm.

The exit window was also machined from a single piece of extruded  Al 7075\:-T651 plate. The window diameter had to be large enough to accept  all of the scattered electrons of interest ($\theta_{\rm{lab}} \lessapprox 14^\circ$) unimpeded. The thickness of this window was optimized to 1) reduce as much as possible the background from beam electrons interacting with the aluminum at the center of the window, 2) maintain the strength required to safely contain the fluid pressure, and 3) provide sufficient window thickness to promote conduction of heat generated by the beam passing through the window. To meet these design requirements, the exit window was composed of three radial zones, as shown in Fig.~\ref{fig:Exit Window Spot}. The outermost zone from 152.4 mm $> r >$ 86.7 mm consisted of a 38.1 mm ($1.5''$) thick annulus with a custom conflat knife-edge machined into the upstream face that mated (using vented Ti bolts and an 1100-O soft aluminum gasket) to the custom conflat knife-edge  machined into the downstream face of the target cell body visible in the Fig.~\ref{fig:Cell-Design} photograph. The second annulus extended from 86.7~mm $> r >$ 7.5~mm and had a convexity of 254 mm with a thickness of 0.51~mm to accommodate the scattered electrons as mentioned above. The final inner section was a  15~mm diameter disk just 0.13~mm thick, through which the electron beam passed. This last section was made as thin as safely possible in order to reduce background from and heat deposition in the aluminum.

To guide the initial design of the window, a simplified model was developed which considered the convex section of the window as spherical such that the stress could be expressed as $S=PR/(2t)$ where $P$ is the pressure load, $R$ is the radius of curvature, and $t$ is the thickness of the shell. From this basic model, thicknesses were varied to optimize the strength, physics, and thermal performance of the window. With a pressure of 0.55 MPa (80 psi), the stress in the domed section of the window was 138 MPa. The maximum allowable stress for the window was determined using the material properties given in ASTM B209 and is the lower of 2/3 the yield stress $S_Y$ or 1/3 the ultimate tensile stress $S_{UT}$, i.e. 175 MPa. Thus, the domed section was deemed suitable for more sophisticated analysis. 

To model the thin central nipple of the window, an expression for the stress $S$ in large deflections (more than $0.1 t$) of thin circular sections from \cite{Roark}  was used:
\begin{equation}
    S=0.423\left( \frac{E P^2 r^2}{t^2}\right)^{1/3} 
\end{equation}
where $E = 71.7$ GPa is the modulus of elasticity, $r$ and $t$ here refer to the radius and thickness of the nipple. 
This gave a stress of 179 MPa which is slightly above the allowable but still deemed acceptable for further analysis.

Similarly,  the deflection at the center was determined~\cite{Roark} as
\begin{equation}
y=0.662 r \left( Pr/(Et)\right)^{1/3} =0.38 \; {\rm mm}
\end{equation}
for the same pressure load used to determine the stress above. This deflection could then be compared to other models and measurements made during testing.

Because of the complex geometry of the window, we ultimately employed a more detailed model using the elastic plastic technique given in the ASME Boiler and Pressure Vessel Code \cite{ASME}. This technique utilized finite-element analysis with an augmented pressure load  at least 2.4 times the expected maximum pressure of 0.69 MPa, which finally enabled us to conclude that the window would be safe to use as designed. 

As a final check, a sample of windows were hydrostatically tested to destruction with failures near 1.7 MPa. This was more than 2.4 times the maximum pressure in the cell, and 8 times more than the typical operational pressure when the target was condensed. The entrance and exit windows were replaced with identical spares about halfway through the experiment.

\subsection{Cooling Power}
\label{sec:Cooling Power}

The Q$_{\rm weak}$ experiment's design requirements included a 180 $\mu$A beam of 1.165 GeV electrons rastered into a pattern no larger than $5\times 5$ mm$^2$ onto a liquid hydrogen target $\approx$35 cm long. The final 34.5 cm target length is corrected for thermal contraction to $T=20$~K and pressure bulging at the nominal operating pressure  $P=220$ kPa. The ionization energy loss associated with the passage of the electron beam through the LH$_2$ is 
\begin{equation}
 P=I_{\rm{beam}} \, L_{\rm{tgt}} \, \rho_{\rm{tgt}} \,  dE/dx  = 2060 \; W,
\label{beampower}
\end{equation}
where the beam current $\rm I_{beam}$ is in $\rm \mu A$, the target length $\rm L_{tgt}$ is in cm, the parahydrogen target density~\cite{NIST} at this temperature $(T)$ and pressure $(P)$ is $\rm \rho_{tgt}=0.0713 \; \rm g/cm^3$, and the energy loss (including the density effect \cite{Sternheimer,Sternheimer2}) is $\rm {dE/dx}=4.653 \; \rm MeV/(g/cm^2)$.
One must also account for the viscous heating of the LH$_2$ (175 W), the heat generated by the submersed LH$_2$ recirculation pump ($\sim$150 W), conductive heat loss to the outside ($\sim$150 W),  reserve heater power for control of the target temperature ($\sim$250 W), and the entrance and exit windows ($\sim$22 W). 
Accordingly, a cooling power of about 3 kW is required.

This far exceeds the cooling power which was then available from the JLab End Station Refrigerator (ESR), which could supply up to 25 g/s of 12 atm, 14.5~K helium coolant for cryogenic targets  (shared between all end stations). 
With coolant returned at $P=3$ atm, this represents a cooling power 
\begin{equation} \label{eq:QismdotCpdT}
Q=\dot{m} C_p  \Delta T
\end{equation}
of only 860 W even if all of the available 15~K coolant were used for the 20~K Qweak LH$_2$ target. To achieve the required 3 kW cooling power,  the available 15~K cooling power had to be increased and augmented with the approximately 20 g/s excess capacity of the 3 bar, 4~K Central Helium Liquifier (CHL) which is normally dedicated to cooling the accelerator's superconducting radio-frequency (SRF) cavities and superconducting magnets in the experimental end stations.  
Use of the CHL excess 4~K helium coolant (in conjunction with the 15~K coolant) 
had three disadvantages. First, SRF operation was strained without the excess capacity margin typically provided. 
Second, since hydrogen freezes around 14 K, use of 4~K coolant was problematic. Finally, the existing vacuum-insulated coolant transfer line infrastructure was not designed for this hybrid situation. 

Although the separate 15~K supply and 20~K return transfer line plumbing was adequate, the existing 4~K supply and its return were co-axial, since the 4K is normally used to cool super-conducting spectrometer magnets in the end-stations which return the coolant at 5 K. Returning the coolant at the 20~K operating temperature of the target required a non-coaxial arrangement. Ultimately this challenge was met by warming up all the superconducting magnets in the Hall C end-station hosting the experiment, hijacking the 4K supply piping for the target, and returning (at 20 K) the coolant supplied at 4~K through the LN$_2$ transfer line shield. This decoupled the 4K supply and return as needed, and improved the effectiveness of the shield. 

\begin{figure*}[!hhhtb] \centering
\includegraphics[width=0.9\textwidth,angle=0]{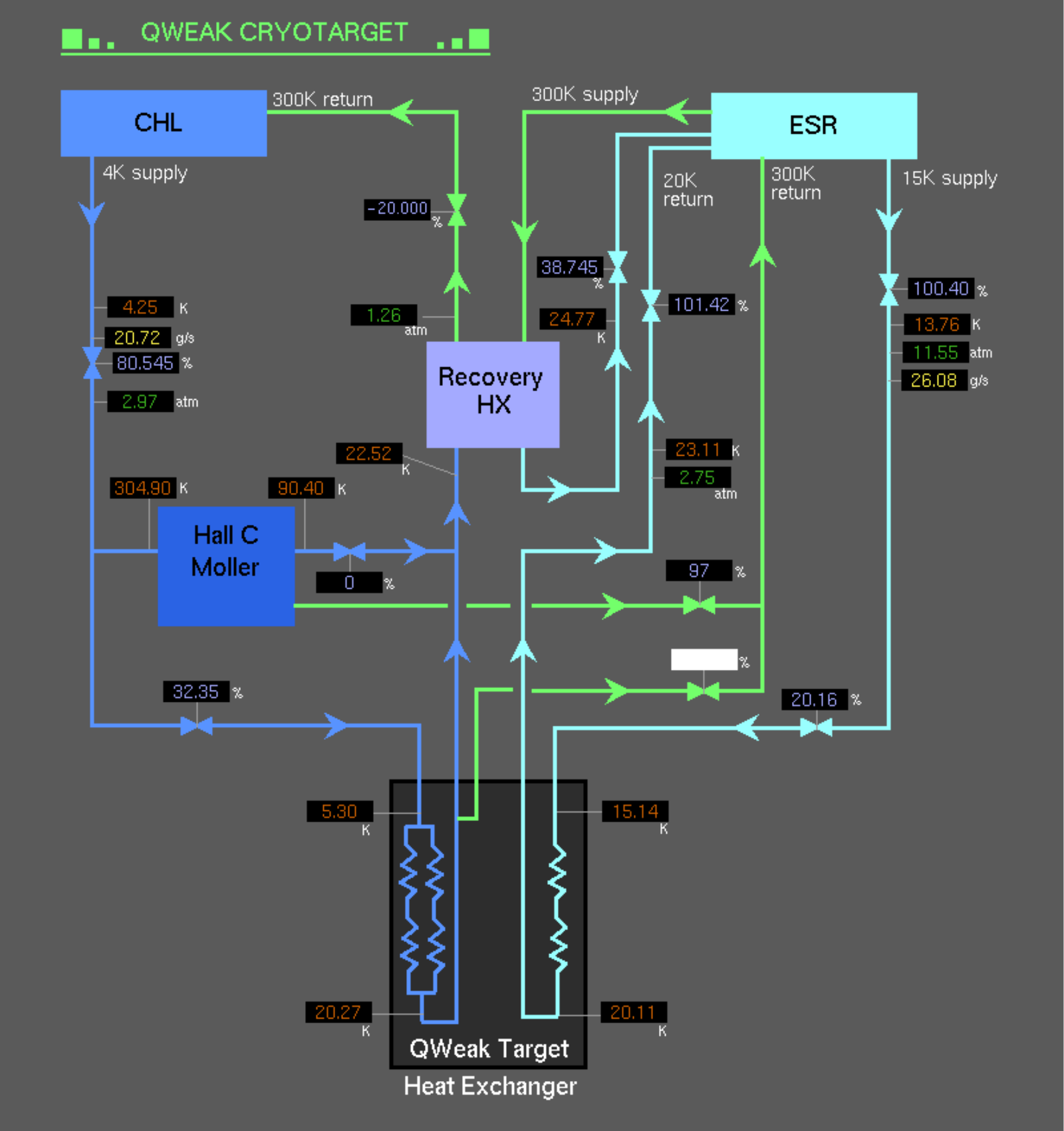}
\caption[ESR]{\label{fig:ESR} Schematic showing the unique configuration of the End Station Refrigerator (ESR) for the Qweak target, taking advantage of both 4K and 15K coolant supplies and reducing wasted enthalpy with a novel recovery heat exchanger. }
\end{figure*}

To further improve the available cooling power for this experiment, a new heat exchanger (HX) was put in place at the ESR which essentially used the remaining enthalpy of the returning  coolant supplied by the CHL to pre-cool the helium being used for the high pressure 15~K supply. This modification doubled the capacity of the 15~K supply. Since the coolant supplied by the CHL had to be returned to the CHL at room temperature anyway, there was no downside to using the CHL return enthalpy for this purpose.
The combined cooling power from both the 15~K ESR and 4~K CHL refrigerators met the unprecedented 3 kW cooling power required for the target. A schematic showing the basic configuration of the CHL and ESR during the \Qweak experiment is shown in Fig.~\ref{fig:ESR}. The Hall C Moller bypass shown in that figure refers to the Moller polarimeter's superconducting magnet which was energized for most of the experiment.

\subsection{Heat Exchanger}
\label{sec:HX}

The purpose of the target's heat exchanger was to use helium coolant from the end station refrigerator to remove heat from the \LH.
The fact that the cooling power required for the target could only be achieved by combining all the 15~K cooling power available from the ESR with all the excess 4~K cooling power of the CHL led to the design of a novel hybrid heat exchanger (HX). 
Combining the 4~K and 15~K HXs into a single (hybrid) HX minimized space, H$_2$ volume, and pressure head loss.

\begin{figure}[!hhhtb] \centering
\includegraphics[width=0.68\textwidth]{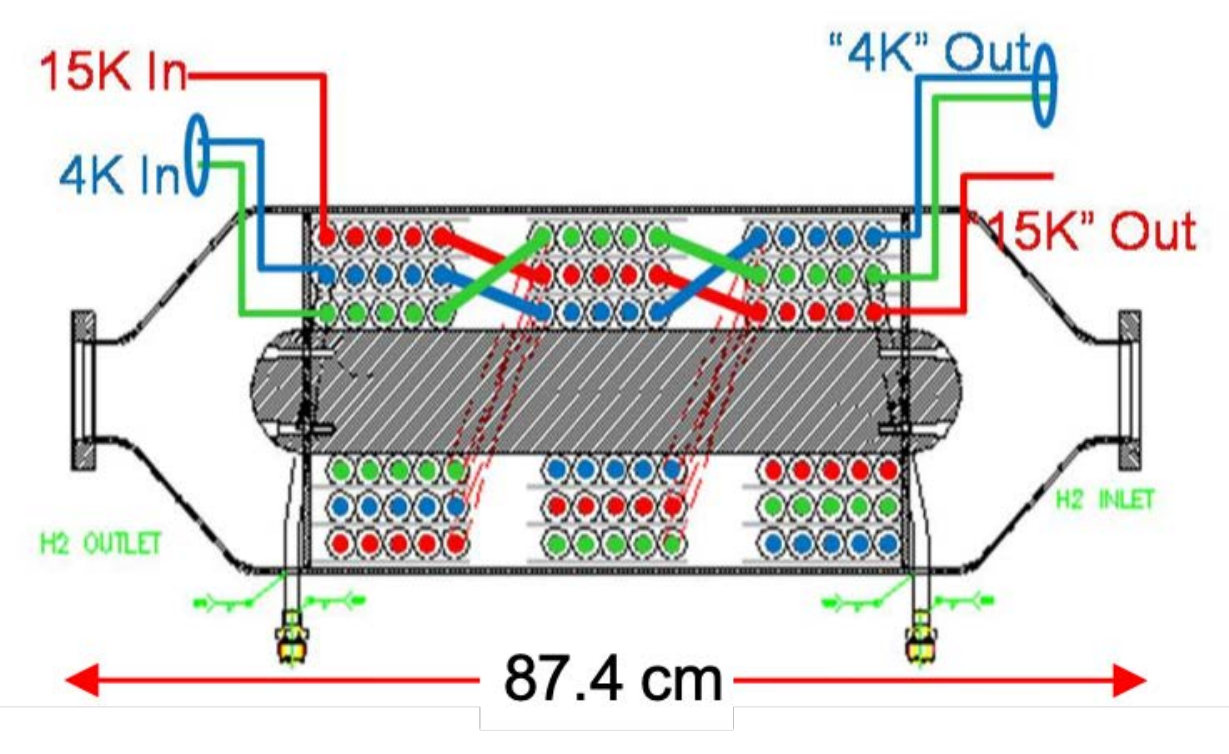}
\caption[HX]{\label{fig:HX} Basic CAD depiction of the hybrid 3 kW heat exchanger. The finned copper-tubing was wound along a cylindrical mandrel which diverted the \LH ~flow through the two 4~K and one 15~K parallel circuits in three sections of alternating radius.}
\end{figure}

This single 3 kW counterflow HX  employed 12.7 mm diameter copper fin-tube with 16 fins per inch. The 0.38 mm-thick fins extended 6.4 mm beyond the copper tube. The HX consisted of 3 adjacent sections, each with 3 radial layers, as depicted schematically in Fig.~\ref{fig:HX}. Each layer in each section was composed of 5 turns of copper fin-tube. Each layer was  separated by a thin perforated stainless-steel sheet. To minimize the pressure drop $\Delta P$ across the HX, no ``rope" was employed to fill the gap between turns as is sometimes done. The fin-tube connections between sections and layers were brazed together as shown in Fig.~\ref{fig:HX} to equalize the pressure drop across the HX for each of the 3 independent helium coolant circuits, two of which were connected in parallel at the inlet and outlet. In other words, each of the 3 fin-tube circuits consisted of an inner layer in one section, a middle layer in another section, and an outer layer in a different section. The 4~K coolant was fed through 2 of the 3 fin-tube circuits, and the 15~K coolant was fed to the third circuit.  A 9.2 cm diameter cylindrical aluminum mandrel occupied the volume inside the inner layer of fin tube. The entire fin-tube assembly was contained in a 27.3 cm-o.d.\ stainless-steel shell 3.4 mm thick and 70.6 cm long (not including the head assemblies at each end) through which the LH$_2$ flowed. The JLab-designed HX was assembled at an ASME shop~\cite{Meyer}.

The most important metrics in the design and operation of a HX are the head it presents and its cooling power. We also used CFD to determine that no local freezing of the \LH ~occurred in the 4~K section of the HX during normal equilibrium operation.

The head loss of the HX was calculated in the CFD model and was consistent with an independent estimate assuming a 15 l/s volume flow of 20~K LH$_2$  using the Darcy-Weisbach formula.  A velocity was obtained  from the volume flow by carefully  estimating the effective flow area  for each layer of fin tube.  The head obtained by this method (1 m) was combined with the head associated with the 7.6 cm $\leftrightarrow$ 27.3 cm abrupt transitions at the ends of the HX to arrive at the predicted overall 2.1 m head loss associated with the HX.

The predicted HX cooling power was studied with CFD and in the design phase by starting with the expression for the heat transfer rate for a HX:
\begin{equation} \label{eq:QisUdTLM}
    Q=U \Delta T_{LM}.
\end{equation}
The log mean temperature difference $\Delta T_{LM}$ for a counterflow HX is expressed in terms of the difference between the coolant and the \LH ~temperatures 
$\Delta T_o$ ($\Delta T_i$) at the outlet (inlet) to the HX:
\begin{equation} \label{eq:dTLM}
    \Delta T_{LM} = \frac{\Delta T_o - \Delta T_i}{\ln{\frac{\Delta T_o}{\Delta T_i}}}.
\end{equation}
The heat transfer coefficient $U$ contains a term to account for the convective heat transfer between the He coolant and the walls of the Cu fin tube, as well as a term to account for the convective heat transfer between the \LH ~and the Cu fin-tube walls. 
Ignoring the thermal resistance of the Cu fin-tube walls,  the overall heat transfer coefficient can be expressed in terms of the heat transfer rate per unit area $h_x$ and the corresponding effective area for heat exchange $A_x^{HX}$ for each fluid $x$  as follows:
\begin{equation} \label{eq:U}
    1/U=\left( \frac{1}{h_{LH} A_{HX}^{LH}} + \frac{1}{h_{He} A_{HX}^{He}}\right),
\end{equation}
where $x$ represents helium or \LH.
For the present case of turbulent flow ($\textrm{Re}(\textrm{LH}_2)\approx1.3\times 10^6$), $h_x$ is \cite{White} 
\begin{equation} \label{eq:h}
    h_x =\frac{0.023 \, C_p \, G^{0.8} \, \eta^{0.2}}{(Pr)^{0.6} \, (D_e)^{0.2} },
\end{equation}
where $C_p$ is the specific heat, $G$ the mass flow rate per unit area ($G=\dot{m}/A_{flow}$), $\eta$ is the viscosity, $D_e$ the effective HX area ($D_e=4$ (tube area)/(heat transfer surface perimeter)), $Pr$ is the Prandtl number $(Pr = \eta C_p / \lambda)$, and finally $\lambda$ is the thermal conductivity.
	
The geometry of the \Qweak HX is summarized in Table~\ref{tab:HXgeometry}, along with the calculated effective areas for heat exchange. The thermodynamic properties of Hydrogen and Helium needed for the coolingpower calculations are listed in Table~\ref{tab:H_He_Properties}.

\begin{table}[!hhtb]
\centering
\begin{tabular}{ r  c  l  }
\toprule
Property & Value & Units          \\ \midrule
Mandrel od	&	3.625 &	in \\
Fin Height	&	0.25 & in \\
Fin Tube diam &		0.5 &	in \\
Spacer thickness &	0.063 &	in \\
Fin thickness &	0.015 & in  \\
Fin pitch	&	16	& fins per inch\\
\# turns		& 5	& turns \\
\# layers	&	3	& layers \\
\# sections	&	3	& sections \\
\LH ~Volume flow	& 15	& l/s \\
Fin Tube Thickness	&	0.035 &	in \\
Total Fin Tube Length 	&	23.79 & m \\ \hline		
Eff. LH2 HX Area &	12.34	& m$^2$ \\
Eff. He HX Area	&	0.816 &	m$^2$ \\
He Flow Area	& 0.94	& cm$^2$ \\
LH2 Flow Area	& 121.6 & cm$^2$ \\ \bottomrule
\end{tabular}
\caption[HX Geometry]{The geometry of  the fin-tube heat exchanger. The Effective HX areas in the table are for all 3 sections. In practice, 1/3 of total Eff. HX areas were used for the 15~K coolant, and 2/3 were used for the 4~K coolant.
}
\label{tab:HXgeometry}
\end{table}

\begin{table}[!hhtb]
\centering
\begin{tabular}{ r  c  c  c  c  l }
\toprule
  &  &  \LH & 15K Coolant & 4K Coolant & \\
 Property & Symbol & Value & Value & Value & Units  \\ \midrule
pressure	 & 	$P$	 & 	35	 & 	175	 & 	22	 & 	psi	 \\ 
temperature	 & 	$T$	 & 	20	 & 	15	 & 	5	 & 	K	 \\ 
density	 & 	$\rho$	 & 	71.3	 & 	49.8	 & 	12.59	 & 	kg/m$^3$	 \\ 
mass flow	 & 	$\dot{m}$	 & 	1.1	 & 	0.0172	 & 	0.0125($\times 2$)	 & 	kg/s	 \\ 
specific heat	 & 	$C_p$	 & 	9384	 & 	5384	 & 	5751	 & 	J/kg-K	 \\ 
viscosity 	 & 	$\eta$	 & 	1.40E-05	 & 	3.96E-06	 & 	2.60E-06	 & 	kg/m-s	 \\ 
thermal conductivity	 & 	$\lambda$	 & 	0.1008	 & 	0.030	 & 	0.018	 & 	W/m-K	 \\ 
Prandtl \# 	 & 	$Pr$	 & 	1.30	 & 	0.7107	 & 	0.8171	 & 		 \\ 
Flow Area	 & 	$A_{\rm flow}$	 & 	121.6	 & 	0.94	 & 	0.94	 & 	cm$^2$	 \\ 
$\dot{m}$/$A_{\rm flow}$	 & 	$G$	 & 	90.45	 & 	183.26	 & 	133.42	 & 	kg/m$^2$-s	 \\ 
\bottomrule
\end{tabular}
\caption[Thermodynamic Properties]{Thermodynamic properties of \LH, the 4~K Helium coolant, and the 15~K helium coolant relevant for the heat exchanger cooling power estimate. Some coolant properties are averages over the pressure and temperature range of each coolant supply. The 25 g/s total 4~K coolant mass flow is split in half in the table to reflect the fact that it was split into two identical layers of the HX (the third of the three layers was used for the 15~K coolant).
}
\label{tab:H_He_Properties}
\end{table}
 
The actual cooling power prediction is now straightforward using Tables~\ref{tab:HXgeometry} and \ref{tab:H_He_Properties} in
Eqn.'s~\ref{eq:QisUdTLM}-\ref{eq:h}. The result is presented in Table~\ref{tab:HXCoolingPower}. In Table~\ref{tab:HXCoolingPower} the hydrogen inlet temperature is set to the operating/outlet \LH ~temperature of 20~K plus the 0.24~K temperature rise expected from a 2.5 kW heat load using Eq.~\ref{eq:QismdotCpdT}. The helium coolant inlet temperature is chosen as 15 or 5~K for the two coolant sources in the hybrid HX. The cooling power result is quite sensitive to the coolant outlet temperature $T^{He}_o$ chosen in the calculation. This temperature cannot exceed the hydrogen outlet temperature of 20 K, and the calculation is at its most conservative if this value is chosen for the helium outlet temperature, as presented in Table~\ref{tab:HXCoolingPower}. With a less aggressive choice of 19~K for $T^{He}_o$, the predicted total cooling power rises from 3066 W to 4864 W. In any case the predicted HX performance seemed capable of serving the requirements of the \Qweak experiment.

\begin{table}[!hhtb]
\centering
\begin{tabular}{ r  c  c  c  l  }
\toprule
  &  &   15K layer & 4K layer & \\
 Property & Symbol  & Value & Value & Units  \\ \midrule 
\LH ~inlet temperature & $T^{LH}_i$	 & 	20.24 & 	20.24 & K	 \\ 
\LH ~outlet temperature	 & 	$T^{LH}_o$	 & 	20	 & 	20	 & 	K	 \\ 
He inlet temperature	 & 	$T^{He}_i$	 & 	15	 & 	5	 & K \\ 
He outlet temperature	 & 	$T^{He}_o$	 & 	20	 & 	20	 & 	K \\ 
Log mean temperature difference  & 	$\Delta T_{LM}$	 & 	1.57	 & 	3.57	 & 	K	 \\ 
He heat transfer rate/area	 & 	$h_{\rm He}$	 & 	2014	 & 	1411	 & 	W/m$^2$-K	 \\ 
\LH ~heat transfer rate/area	 & 	$h_{\rm LH}$	 & 	1786	 & 	1786	 & 	W/m$^2$-K	 \\ 
heat transfer coefficient & 	$U$	 & 	510	 & 	365	 & 	W/K	 \\ 
efficiency estimate	 & 	effi	 & 	90\%	 & 	90\%	 & 		 \\ 
HX cooling power/layer	 & 	${Q^{\rm {eff}}}$	 & 	721	 & 	1172	 & 	W	 \\ 
Cooling power both 4K layers	 & 	$Q{\rm{_{4K}}}$	 & 	-	 & 	2345	 & 	W	 \\ \midrule
Total HX cooling power	 & 	${Q\rm{^{HX}_{tot}}}$	 & \multicolumn{2}{c}{3066}		 	 & 	W	 \\ 
\bottomrule
\end{tabular}
\caption[HX Cooling Power]{Predicted cooling power  for the \Qweak counterflow HX.
}
\label{tab:HXCoolingPower}
\end{table}

During the experiment, the HX performed well and handled total heat loads as high as 3.2 kW. The only operational difficulties  had to do with the tendency to start making hydrogen slush (partially frozen hydrogen)  during cooldown, due to the use of 4~K coolant in the HX. This was dealt with by adding a resistive heater (described in Sec.~\ref{sec:aux}) to the 4~K supply line in the scattering chamber, to more quickly and forcefully react to sudden drops in the 4~K return temperature during the infrequent $\approx$ 8-hour-long cooldowns required to condense the hydrogen. This was preferable to closing the 4~K supply valve, which had a negative impact on the ESR as well as the coolant transfer line. 

\subsection{Heater}
\label{sec:Heater}
The High Power Heater (HPH) was used to replace the beam heat load when the beam was off, as well as to regulate the loop temperature within about 10 mK.  As with the other target components, both CFD as well as analytical tools were used to design the HPH. 

The heater was initially powered by a Sorensen 3 kW 60 VDC power supply requiring an ideal resistive load of 1.2 $\Omega$.  Unfortunately, the total resistance of the heater with power leads was about $R_H = 1.33\ \Omega$.  During the second half of the experiment, when we were operating with the maximum beam current, the heater power plateaued around 2700 W with the 3 kW power supply.   During beam trips at these conditions, we required more dynamic range to minimize the temperature oscillations, hence we replaced the 3 kW power supply with a 4 kW 80 VDC power supply. 

\begin{figure}[!hhhtb] \centering
\includegraphics[width=0.5\textwidth,angle=0]{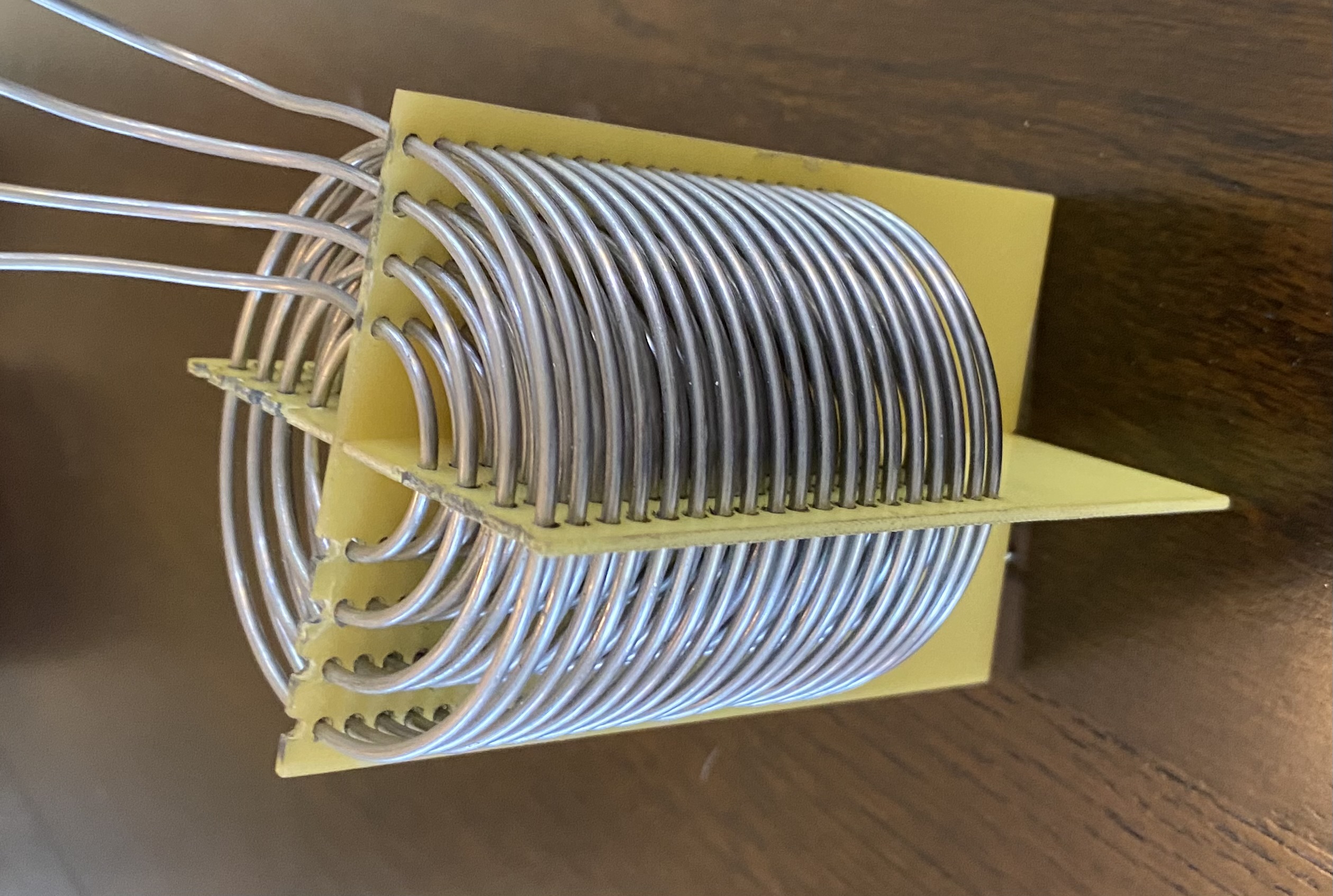} 
\includegraphics[width=0.75\textwidth,angle=0]{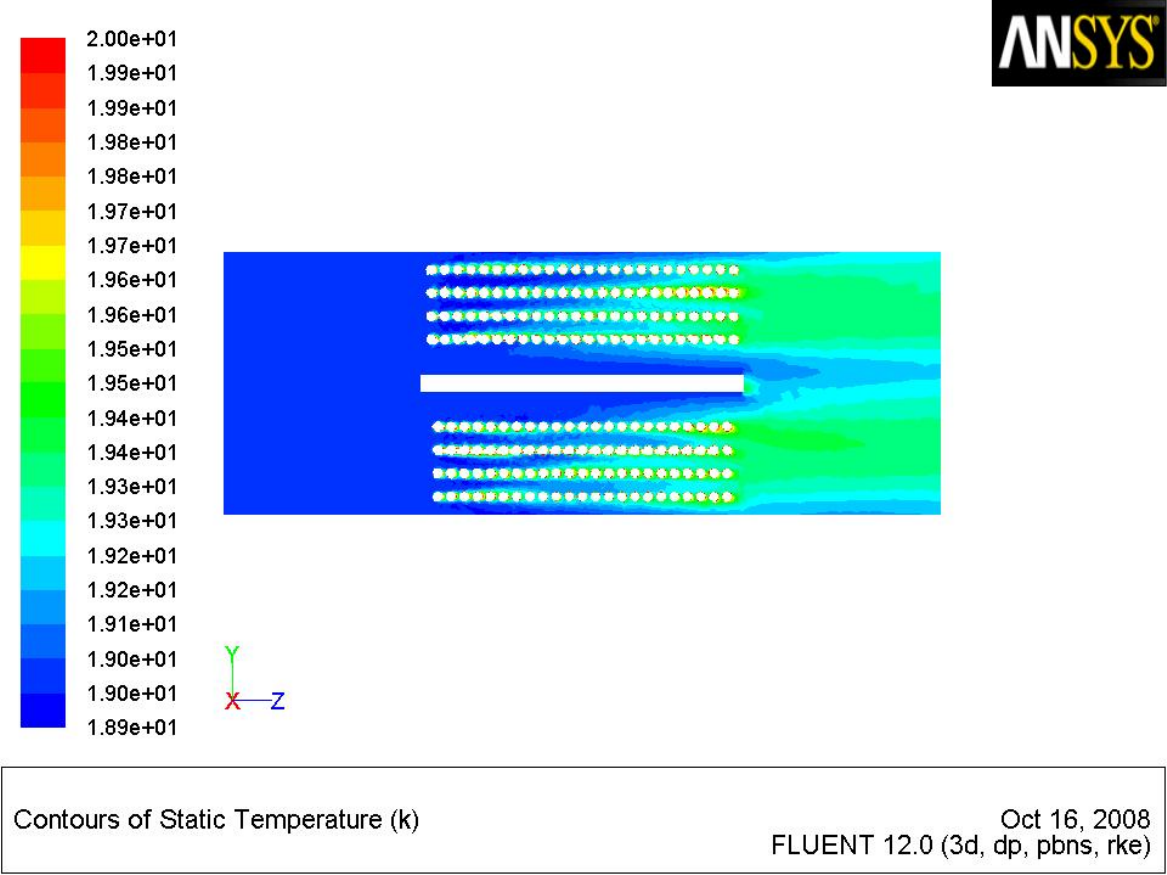}
\caption[Heater]{\label{fig:Heater} The upper figure shows a photo of the four-layer concentric heater wound onto a crossed G10 frame before insertion into a dedicated $3''$ diameter spool piece. LH2 flowed along the axis of the NiCr-A windings as evident in the lower figure, which shows a CFD simulation of the equilibrium LH$_2$ temperature in degrees K.}
\end{figure}

The heater consisted of four layers of 13 AWG Nichrome wire wrapped through holes in crossed G10 boards 1.59 mm thick.  The heater resided in a 27.94 cm long section of 7.62 cm loop pipe with conflat flanges.  Heat transfer calculations were done assuming one can treat the heater as an array of in-line cylinders or tubes in a crossflow ~\cite{Zuk}.  CFD simulations were performed to confirm these calculations. The wound heater and the CFD simulation is shown in Fig.~\ref{fig:Heater}.
The wire had a diameter $D=1.83$ mm, 
resistance per 
meter of 0.420 $\Omega$/m, and total length of about 11.5 m.  With a heat load of 2500 W, the calculations required about 8 meters of wire to keep the surface temperature below 23.6~K (boiling point for hydrogen at 35 psi), the extra length provided a safety margin and added resistance to get closer to the optimal resistance.  

Of course, when the beam was on and the experiment was acquiring data, the heat load from the heater dropped to a few hundred Watts. Strictly speaking, it was only necessary to avoid boiling during these less demanding conditions,  but the heater was designed to avoid boiling when the beam was off and the heater was on at full power. 

The longitudinal spacing between the rows was $X_l =2D$ and there were 23 rows.  The transverse spacing was $X_t = 3.6 D$.  Layers 1 and 4 were connected in series as were layers 2 and 3, providing two segments of wire of roughly equal length.  These two segments were then connected in parallel to produce the proper resistance.  This resistance was determined from the current versus voltage data taken while the heater was submerged in a bath of liquid nitrogen and was 1.3 $\Omega$.   

An inline array of wires was used rather than a staggered array to minimize the pressure drop through the heater.  The pressure drop was calculated with CFD, and found to be in agreement with analytic estimates using 
\[\Delta P = \frac{f N_L G_{max}^2}{2 g_c \rho_{H2}} \approx 1.86 \; \mbox{kPa} 
\]
where $f$ is the friction factor, $G_{max} = U_{max}\rho_{h2}$, $N_L$ is the number of rows in the heater, and $g_c = 1\ \mbox{kg}\cdot \mbox{m}/\mbox{N}\cdot \mbox{s}^2$.  The friction factor~\cite{Jakob} has the form:
\[f=\left[0.176 +0.32\frac{X_l}{D}\left(\frac{X_t}{D}-1\right)^{-n_f}\right]Re_f^{-0.15}\]
where the exponent is $n_f=0.43+1.13D/X_l$.  For our flow speeds, the friction factor had a value of 0.084. The Reynolds number was evaluated at the maximum average flow velocity of the fluid, $U_{max}$, and has the form
\[Re_f= \frac{ U_{max} {\rho_{h2}}_f D}{\mu_f}\] where $\mu_f$ is the viscosity of the hydrogen and $\rho_{h2}$ is the density.  For our geometry, $U_{max} = 5.06$ m/s and the Reynolds number was about $4.8 \times 10^4$. 

\subsection{Circulation Pump}
The purpose of the pump is to circulate \LH ~around the target loop, which contains elements that add heat (the heater and the beam) as well as elements that remove heat (the heat exchanger). In general, pumping \LH ~faster across the beam axis reduces heating from the beam and mitigates boiling, but also results in increased  heating from friction with the loop surfaces. In most cryotargets, the recirculation pump is the component most prone to failure.

A custom LH$_2$ recirculation pump was built at Jefferson Lab. The required pump head  was determined by adding the head from the target loop and all its components. The capacity was determined by scaling up the performance of the G0 target, as described in Sec.~\ref{sec:Performance_Scaling}. The design head $H=11.4$~m  (\LH) and capacity $Q=0.015$~m$^3$/s at the nominal 30 Hz shaft rotation determines the dimensionless specific speed $$\Omega_s = \frac{N(rpm) \sqrt{Q(m^3/s)}}{52.9 \left[ H(m) \right]^{3/4}} = 0.671$$ ($N_s=1835$ in US units). This suggests a centrifugal pump geometry capable of providing a large head and moderate capacity~\cite{avallone1996marks}. Since 2-axis motion was a design requirement, an in-line, submersible pump design was chosen over one with an external motor. 

\subsubsection{Required Head and Capacity}
\label{sec:Pump HQ}

Assuming a capacity of 15 l/s, the head associated with each of the major elements of the loop was determined as described above in Sec.'s~\ref{sec:Cell_Design}, \ref{sec:HX}, \& \ref{sec:Heater}. 
The head associated with the heat exchanger (2.1 m) 
and the heater (3.0 m)  
was calculated analytically and checked using CFD simulations.  The head associated with the detailed cell design was obtained from CFD simulation alone (2.5 m). 
The head associated with the loop plumbing (straight pipe, flex hose, elbows, enlargement and contraction of the piping where required) was calculated analytically (3.8 m)~\cite{Crane}.  
The total estimated head was 11.4 m 
for the entire loop.

Larger capacity results in faster fluid flow across the beam axis, less density reduction, and smaller target noise~\cite{G0tgt}. However, there is a practical constraint imposed by the frictional heat of the fluid in the system competing against the finite cooling power available for the target.  This viscous heating is given by 
\begin{equation}
P_{\rm{viscous}}= 0.72 \; \rm{Capacity(l/s) \; Head(m)}\, /\, \epsilon,
\label{eq:viscous}
\end{equation}
where the pump efficiency $\epsilon$ is estimated to be 72\%~\cite{Tuzson}. Since the capacity is proportional to the fluid velocity, and the head is proportional to the square of the velocity, the frictional heating increases with the velocity cubed. A design constraint imposed on the pump was to limit the viscous heating to  $\leq$10\% of the beam power deposited along the length of the target due to ionization energy loss of the beam  (2.1 kW). With an 11.4 m head and 15 l/s capacity, $P_{\rm{viscous}} = 177$~W.  The nominal pipe diameter in the target loop was chosen to be $3''$ in order to slow down the fluid velocity to 3.3 m/s and meet the viscous heating constraint.

The expected torque can be estimated from $$\tau(Nm) = V_s \Delta P/(2 \pi)=168  \textrm{ oz-in,}$$  assuming a 60\% pump efficiency, where $V_s$, the volume displacement per revolution, is 1/2 l for a 30 Hz pump with a 15 l/s capacity, and the head $\Delta P$ is 1.3 psi.

The system pressure ($P=220$ kPa) was chosen to be well above the parahydrogen vapor pressure~\cite{VP}  ($P_{vp}$=94 kPa at 20 K) in order to mitigate cavitation. The net positive suction head
$$ \textrm{NPSH} = \frac{P}{\rho g} + \frac{v^2}{2g} - \frac{P_{vp}}{\rho g} =175 \; \rm{m}.$$

\subsubsection{Pump Fabrication}
\label{sec:Pump Design}

\begin{figure}[!ht] \centering
\includegraphics[width=0.45\textwidth]{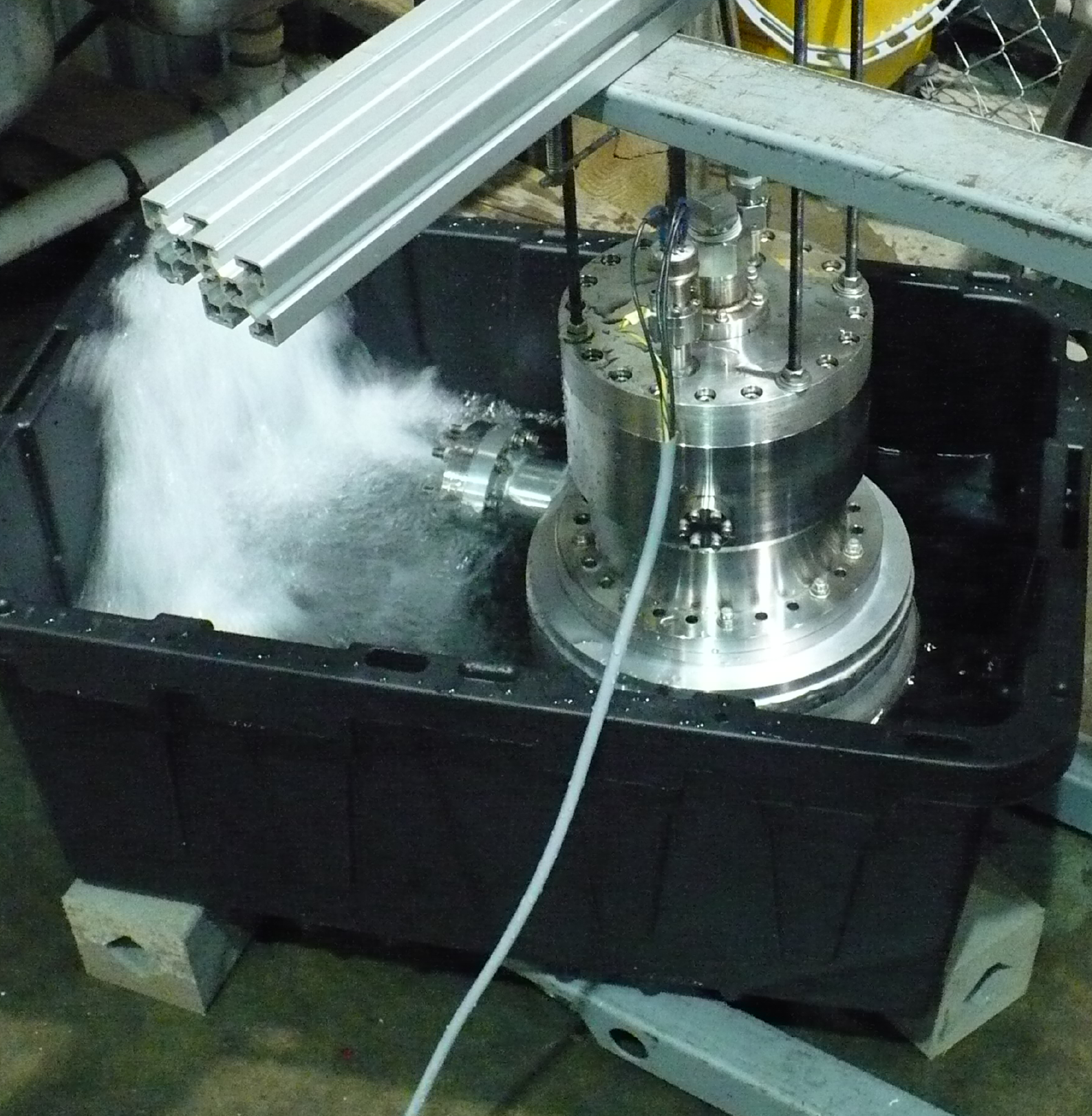}
\includegraphics[width=0.45\textwidth]{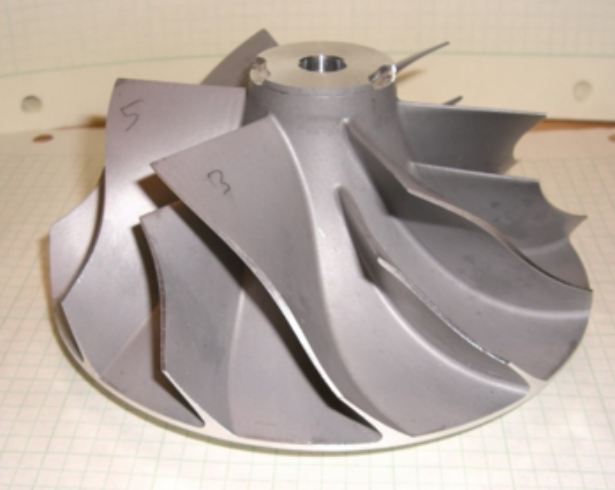}
\caption[Pump Components]{\label{fig:impeller} (Left) The pump is shown being tested in a water bath. The 1 hp pump motor is on top. The pump volute is below the motor, with the suction side submerged. (Right) The impeller used for the centrifugal \LH ~circulation pump is shown. The impeller radius was 7 cm, the height was 6.5 cm, and there were 12 blades. }
\end{figure}

The pump was adapted from a commercial (Garrett Motion, Inc.) A356.0 cast aluminum automotive turbocharger (see Fig.~\ref{fig:impeller}). Conflat flanges (Al 2219-T851) were welded to the pump volute to connect to the target loop. The inner diameter of the flanges was 14.0 cm (outlet) and  7.3 cm (inlet). A third flange on top of the volute with a 14.9 cm i.d.\ was used for the motor housing. The pump and motor assembly was 46.5 cm high and 27.0 cm in diameter not including the outlet flange. The impeller was custom cast and balanced for our application by Turbonetics, Inc.

The most relevant features of the impeller geometry are the outlet radius ($r_2$=7.1 cm), the height at the outlet radius ($h_2$=1.4 cm), the inlet radius ($r_1$=5.4 cm),  the angle of the impeller blades to the tangent of the outer/inner impeller circumference ($\beta_2$=50$^\circ$/$\beta_1$=5$^\circ$), and the number of blades (Z=12). Using the expected capacity ($Q=0.015$ m$^3$/s) and rotational speed (30 Hz or $\omega$=188 radians/s), the slip $$\sigma=1 - \frac{\sqrt{\sin{(90^\circ - \beta_2)}}}{Z^{0.70}}=0.86,$$ and the expected head 
$$ H = \frac{\sigma (\omega r_2)^2 - r_2 \omega Q \tan{(90^\circ - \beta_2)} }{2 \pi g r_2 b_2 } = 12.5 \; \rm{m}$$
were predicted~\cite{Tuzson}, where $g=9.8$ m/s$^2$.

The pump efficiency was not directly measured. It was estimated~\cite{PumpHandbook} from the measured capacity, shaft speed, specific speed, and estimated surface roughness of the loop:

\begin{equation}
 { \eta = 0.94 - 0.08955 \left[ \frac{Q({\rm{gpm)}}}{N({\rm{rpm}})} \left( \frac{3.56}{\epsilon(\mu {\rm{m}})} \right)^2      \right]^{-0.21333} - 0.29\left[ \log\left( {\frac{2286}{N_s}} \right) \right]^2  }.
\end{equation}
\label{eq:pumpeffi}

With $\epsilon \sim 10 \; \mu$m, $Q=230$ gpm, $N=1800$ rpm, and $N_s$=1835, the pump efficiency was expected to be 72\%.

The pump motor was a nominally room temperature 1 hp AC induction inverter duty 230 V, 2.8 A explosion-proof Baldor motor. The commercial motor housing was replaced with a stainless steel 316L custom housing to adapt it to the pump volute. The pump shaft drove the pump impeller on one end, and a small tachometer magnet on the other. 
The motor shaft was slightly resized to accommodate  15.9 mm \& 22.2 mm
diameter cryogenic bearings with stainless steel balls and races, a vespel retainer and molybdenum disulfide dry lubricant. 
The motor was controlled with an Elite microsystems drive controller~\cite{elite}.  A network of high power resistors was employed between the motor and the controller to provide a load when the motor was cold. These were optimized during tests of the pump in water (see Fig.~\ref{fig:impeller}) and full immersion tests in liquid nitrogen prior to installation of the pump in the target loop. The LN$_2$ tests led to adjustments in the impeller-volute clearance, minor bearing problems, and controller problems which were solved with the resistor network. Eventually the pump was declared ready after running in an open loop at 45 Hz in LN$_2$, which has 11 times the density of LH$_2$.

To help keep the nominally 750 W motor cold, several turns of 6.4 mm diameter copper tube were wrapped around the outside of the pump motor housing.  This tubing carried 20~K helium coolant returning to the ESR from the 4K helium supply circuit in the target heat exchanger. To obtain a rough estimate the 20~K helium mass flow in these windings, we scaled the overall 16.6 g/s 4K supply mass flow to the target by the ratio of the cross-sectional areas of the pump motor tubing and the two layers of 4~K fin-tube in the hydrogen HX to obtain an estimated 0.8 g/s. This small helium flow was returned to the ESR via the existing ESR warm return as shown in Fig.~\ref{fig:ESR}.

Unfortunately, because the pump motor housing was stainless steel, this technique was not very effective at removing heat from the pump motor. The Baldor motor was positioned at the high point of the \LH ~loop.  This essentially isolated the motor housing providing a conducive environment for vapor-lock to occur.  Without direct cooling from \LH ~or LHe, the bearings overheated causing the race to fail and ultimately the failure of the bearing assemblies. 

In response to this setback, two changes were made. First, new (440C) stainless steel races with Si$_3$N$_4$ ceramic balls and a torlon retainer  were used, with tungsten disulfide, a dry lubricant. These bearings provided a less effective seal between the pump motor and the volute, which was useful for the second improvement: a short $1/4''$ bypass tube was added from the top of the motor housing to the suction side of the pump providing a circulation path and preventing H$_2$ vapor from collecting in the stator housing. 
The determination of the additional 150 W heat load associated with this hydrogen bypass is presented in Sec:~\ref{sec:cooling_power_budget}. The reduction in pump head was negligible. By scaling the target \LH ~flow by the ratio of the cross sectional areas of the $1/4''$ hydrogen bypass and the $3''$ pipe used for the target's recirculation loop, we estimate about 1 g/s of hydrogen was diverted through the bypass, out of the total 1100 g/s circulating in the target loop.
This was enough to keep the pump and the target operational for the remainder of the experiment.

\subsection{Auxiliary Systems}
\label{sec:aux}

Several auxiliary systems were implemented to improve the operational performance and safety of the target. The most important of these was a 500 W resistive heater clamped to the 4~K helium supply line in the scattering chamber, just before the HX. This heater consisted of a nichrome ribbon sandwiched between 2 layers of kapton, clamped to the 4~K supply pipe with large copper blocks. The 4~K  heater was connected to a Power Ten DC power supply which was controlled by a feedback (PID) loop using one of the LH$_2$ thermometers as input. If the LH$_2$ temperature fell below a threshold value (typically 17 K) then up to 500 W of power was automatically applied to the 4~K heater to arrest the fall in LH$_2$ temperature. 
This heater was especially useful during  cooldown of the target, before the H$_2$ was condensed, to help prevent H$_2$ ice from forming on the 4~K sections of the HX. However it also proved useful on several occasions when compressor trips in the ESR resulted in  sudden drops in the nominally 15~K coolant temperature, which had the potential to lead to a dangerous freezing of the H$_2$.

Another system was put in place to help with the difficult cooldown of the target. This consisted of a 4~K bypass valve in an external cold box which could be used to shunt the 4~K coolant supply to its return path prior to the target. This facilitated greater 4~K coolant flow on the supply side, essential to cooling the transfer lines, without overwhelming the target's 4~K HX section during cooldown.

Another 4~K PID feedback loop acted as deep fallback to prevent the hydrogen from freezing in the target during off-normal events. Although a trained target operator was always present when the target was condensed, this 4~K PID loop was meant to act if the target operator did not.  A temperature sensor in the LH$_2$ flow path provided the input to a PID loop controlling the 4~K  Joule-Thompson (JT) supply valve. If the target temperature fell below 15 K, the PID would automatically step the 4~K supply valve closed.

Finally, the 2-axis motion system for positioning the target on the beam axis relied on glides and slides that were lubricated with vacuum grease. The temperature of the motion system therefore had to be maintained near 300~K  in the scattering chamber vacuum, and was monitored with platinum resistors. To overcome the thermal conduction from the 20~K target to the motion system, another Power Ten DC power supply was used to supply $\sim$40 W to resistive heaters clamped to the motion system assembly.

\subsection{Motion System}

A motion system was implemented in order to position the LH$_2$ target on and off the beam axis, to study and determine the experiment's optimum neutral axis, as well as to position a large number of solid targets on the beam for background measurements and diagnostics.

To set the initial pitch, roll, yaw, and position along the beam line, a ``cell adjuster" was employed which facilitated the positioning of the  target cell and solid target ladder  by hand over a limited range when the $\sim 61$ cm diameter scattering chamber access ports were removed (with the target at STP). 
Use of flex hose~\cite{boa} to connect the cell to the  loop allowed these adjustments to be made independently of the rest of the loop. 
A laser tracker was used to determine the target's coordinates from pre-fiducialized tooling ball locations on the LH$_2$ cell as well as the solid target ladder, in conjunction with long-established survey points in the experimental hall. The cell adjuster was tweaked in an iterative process to achieve the desired results. 

A dynamic 2-axis motion system was built to remotely position the target  on the beam axis vertically and horizontally while the target was cold via the following basic arrangement: The shaft of a precision linear actuator, from Danaher Motion, penetrated through the top plate of the scattering chamber via a differentially-pumped sliding vacuum seal. This actuator shaft attached to a horizontal stainless steel  plate 
which was fixed at both ends to vertical guide rails. The plate was thus constrained horizontally, and could only move vertically as the electric cylinder was extended or retracted. A second stainless-steel plate was hung from three guide rails which were affixed to the bottom of the first plate. These 3 horizontal rails were oriented perpendicular to the beam axis. The lower plate was welded to the top of a 1.57 m long, 20.3 cm diameter, 3.2 mm thick  stainless steel pipe which supported the target loop at its lower end. When the lower plate was moved horizontally, it carried the target with it horizontally, and when the upper plate was lifted, it carried the lower plate and the target with it vertically. The heat lost to the environment was only $Q=A/l\int_{20\,K}^{300\,K} k\, dT = 4 W$, where $l$ and $A$ denote the length and cross-sectional area of the pipe.

\subsubsection{Vertical Motion System}

The vertical motion system was formed from two vertical THK LM guide rails, model number HSR85-A of length 99 cm, which supported a horizontal stainless-steel plate between them.
These rails along with a pair of vertical steel I-beams were connected at their ends to a top and bottom ring (see Figure 31).  The top ring was fixed to the bottom of the top plate of the scattering chamber.  The rails were packed with vacuum grease.  

The vertical motion was achieved using a Thomson TC5 series electric cylinder (TC5-T41V-100-10B-600-MF1-FS2-B) with a T41V stepper motor from Danaher.  This actuator has a 600 mm stroke with a 24 VDC brake on the ball screw, a thrust load capacity of 25,000 N, and a quoted repeatability of $\pm 0.013$ mm.  The ball screw has a pitch of 10 mm/revolution  and a 10:1 gear reduction. 
To minimize any side loading of the cylinder, it was connected to the upper stainless-steel plate via a sliding horizontal disk riding on ball bearings packed with molybdenum disulfide vacuum lubricant. 
 
The electric cylinder was positioned above the center of the top plate of the scattering chamber (in air). The actuator shaft penetrated the scattering chamber via a sliding seal fitted to the top plate.  The sliding seal had two pairs of O-rings  and was differentially pumped between the O-ring pair.  The actuator was controlled by an IDC S6961 stepper motor drive controller which employed two pairs of end-of-travel limit switches and a home switch.  In addition to the limit switches, each guide rail was fitted with hard stops at the top and bottom, set at the extreme limits of travel.

\subsubsection{Horizontal Motion System}

As mentioned in the introduction to this subsection, the plate hung from the long stainless pipe supported two rails to allow $\pm 5$ cm of  horizontal motion transverse to the beam direction.  Two THK LM guide rails, model number HSR35-M1A of length 34.3 cm were attached underneath the table. 
Each rail had two blocks which attached to the plate.  These rails had a basic load rating of 37.1 kN dynamic and 61.1 kN static. 
In addition to the rails, there was a THK LM guide actuator model number KR46 with a 10 mm lead on the ball screw to move the plate.  The ball screw was attached to a $90^\circ$, 10:1 gear box which had been repacked with vacuum grease.  

The gearbox was attached to 
a Phytron VSS-UHVC Cryo stepper motor.  It was designed to operate in an ultra-high vacuum environment.   A 24 VDC brake was attached to the ball screw and there were also end-of-travel limit switches and a home switch.  The Phytron motor was controlled with the IDC S6961 drive. 

\subsection{Scattering Chamber}

The \Qweak scattering chamber  contained and supported the cold target loop in an insulating vacuum. It was composed of a rectangular lower half, a cylindrical upper half, and a short transition piece in between. The upper and lower pieces were reused from previous experiments. The chamber was about 3.3 m high, with an inside width of 81 cm along the beam axis. The vacuum in the scattering chamber was typically around $8 \times 10^{-7}$ Torr when hydrogen was condensed in the target.

The electron beam passed through  large 51 cm ports on the lower  half of the chamber. 
The upstream flange was equipped with a fast-acting gate valve (GV). The downstream flange was equipped with a custom made, explosion proof, all Aluminum 40.6 cm diameter extended stroke GV with a 5 s closing time. 
The extended stroke was used to retract the gate from the small angle scattering region in order to improve the lifetime \cite{Lee} of the ethylene propylene diene (EPDM) seals on the gate. Lead shielding provided in the region where the gate sat when retracted  further improved the lifetime, according to simulations. 
The beamline flanges were equipped with metal o-rings. Both valves were vacuum-interlocked.

The  41 cm GV was  closed whenever personnel were in the hall and the target hydrogen was condensed. The scattering chamber window was downstream of the GV- thus when the GV was closed, the target effectively had no thin windows. This improved personnel safety in the hall. 

When the GV was open during data-taking, all the scattered electrons which fell into the acceptance of the experiment passed through the open throat of the GV and through eight 0.89 mm-thick Aluminum 2024-T4 vacuum windows arrayed in a spoked, wagon-wheel configuration (matching the experiment's acceptance) downstream of the GV. 
The ultimate tensile strength (UT) of this material is 469 MPa- the window design is allowed to go to 50\% of this value.  
Finite element analysis calculations predicted the stress in the window is 186 MPa  
when the differential pressure is 1 atmosphere in either direction, only 80\% of the allowable stress. 

Although nominally a vacuum window, the window was designed to withstand this stress in either direction, since in the event of a cell rupture the pressure inside the scattering chamber could go as high as 198 kPa.

Another integral part of the scattering chamber was the dump tank, 
a 4013 liter steel tank connected to the scattering chamber via a short length of 15.2 cm diameter pipe. Although equipped with its own vacuum pump, it was part of the same vacuum system as the scattering chamber. 
The dump tank was meant to mitigate the pressure rise from the isothermal liquid-gas phase transition that would take place in the event of a target cell rupture. In that accident scenario, 
the \LH ~would suddenly find itself in the vacuum of the scattering chamber. The transition from liquid to gas and corresponding pressure rise could happen too quickly for the vent system to handle. So we assumed this transition is instantaneous, and provided enough passive volume to handle the pressure rise associated with the phase transition, keeping the system  below half an atmosphere. 
The vent system could then handle the relatively slow pressure rise associated with the warming of the vapor due to convective heat transfer with the walls of the system.

\subsection{Gas Handling System}
The Hydrogen gas connections were made on either side of the pump. The pump head is the measured differential pressure between these (divided by the specific gravity). At the outlet side of the pump, a $1.5''$ flex line was connected to a feedthrough on the top plate of the scattering chamber which led back to the target gas panel via  $1''$ tube. This was the target supply line. On the suction side of the pump, between the pump and the HX at the top corner of the loop, a $3''$ 
tee provided a cold $3''$ 
relief tube to the outside of the top plate. From there a $2''$ tube was used to the gas panel. To accommodate the full $\approx 2''$ range of the target's vertical motion, there was a $180^\circ$ fitting midway along both the supply and return lines, such that the lines were mostly horizontal when the target was raised, and mostly vertical when the target was lowered. 
The top of the relief tube was warm, and connected through a short flex hose to hard piping leading back to the target gas panel, and on to the Hydrogen ballast tanks 
(22712 STP liters total) stored outdoors. The 4K and 15K helium coolant supply and return lines were implemented in a similar fashion. 

Whenever the target was condensed, the target H$_2$ gas supply and return were connected through a $2''$ check valve which allowed gas flow to the outdoor H$_2$ ballast tanks  located $220'$ away via $2''$ pipe.  A small $1/4''$ solenoid valve was kept open between the ballast tanks and the target to insure the pressure  in the tanks and the target was the same. When the target was being filled the $2''$ check valve was bypassed. 
When the target was at room temperature the pressure in the system was typically about 60  psia, and when condensed about 33 psia. Since the ballast tanks were outdoors, there was a diurnal pressure variation with the outdoor temperature of $\pm 1$ or 2 psi, and a slower response with the season. Because the \LH ~in the target can be considered an incompressible fluid, these changes in pressure have negligible effect on its density.

Independent primary and secondary relief paths were implemented.  The $2''$ primary relief path was inerted with 1 psig of helium to an elevated parallel plate relief valve located outdoors $150'$ away from the gas panel in Hall C. It was connected through check valves to the exhaust of the mechanical pump that served the gas panel, and the pumps that provided the insulating vacuum in the scattering chamber. It was also connected to the target's  $31''$ H$_2$ supply and $2''$ return through a 60 psig $2''$ relief valve in parallel with  25 psig $2''$ burst disk. A check valve separated this relief tree with the parallel plate relief outdoors.

The secondary containment for the H$_2$ in the target in the event of a cell rupture accident  was the scattering chamber, isolated from thin windows by the vacuum interlocks on the fast-acting gate valves upstream and downstream. A secondary relief system was therefore provided to deal with this kind of accident scenario. The 1060 gallon dump tank discussed earlier would limit the pressure rise associated with the H$_2$ phase transition from liquid to gas in the scattering chamber's former vacuum space. A $4''$ relief tree consisting of three $2''$ check valves and an 8 psig $4''$ rupture disk acted as this secondary relief. It connected the scattering chamber via a dedicated long  $4''$ diameter nitrogen-inerted vent line to a parallel plate relief vent outdoors. Finally, the H$2$ supply and return lines were also each connected to this same secondary parallel plate relief valve through independent 80 psig relief valves. 

A vacuum switch controlled by a scattering chamber vacuum pressure transducer was used to shut down all the relevant electronics which could act as potential ignition sources, and closed a solenoid valve to isolate the H$_2$ ballast tank and prevent the large outdoor H$_2$ inventory from being dumped into the scattering chamber.

\subsection{Loop Instrumentation}

Ten ports with 7 cm conflat flanges were provided on the top plate of the scattering chamber to bring signals or power in and out of the vacuum space of the scattering chamber. Pressure transducers were located on the target gas panel about 30 m from the target itself.
Pairs of temperature sensors were located at 5 positions in the loop. Going clockwise around the loop looking downstream (see Fig.~\ref{fig:MainGUI}), these were the  pump outlet/heat exchanger inlet, the heat exchanger outlet/cell inlet, the cell outlet, the heater inlet,  and the heater outlet/pump inlet. 

These temperature sensors (TS) were calibrated negative temperature coefficient thin-film 4K-100K Cernox CX 1070 SD 4D resistors (4K-100K) or CX1070 SD-4L (4K-325K), mounted on a G10 stalk affixed to a ten pin CeramTec 10236-02-CF feedthrough. The feedthrough was mounted to a standoff on the loop via a 3.4 cm 
mini-conflat seal. These seals are rated for 3.4 MPa, 
2 kV, and 7 A per 1.57 mm 
diameter pin. Two resistors were mounted on each stalk for redundancy. A standard four wire connection was made for each TS to eliminate the resistance of the lead wires from the measurement. The stalk put the resistors well into the flow space of the loop. One of the five TS was accurate at room temperature, and was used to monitor the cooldown and warmup processes. It was situated in the top right corner of the loop (pump outlet) where it  also indicated when the target was full at the end of a cooldown. This layout provided redundant thermometry across each major element of the loop: cell, heater, pump, and HX. The TS at the cell entrance was nominally used to control the target temperature; however, in principle, any of the other locations would serve this purpose equally well.  

In addition to the TS's employed in the \LH ~loop, three Cernox TS's were used to monitor each of the two coolant circuits. In each case a TS  monitored the coolant supply temperature before the Joule-Thompson (JT) valve, after the JT, as well as the coolant return.

Besides the Cernox resistors, generally considered accurate to 20 mK, a number of uncalibrated, less accurate PT-103 platinum resistors were also  employed at  the horizontal motion motor, the dummy target frame in several locations, and the lifter plate. 

The 60V, 50A Sorenson high power heater power supply cable was brought to a CeramaTec 18099-08-CF 4 pin, 500V, 46A/pin 3.4 MPa,  
7 cm conflat feedthrough on the top plate of the scattering chamber. From there heavy gauge wire brought the power through the vacuum of the scattering chamber to a CeramTec 17069-08-CF 4 pin, 55 A 2.4 mm 
diameter pin, 10,000 V, 10.3 MPa 
feedthrough on a 7 cm 
conflat. Inside the loop, the connection to the four heater coils (designed to be arranged as two independent heaters in parallel) was made with a welded connection.

There were three pump leads plus a dedicated ground. The vacuum feedthrough used was a 10094-09-CF700V, 7A/pin 3.4 MPa 
10 pin feedthrough. The pump tachometer provided two signal lines. 

The horizontal motion Phytron stepping motor required five leads. The vacuum penetration for these was a ten pin, CermaTec 3.4 MPa  
700V, 7 A per 1.6 mm 
diameter pin feedthrough. The leads connected directly to the motor. The two wires from the 24 V brake on the horizontal motion gear reducer shaft  also used this feedthrough. Signals from the several limit switches associated with the horizontal motion system fed through one of several 35 pin vacuum feedthrough connectors on the top plate of the scattering chamber.

\subsection{GUIs}
The target was controlled with a number of Graphical User Interfaces (GUIs). The main GUI is shown in Fig.~\ref{fig:MainGUI} along with typical temperature, pressure, heater power, pump rotation frequency, coolant supply parameters, beam current, raster size, vacuum, as well as target selection and position vertically and horizontally. The parameters shown in Fig.~\ref{fig:MainGUI} represent conditions with 180 $\mu$A of $4\times 4$ mm$^2$ beam rastered on the \LH ~target. All the parameter values shown in the GUI were color coded (green, yellow, red, white) to indicate whether they were (respectively) within a their pre-determined safe range, slightly outside the safe range, well outside their safe range, or if their readout had failed. In addition to the font color, an audible alarm sounded when any of these parameters was not within its safe range.

From the main GUI all the secondary GUIs could be launched. These covered summaries of the temperature and pressure sensors, predetermined target position values, details of the IOCs, heater power, pump, ESR status, JT valve status, and safe beam current and size parameters for the \LH ~target and each of the 24 solid targets, as well as the alarm handler system, and stripcharts for all of the most relevant parameters to monitor during the experiment.

\subsection{Solid Target System}

\begin{figure}[!ht] \centering
\includegraphics[width=0.75\textwidth,angle=0]{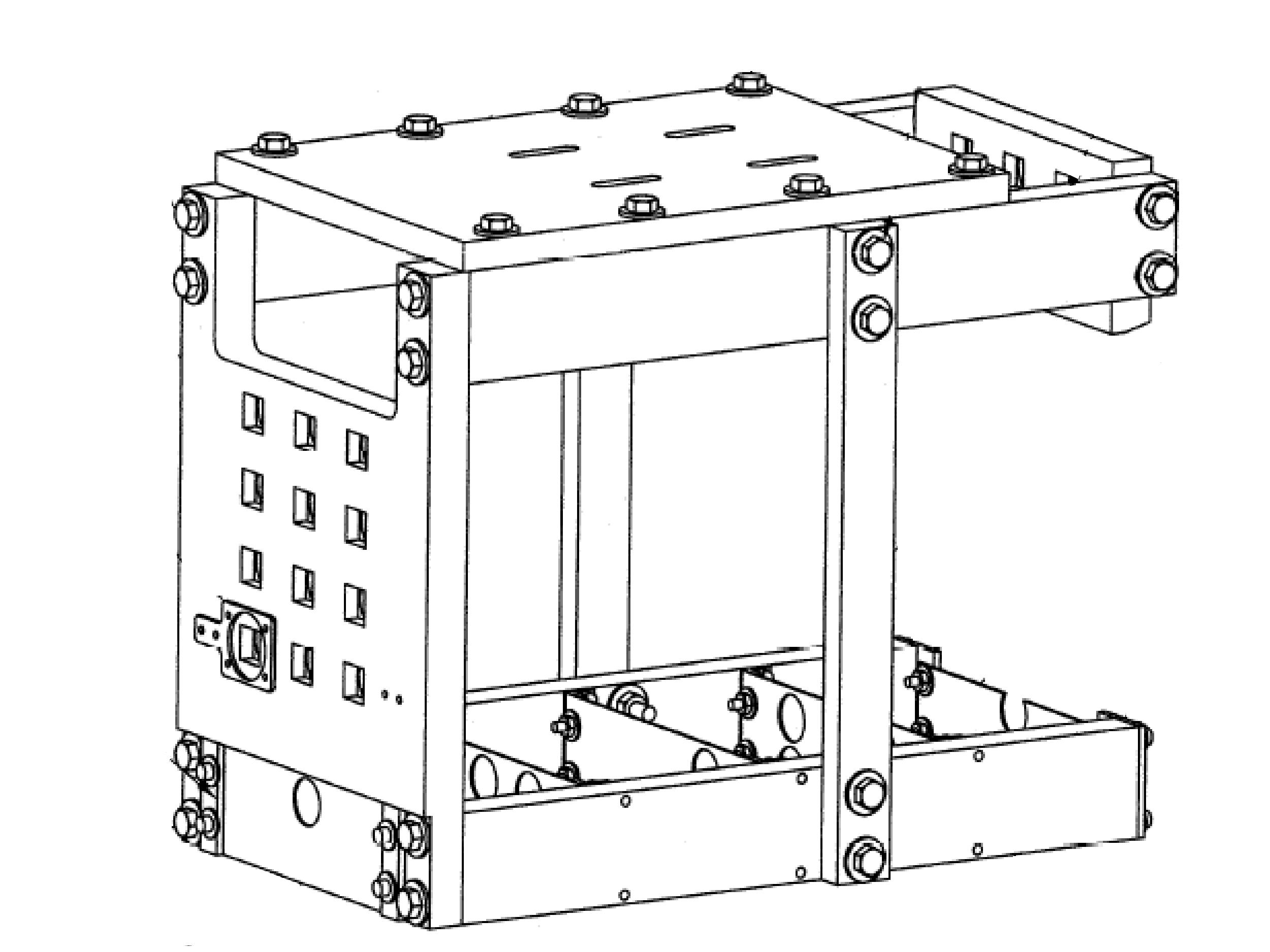}
\caption[Solid Target Ladder Drawing]{\label{fig:SolidTgtLadder} The solid target ladder. There are four rows and 3 columns of upstream positions  on the left of the figure. Each of the 12 square openings visible on this (upstream) face of the 1.9 cm thick frame was $15\times 15$ mm$^2$. There are a further two rows and three positions of different patterns of five foils on the bottom of the ladder. The two rows and three columns of downstream target positions are behind the frame in the upper right of the figure.}
\end{figure}

An extensive system of 24 solid targets was contained in an assembly (see Fig.~\ref{fig:SolidTgtLadder}) attached to the bottom of the LH$_2$ target cell.

These targets were arranged in three arrays. One array was composed of various combinations of foils in 2 rows and 3 columns at 5 ($z$) positions along the beam axis between the upstream (entrance) and downstream (exit) LH$_2$ cell windows. The combinations of ``optics targets" in this  array were used to aid the development of vertex reconstruction algorithms. 

An second, upstream array of 12 targets arranged in 4 rows and 3 columns was situated at the same ($z$) location along the beam axis as the upstream window of the target cell. Likewise, a downstream array of 6 targets arranged in 2 rows and 3 columns was located at the $z$ of the exit window of the LH$_2$ cell. These  two arrays  were used for background subtraction of the upstream and downstream aluminum cell windows of the LH$_2$ target. Different thickness aluminum background targets were provided in both the upstream and downstream matrices to get a handle on radiative corrections. Targets of pure aluminum, thick and thin carbon targets, and beryllium were also provided. Other targets in these arrays were used to measure the relative location of the beam and the target system using a BeO viewer, and thin aluminum targets with various size holes in their centers. 

\begin{figure}[!htb] \centering
\includegraphics[width=.7\textwidth,angle=0]{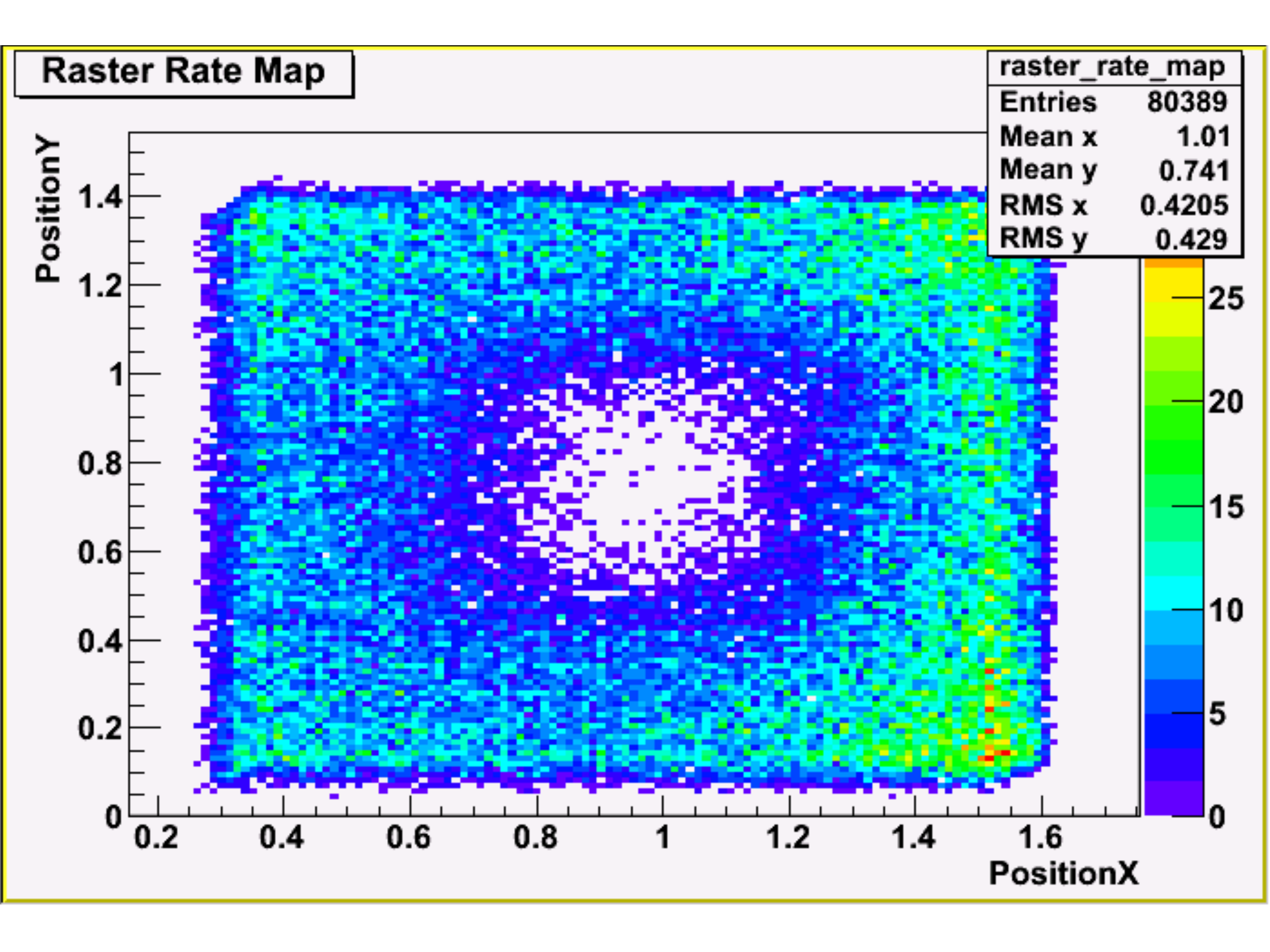}
\caption[Beam Profile]{\label{fig:holetgt} Profile of the beam position on the hole target. The central area devoid of events represents the 2 mm $\times$ 2 mm hole in the target illuminated by a 4 mm $\times$ 4 mm dithered beam.}
\end{figure}

These latter targets, in particular, were crucial to establishing the optimal horizontal and vertical position of the target system with respect to the beam. Thin aluminum ``hole targets" with 2 mm $\times$ 2 mm square holes punched out of their centers were moved into the beam. The beam position was dithered typically in a 4 mm $\times$ 4 mm pattern at the target. The current in the dithering magnets was digitized so the beam position inside this pattern was known at any given point in time. Beam electrons which passed through the holes in these targets created no triggers in the experiment's detectors. However, electrons which missed the hole could be scattered into the detectors, creating an event trigger and thus a 2-dimensional profile of their position at the target using the dithering magnet currents. These profiles provided precise maps of the shadows left by the target hole relative to the dithered beam position such as shown in Fig.~\ref{fig:holetgt}.  By measuring the hole profiles at both the upstream and downstream $z$ locations, the $x$, $y$, pitch, roll, and yaw of the extended target could be accurately determined. Offsets in $x$ and $y$ could be corrected in real time using the two-axis motion system, but due to the extended target length of the LH$_2$ cell, pitch and yaw corrections were problematic.  Indeed, the hole target measurements made after the initial cooldown of the target revealed an unexpected 4 mm pitch which occurred during cooldown. Prior to subsequent cooldowns, the target was pre-pitched in the opposite direction by this amount, and subsequent hole profiles revealed the cold pitching had been successfully corrected. The success of the target positioning achieved using the hole targets was confirmed after the experiment by inspection of the spots left by the beam on the target cell windows as well as the solid targets, which were in all cases within 1 mm of the center of each respective target.

An extensive effort went into the design of the solid targets and their frames  to optimize heat conduction  in order to use as much beam current as possible for the background measurements, and to ensure that the acceptance associated with the background targets in the upstream frame was not obstructed by the optics targets or by the downstream target frame. CFD simulations augmented analytical calculations which optimized the heat transfer between the center of each target where the beam heating occurred, and the cold reservoir of the LH$_2$ cell. 

Each of the 18 (non-optics) targets in the upstream and downstream frames was 2.54 cm square, and was dropped into 2.71 cm square pockets machined just 1.3 cm deep in 1.9 cm thick aluminum frames. By providing  a smaller 1.5 cm square opening only 0.6 mm deep in the opposite face of each pocket, 2/3 of the surface area of one face of each target was in thermal contact with the frame. The  side of the frame with the larger pockets was threaded 12.7 mm deep ($1.25''$-12 UNF) to accept 31.8 mm diameter aluminum threaded pipe (22 mm i.d.) which pushed each target into its pocket against the lip at the boundary of the two different-sized squares. This lip provided the  mechanical contact necessary for good heat conduction from each target to the frame. The heat transfer from the center of each target was studied using CFD simulations, and benchmarked against measured temperatures at various locations in the target ladder assembly as the beam current was raised on each target.

\section{ Performance}

\begin{figure*}[ht]
\centering
\includegraphics[width=\textwidth,angle=0]{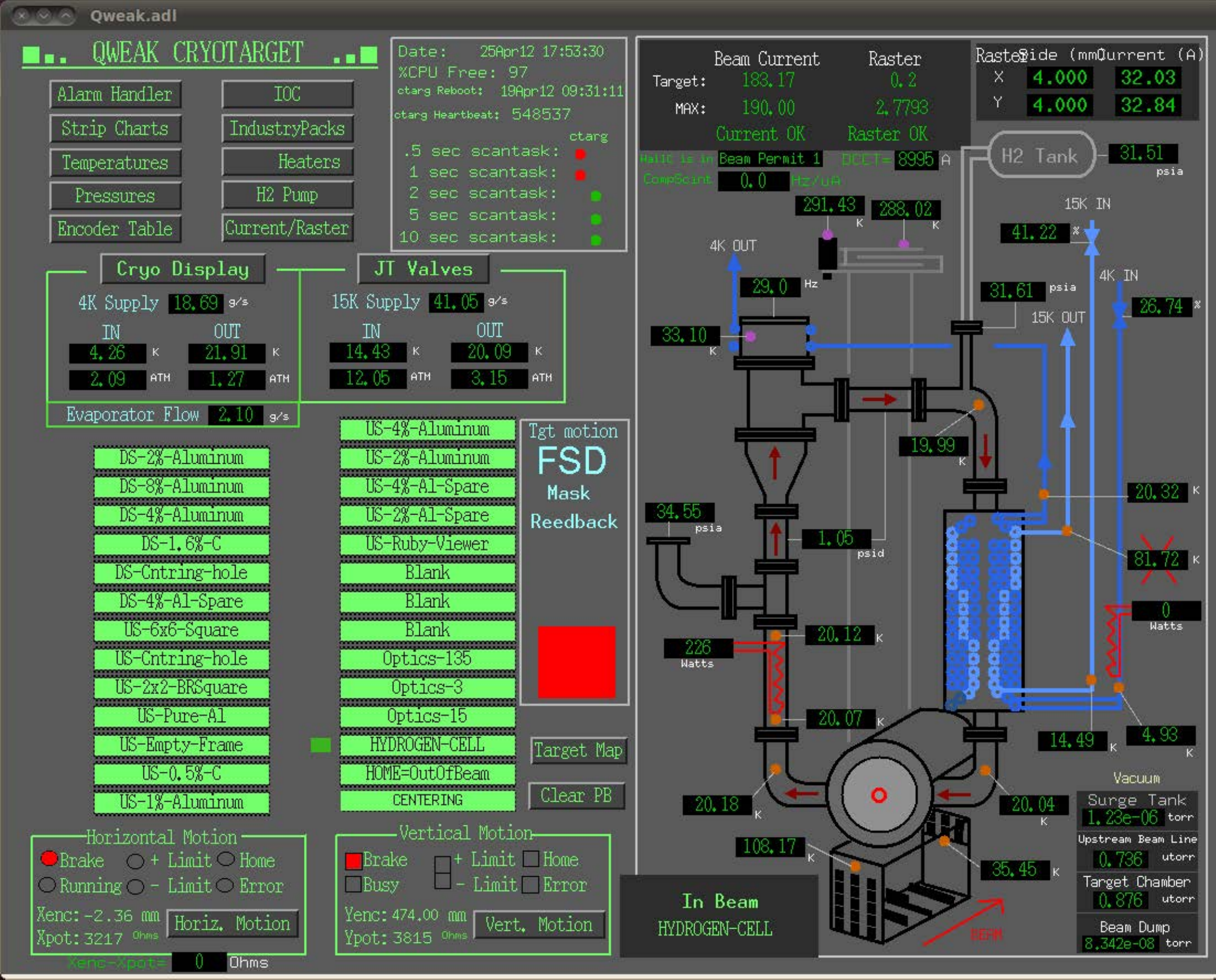}
\caption[Target Control Main GUI]{\label{fig:MainGUI} A typical snapshot of the  target control graphical user interface (GUI) during the experiment, showing the coolant parameters as well as some of the instrumentation values around the hydrogen recirculation loop.}
\end{figure*}

The equilibrium performance of the target with 183 $\mu$A of beam is summarized in Fig.~\ref{fig:MainGUI}. 

\subsection{Cooling power budget}
\label{sec:cooling_power_budget}

The cooling power was measured with Cernox thermometers in the coolant flow at the inlet and outlet of the HX, downstream of the JT valves used to control the flow of each supply. The 4K and 15K coolant massflows, supply and return pressures were measured from instrumentation at the JLab End Station Refrigerator (ESR).  Typical conditions are summarized in Table~\ref{tab:coolant}. These correspond to a cooling power of 1486 W on the 15~K circuit, and 1739 W on the 4~K circuit, for a total cooling power of 3225 W.

\begin{table}[hhtb]
\centering
\begin{tabular}{ r  c  l  }
\toprule
Property & Value & Units          \\ \midrule
4K Supply $P$ & 2.21 & atm \\
4K Supply $T$ & 5.02 & K \\
4K Return $P$ & 1.38 & atm \\
4K Return $T$ & 20.31 & K \\
4K Mass flow & 16.6 & g/s \\
4K $\Delta H$ & 104.7 & J/g \\
4K Cooling Power & 1739 & W \\ \hline
15K Supply $P$ & 12.1 & atm \\
15K Supply $T$ & 14.8 & K \\
15K Return $P$ & 3.09 & K \\
15K Return $T$ & 20.31 & K \\
15K Mass flow & 40.5 & g/s \\ 
15K $\Delta H$ & 36.7 & J/g \\ 
15K Cooling Power & 1486 & W \\ \hline
Total  Cooling Power & 3225 & W \\ \hline
\bottomrule
\end{tabular}
\caption[Coolant Properties]{Coolant properties obtained after the Moller polarimeter superconducting solenoid was offline for several days. Therefore the parameters in the table reflect those for the \LH ~target only. The HPH was 2200 W with the beam off, and 260 W with the beam on at 180 $\mu$A.  $P$ and $T$ refer to pressure and temperature, $\Delta H$ refers to enthalpy change. 
}
\label{tab:coolant}
\end{table}

The other side of the ledger consists of the various heat loads on the target, which are summarized in Table~\ref{tab:heatload} for conditions when there was 180 $\mu$A of beam on the target. The target pump was running at 29.4 Hz, producing a head of 10.5 m and a LH$_2$ mass flow of  1.23 kg/s. 

\begin{table}[ht]
\centering
\begin{tabular}{ r  r  }
\toprule
Source & Value  \\ \midrule
Beam Power in LH$_2$ & 2075 W \\ 
Beam Power in Cell Windows & 23 W \\ 
Viscous heating & 177 W\\
Radiative losses & 150 W\\
Table Heater & 40 W \\
Pump Motor \LH  ~bypass  & 150 W \\
Reserve heater power & 260 W \\ \midrule
Total Heat Load & 2875 W \\ \bottomrule
\end{tabular}
\caption[Heat Loads]{Target heat loads. Some were measured, others were estimated. 
}
\label{tab:heatload}
\end{table}

The beam power (see Eq.~\ref{beampower}) of 2075 W accounts for the ionization energy deposited in the 34.5 cm long LH$_2$ target determined for a 180 $\mu$A electron beam, accounting for the density effect. It uses the density $\rho$=71.8 kg/m$^3$ at the operating conditions of the target (20 K, 32 psia). The power deposited by the beam in the thin aluminum windows of the target cell ( 0.23 mm combined thickness) was only 23 W.  The 177 W viscous heating (see Eq.~\ref{eq:viscous}) was determined from the measured pump head (10.5 m) and capacity (1.2 kg/s). The 40 W table heater, discussed in Sec.~\ref{sec:aux} kept the components of the motion system at room temperature. The heat loss associated with conductive and radiative losses to the outside environment were estimated  from the amount of time ($\sim$2 days) the target took to warm up to room temperature from 50~K once the coolant supplies were shut off. Together with the estimated cold mass (300 kg), and an average value for the heat capacity (366 J/kg-K), the losses were $ Q = m C_p \Delta T /{\rm time} = 160 $~W. The 260 W average reserve heater power was maintained at all times to control the target temperature when the beam was on.
The \LH ~bypass heat load used to help cool the pump motor was discussed in Sec.~\ref{sec:Pump Design}.

\begin{figure}[!htb]
\centerline{\includegraphics[width=.7\textwidth,angle=0]{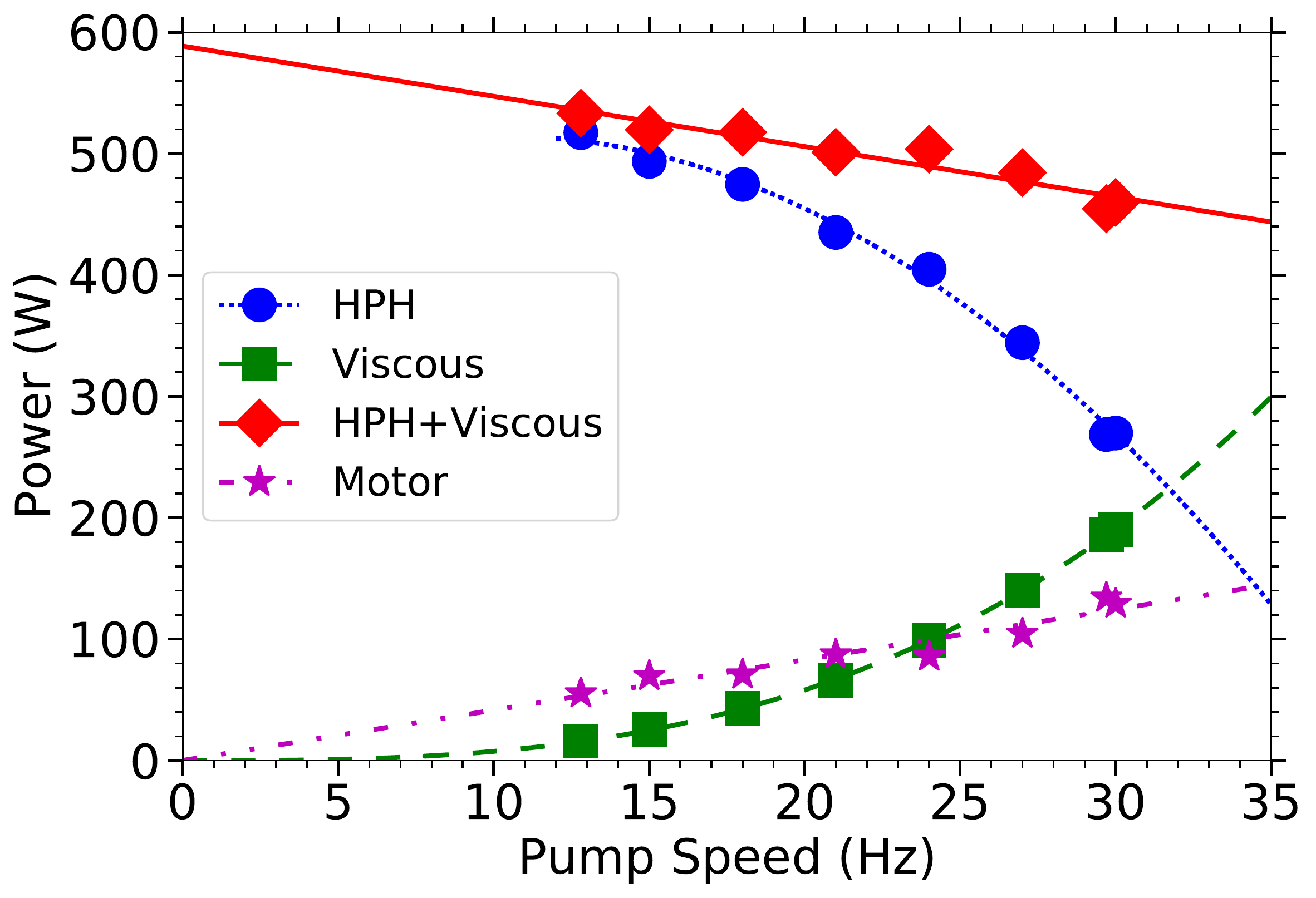}}
\caption[Pump heat load vs Pump Speed] {\label{fig:pumpheat} Measurements of the heat load associated with the pump as a function of the pump speed. The blue circles (fit: dotted line) are the high-power  heater (HPH) power measured with 180 $\mu$A of beam current rastered $4\times 4$ mm$^2$ on the target as the pump speed was varied. The green squares (fit: dashed line) are the viscous heat load calculated from the measured pump head and linearly-scaled mass flow at each pump speed. The red diamonds (fit: solid line) represent the sum of the heater power plus the viscous heat load. Finally, the purple stars (fit: dashed-dotted line) are subtracted from the intercept of the latter curve, yielding the pump heat-load without the effect of viscous heating. At the operational frequency of 30 Hz, the pump heat load was about 150 W. }
\end{figure}

The  pump heat load associated with the $1/4''$ hydrogen bypass discussed in Sec.~\ref{sec:Pump Design} was determined from measurements of the heater power as the pump speed was varied. Since the target temperature was kept fixed at 20.00~K by a PID loop, the heater power changed automatically to compensate for the changes in the pump motor heat load and the viscous heating in the loop. The viscous heating can be calculated from the measured volume flow and pump head at each pump speed, so it can be subtracted from the observed changes in the heater power to arrive at the pump motor heat load. Typical results are shown in Fig.~\ref{fig:pumpheat}. They indicate the heat load from the pump motor at 30 Hz is about 150 W. 

To check this result, the $\Delta T$ across the pump was used to calculate the pump power. This method has a large uncertainty, because the changes in $\Delta T$ are only of order 10 mK. With that caveat, however, by averaging both the pump inlet and outlet temperatures the pump motor power obtained with this more uncertain method was also about 150 W at 30 Hz.

The estimated total cooling power presented in Table~\ref{tab:coolant} is within 350 W of the estimate shown in Table~\ref{tab:heatload} for the heat loads associated with the target. Some of the entries in these tables were estimated. Massflows, in particular, are not considered very reliable, so the $\approx 10$\% agreement between the cooling power and heat load totals is reasonable.

\subsection{Mass Flow Measurements}

The \LH ~mass flow was determined by measuring the temperature difference $\Delta T$ across the heater. The target loop was designed with this measurement in mind, so pairs of thermometers were situated on opposite sides of the heater (as well as before and after the heat exchanger, and after the cell). The mass flow $\dot{m}$ can be derived from the relationship 
\begin{equation} \label{eq:mdot=QCdT}
  \dot{m}({\rm kg/s}) = \frac{Q({\rm W})}{C_p ({\rm J/kg} \mbox{-} {\rm K}) \; \Delta T({\rm K})}. 
\end{equation}
The specific heat of LH$_2$ at 20~K and 221 kPa is 9425 J/kg-K (note that $C_P \approx 5.2$ J/g-K for helium in our thermodynamic range). The heater power Q was determined from the output current and voltage  of the heater power supply, which is assumed known to 10\%. The thermometry consisted of negative temperature coefficient thin-film zirconium oxy-nitride semiconductor diodes (Cernox resistors~\cite{Cernox}). The stability of the Cernox resistor temperature measurements is 0.08\%~\cite{Cernox_stability}, and dominates the uncertainty in the mass flow measurement. To eliminate potential offsets in the temperature and power measurements, the mass flow was determined from the difference of measurements obtained at two different power levels (Q$_1$=2261 W and Q$_2$=274 W) via
\begin{equation}
  \dot{m} = \frac{Q_1 - Q_2}{C_p \left( T_1^{\rm \;in} - T_1^{\rm \;out} -T_2^{\rm \;in} +T_2^{\rm \;out} \right)}, 
\label{eq:massflow}
\end{equation}
where the superscripts in and out denote temperature measurements before and after the heater. The temperature factor in parentheses in Eq.~\ref{eq:massflow} amounted to only 170 mK.
The mass flow determined from the average of many such measurements was $1.2 \pm 0.3$ kg/s. The volume flow corresponding to the LH$_2$ density of 71.3 kg/m$^3$ was $17.4 \pm 3.8$ liters/s. The pump speed during these measurements was approximately 29.4 Hz.

\subsection{Pump Head}

\label{sec:PumpHead}

The head was directly measured to be 7.6 kPa  at the nominal pump speed of 30 Hz using a differential pressure gauge across the hydrogen supply and return lines. These lines  connected to the target loop on opposite sides of the hydrogen pump. An Orange Research  1516-S1073 0-5 psid differential pressure gauge provided an analog readout of the pump head at the target gas panel.

An Omega Engineering PX771A-300WCDI differential pressure transducer provided an output that was digitized and monitored during the experiment. In addition, the electron beam was  automatically shut off if the pump head dropped below a preset threshold. This protection was put in place due to the concern that in the event the hydrogen pump tripped off, the convective cooling at the windows of the target cell could be insufficient to prevent the beam from eventually melting through the windows, even though the power deposited by the 180 $\mu$A beam in the 0.127 mm thick cell exit window was only 13 W.

\subsection{Bulk Density Reduction }

Bulk density reduction characterizes the dynamic equilibrium density reduction (effective thickness) of the target due to heating of the LH$_2$ by the beam in the interaction region. Localized heating can form bubbles of hydrogen vapor in the beam interaction region. 
Non-localized heating of the LH$_2$ can also contribute. A good rule-of-thumb is that  a 1~K increase in the average temperature (near our operating conditions) corresponds to a density reduction $\Delta \rho/\rho \sim 1.5\%$.  This effect increases the running time required to reach a given statistical goal for an experiment.

\begin{figure}[hhhtb]
\centerline{\includegraphics[width=0.7\textwidth,angle=0]{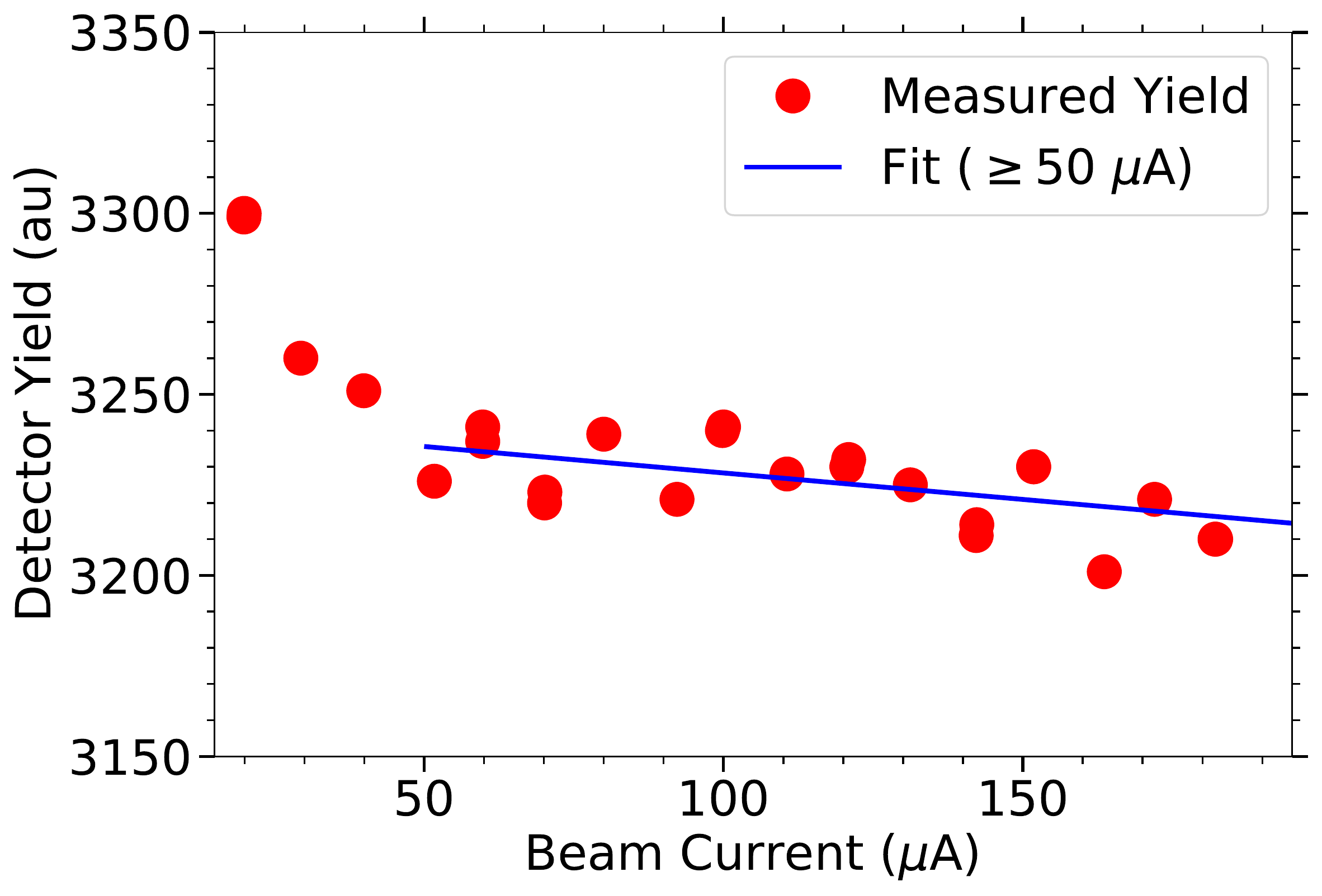}}
\caption[Detector Yield vs Beam Current] {\label{fig:BulkBoiling} 
The charge-normalized detector yield (red circles) in arbitrary units measured at different beam currents during a beam current calibration. The beam current monitors were linear above about 50 $\mu$A, where the slope could be fit (solid blue line) to characterize the density reduction as the beam current was raised. From the fit the relative change in yield between 0 and 180 $\mu$A is 0.8\%.
}
\end{figure}

Consistent measurements of the bulk density reduction were difficult to obtain over the large range of beam currents used in this experiment. Detector and BCM non-linearity, as well as pedestal shifts, contributed to this difficulty, especially below 50 $\mu$A.  Results were obtained parasitically during BCM calibrations in which the beam current was raised and then lowered in $\sim 20 \: \mu$A, 1-minute-long steps between $20-180 \: \mu$A alternated with 1-minute-long beam-off periods, shown in Fig.~\ref{fig:BulkBoiling}. These provided an estimate for the bulk boiling of $0.8\% \pm 0.8\%$ per 180 $\mu$A- the 100\% uncertainty accounts for the inconsistencies. 

To check this result, detailed CFD simulations  were performed, which were first  benchmarked using the G0 target  \cite{G0tgt} geometry. The simulations predicted the G0 target density reduction should be $1.1 \pm 0.2$\% over 40 $\mu$A. The measured result reported in Ref.~\cite{G0tgt} was $1.0 \pm 0.2\%$ at 40 $\mu$A, in excellent agreement with the CFD simulation. Therefore  the bulk density reduction  of the \Qweak target (and its uncertainty) was taken as the CFD prediction using the \Qweak target geometry:  $0.8\% \pm 0.2\%$ at 180 $\mu$A.

\subsection{Transient density changes}
Transient changes in density occur in response to a loss of incident beam when an accelerating cavity trips off, for example. These trips occurred typically 5 times per hour. The proportional-differential-integral (PID) feedback loop that constantly adjusted the resistive heater to maintain the target temperature at 20.00~K raised the heater power to compensate for the loss of heating from the beam, and reduced the heater power when the beam returned. Over 2 kW were shuffled between the beam and the heater when the beam tripped off from 180 $\mu$A. To improve the target temperature response to such a large change in conditions, the PID feedback also looked at the beam current, and increased the PID heater power step size when big changes were observed in the beam current.

\begin{figure}[!hhhtb]
\centerline{\includegraphics[width=0.8\textwidth,angle=0]{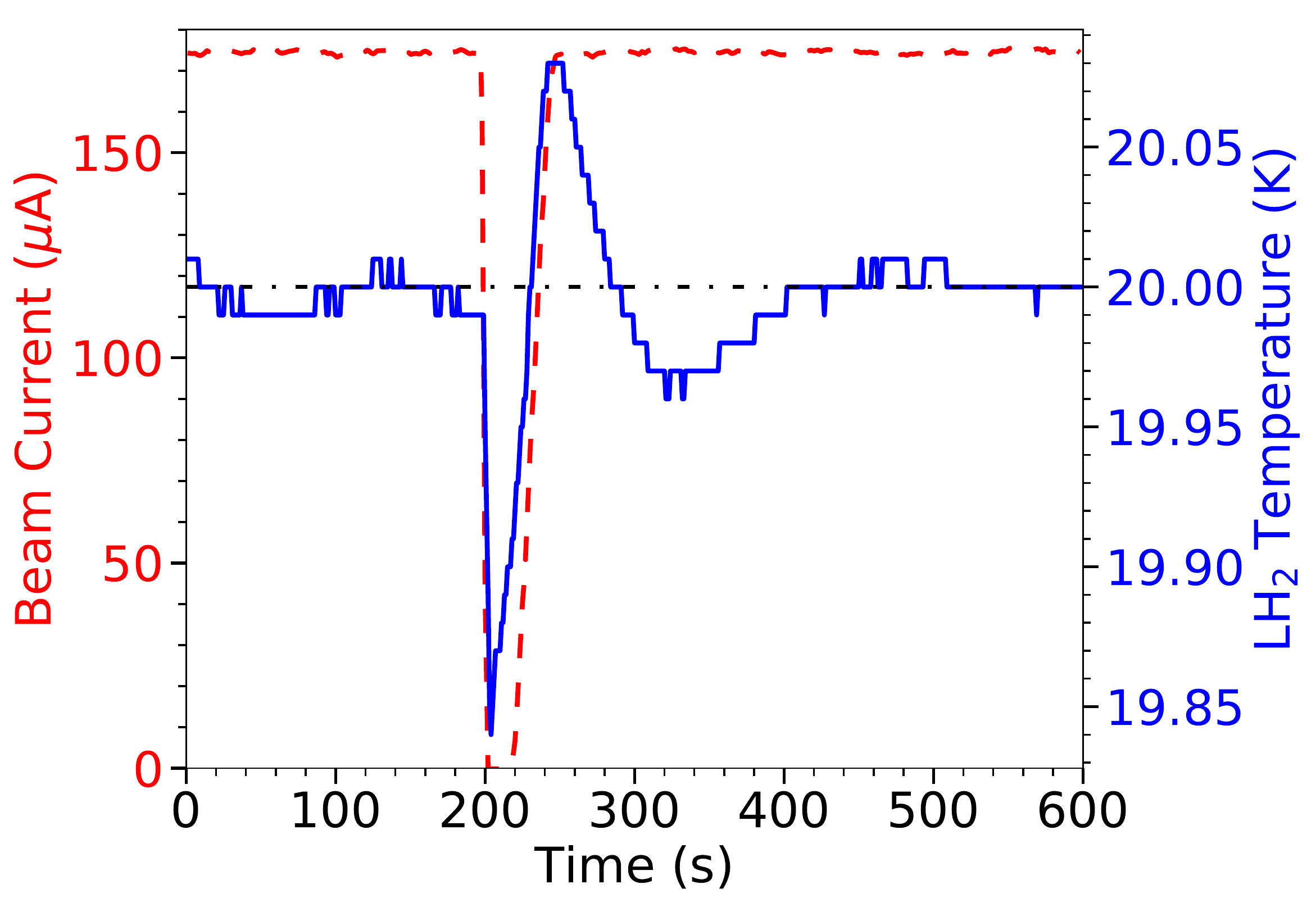}}
\caption[Target Temperature and Beam Current vs Time] {\label{fig:BeamTrip} 
Target temperature (solid blue line, right axis) response to a sudden trip in the beam current (dashed red line, left axis)  as a function of time.  
The damped temperature oscillation settles out to within 30 mK of the 20.0~K goal temperature (black dash-dot line) within about 20 s of full beam restoration to 174 $\mu$A. }
\end{figure}

The temperature response to a typical beam trip (174 $\mu$A, pump at 30 Hz, raster $4\times 4$ mm$^2$) is shown in Fig.~\ref{fig:BeamTrip}. The maximum temperature excursion reached when the beam was fully restored was about 80 mK. The magnitude of this excursion dropped to 30 mK within about 20 s of full beam restoration, and took another 120 s to completely subside.  The temperature excursion was about 160 mK when the beam tripped off, but this is irrelevant of course since without beam, no data were recorded. In fact the event analysis only occurred when the beam current was above a threshold (typically 130 $\mu$A) close to the nominal operating current at the time. These small beam-trip temperature excursions were  slow compared to the helicity reversal frequency, so  they contributed only marginally to the asymmetry width and were not considered a problem beyond the loss of data-taking efficiency they represented.

\section{Target Noise}
\label{sec:Boiling}

\begin{figure}[!hhhtb]
\centerline{\includegraphics[width=0.8\textwidth,angle=0]{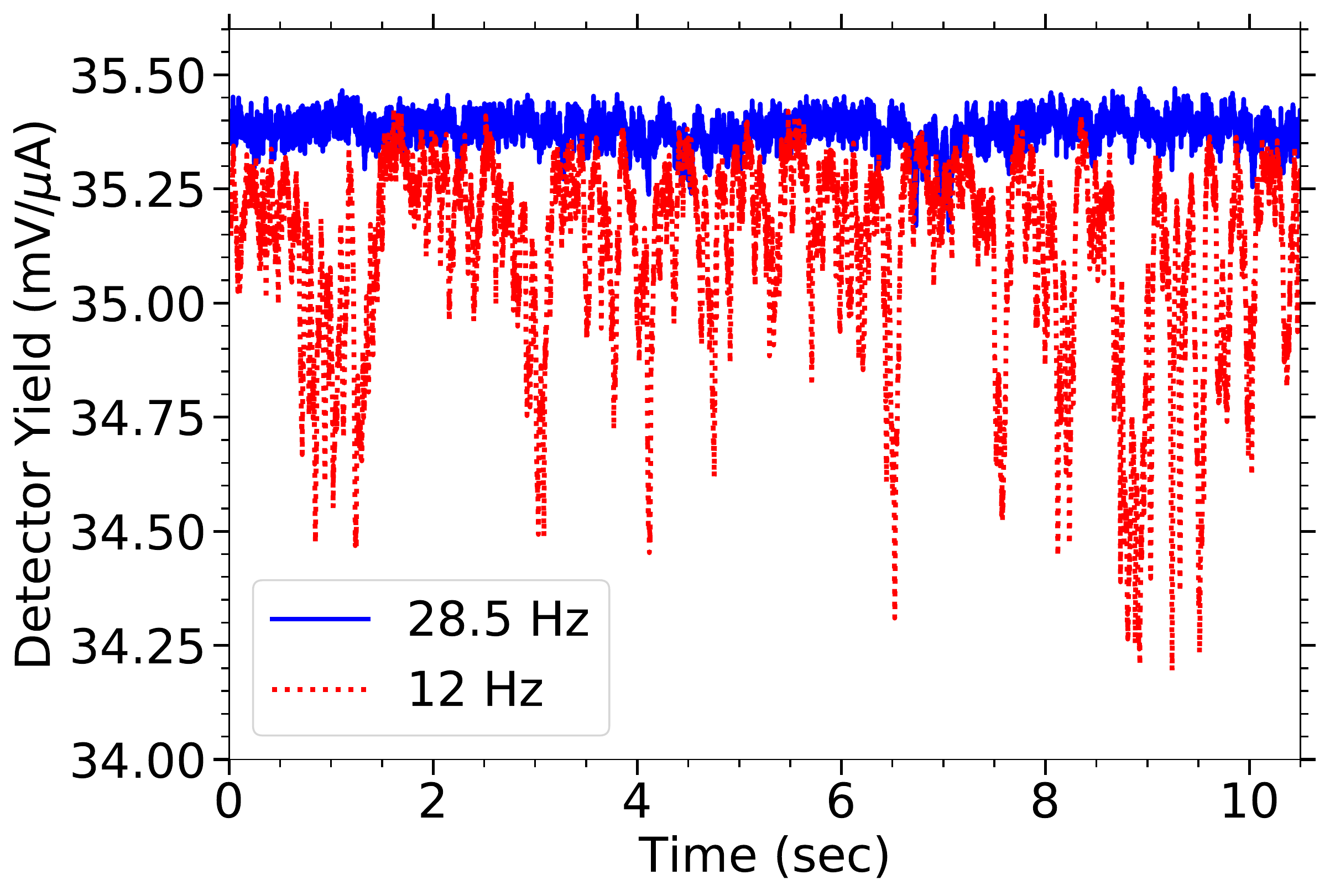}}
\caption[Detector yield vs time at 2 Different Pump Speeds]{\label{fig:Bubbles}  Behavior of the charge-normalized detector yield over 10 seconds  for data acquired at the nominal 28.5 Hz \LH ~pump speed (solid blue line) and at 12 Hz (dotted red line). At the nominal 28.5 Hz rotation frequency of the \LH ~recirculation pump,  the detected scattered electron yield is reasonably constant with time. 
At the lowered pump rotation frequency of 12 Hz, the effects of target noise (density fluctuations) appear as significant drops in detected yield (as much as $\approx$ 3\%) with time. The beam current and raster size were the same for each of the two plots in this figure (180 $\mu$A, $4\times 4$ mm$^2$). }
\end{figure}

Density fluctuations in the LH$_2$ that take place near the beam helicity reversal frequency (960 Hz) are called target noise, or more loosely, target boiling. This phenomenon can be clearly seen in the variation of the (charge-normalized) detected scattered electron yield with time. In Fig.~\ref{fig:Bubbles} the time-dependence of these yields is plotted at two different rotation frequencies of the liquid hydrogen pump, in other words, at two different average LH$_2$ flow velocities in the interaction region. 
At the higher flow velocity used during normal operation of the target in Fig.~\ref{fig:Bubbles}, boiling is reduced relative to the lower velocity used in Fig.~\ref{fig:Bubbles} which moved the LH$_2$ more slowly across the beam axis, allowing it to warm up more.  The  brief $\sim 0.1$ s, $\sim 2$\% drops in the 12 Hz yield visible in Fig.~\ref{fig:Bubbles} are associated with 
density fluctuations forming along the path of the electron beam in the liquid hydrogen. 

The following sections describe several independent methods used to measure the target noise \dAtgt \hspace{-0.3em}, which all yield consistent results. We nominally  
change one independent variable at a time (beam current, LH$_2$ recirculation pump speed, beam raster size, etc.) and observe the change in the dependent variable \dAqrt, the  asymmetry width measured over helicity quartets. The latter is assumed to be comprised of the sum in quadrature of a fixed component and the target noise contribution \dAtgt \hspace{-0.3em}.

\subsection{Current scan}
\label{sec:CurrentScan}

The most difficult method used to determine the target noise \dAtgt involves changing the beam current, because of course both the statistics and the Beam Current Monitor (BCM) resolution also depend on beam current. We model the beam current dependence of the measured helicity-quartet detector asymmetry width  as follows:

\begin{align} \label{eq:current_scan}
    \Delta {{A_{\, qrt}}} 
    & = \sqrt{\left ( \frac{a}{\sqrt{I}} \right)^2 + 
    \left ( \frac{b}{I} \right)^2 + 
    \left ( cI^e \right)^2 + d^2} \nonumber \\
    & = \sqrt{\left ( \Delta {{A_{\, stat}}} \right)^2 + 
    \left ( \Delta {{A_{\, BCM}}} \right)^2 + 
    \left ( \Delta {{A_{\, tgt}}} \right)^2 + \left ( \Delta {{A_{\, excess}}}  \right)^2}.
\end{align}

The coefficients $a$, $b$, $c$, and $d$ represent the counting statistics, BCM noise, target noise, and other fixed (current-independent) excess contributions, respectively, to the quartet asymmetry width \dAqrt \hspace{-0.3em}.   The functional form of Eq.~\ref{eq:current_scan} reflects the usual 1/$\sqrt{N}$ counting statistics,  BCM noise inversely proportional to beam current, and through the additional parameter $e$ the unknown  exponent governing the dependence of target noise on beam current. 
The five parameters were determined by fitting the measured \dAqrt at eight different beam currents $I$ from 50-169 $\mu$A. The measurements were performed with the LH$_2$ recirculation pump speed fixed at 30 Hz and the raster dimensions fixed at $3.5 \times 3.5$ mm$^2$ at the target. Note however that for most of the experiment, the raster dimensions at the target were $4.0 \times 4.0$ mm$^2$. 

\begin{figure}[!hhhtb]
\centerline{\includegraphics[width=0.9\textwidth,angle=0]{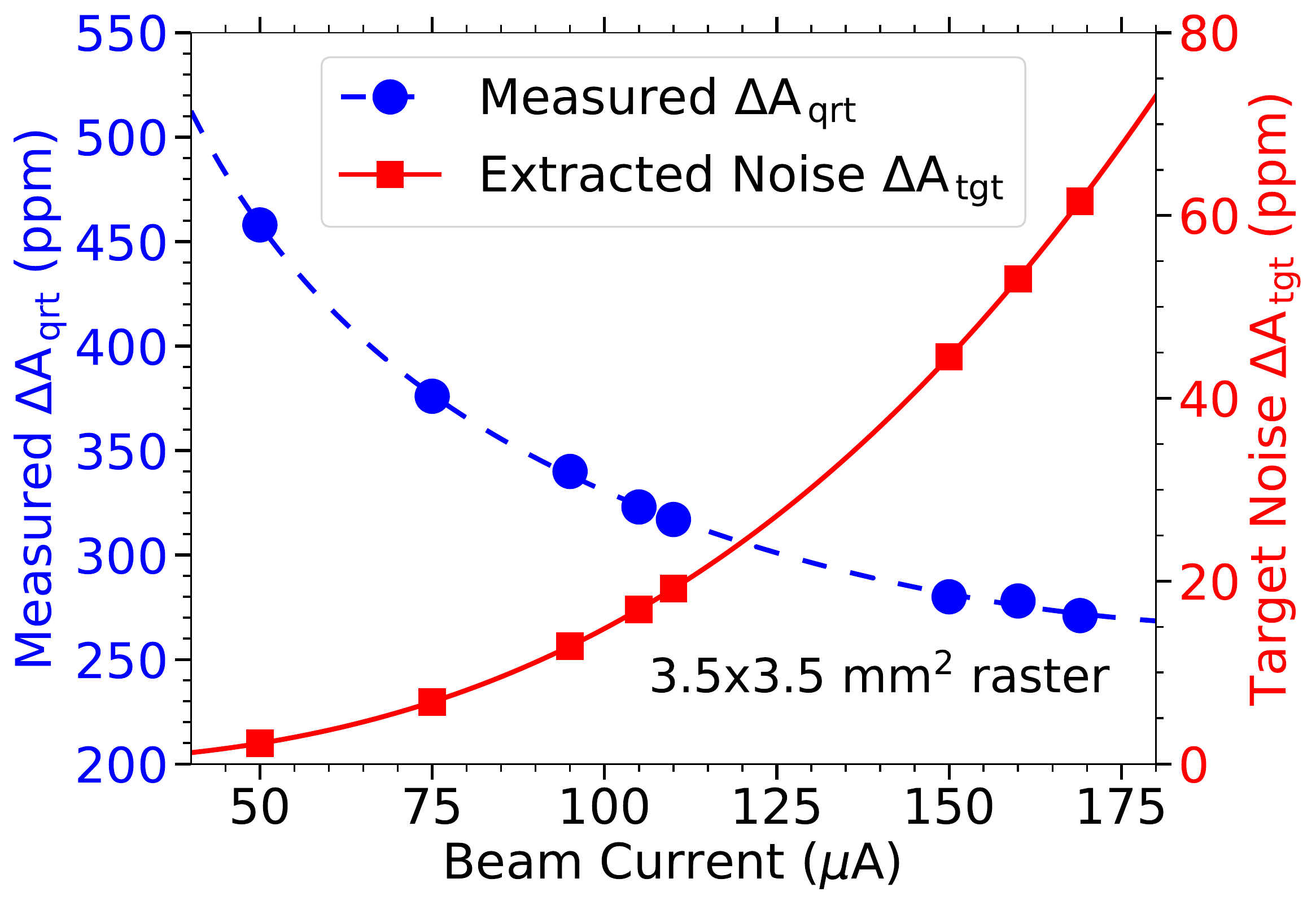}} 
\caption[Target noise vs Beam Current] {\label{fig:Current_Scan} 
The detector asymmetry width \dAqrt measured over helicity quartets at different incident beam currents (blue circles, left axis) with the LH$_2$ recirculation pump at 30 Hz, and a $3.5 \times 3.5$ mm$^2$ raster. The dashed blue line is a fit to these data using Eq.~\ref{eq:current_scan}. The target noise term \dAtgt extracted in quadrature from the fit at each beam current is shown as the red squares (right axis), along with the fit representing that term (solid red line). }
\end{figure}

The \dAqrt measurements and the five-parameter fit are shown in Fig.~\ref{fig:Current_Scan}, along with the target noise term \dAtgt extracted from Eq.~\ref{eq:current_scan}. The coefficient of determination ($R^2$) of the  fit is 1.00. 
The fit coefficients are $a=2996.5$, $b=5995 $, $c=5.49 \times 10^{-5} $, and $d=125.47 $ ppm with $I$ in $\mu$A. The fitted exponent $e$ is $2.715 $. 
Reasonable fits can also be obtained with $e=2$ or $3$. 
Extrapolating the fit to  180 $\mu$A,  the statistical width \dAstat is 233 ppm, and the BCM noise \dAbcm $ = b/I = 33$ ppm is reasonably similar to the $\sim40$ ppm determined independently from the BCM double difference method described in Ref.~\cite{QweakNIM}.  The target noise component \dAtgt $= cI^{e} = 73.1$
ppm at 180 $\mu$A, 30 Hz (pump), and $3.5 \times 3.5$ mm$^2$ (raster). Scaled quadratically (using the results obtained in Sec.~\ref{sec:Raster_Scan}) to the $4.0 \times 4.0$ mm$^2$ raster area used for most of the \Qweak measurement, the predicted 30 Hz, 180 $\mu$A target noise would be \dAtgt $= 56 $ 
ppm.

\subsection{ Raster scan}
\label{sec:Raster_Scan}
The nominally 250 $\mu$m diameter incident electron beam was rastered (dithered) in both the horizontal and vertical directions to reduce the power density at the target. The raster ~\cite{QweakNIM} consisted of 2 pairs of air-core coils, (two horizontal, and two vertical), which produced paraxial displacements of the beam up to $5\times 5$ mm$^2$. The raster magnets were driven at $\approx 25$ kHz, with 960 Hz difference between the $x$ \& $y$ excitations so the raster completed one full pattern every 960 Hz helicity window. 
Increasing the area of the beam at the target reduces 
boiling associated with the beam's ionization energy loss in the aluminum entrance and exit windows as well as the temperature rise in the $LH_2$ in the interaction volume. The raster size thus presents another knob which can be turned to determine the target noise \dAtgt \hspace{-0.3em}, independently of the beam current. 

\begin{figure}[!hhhtb]
\centerline{\includegraphics[width=0.7\textwidth,angle=0]{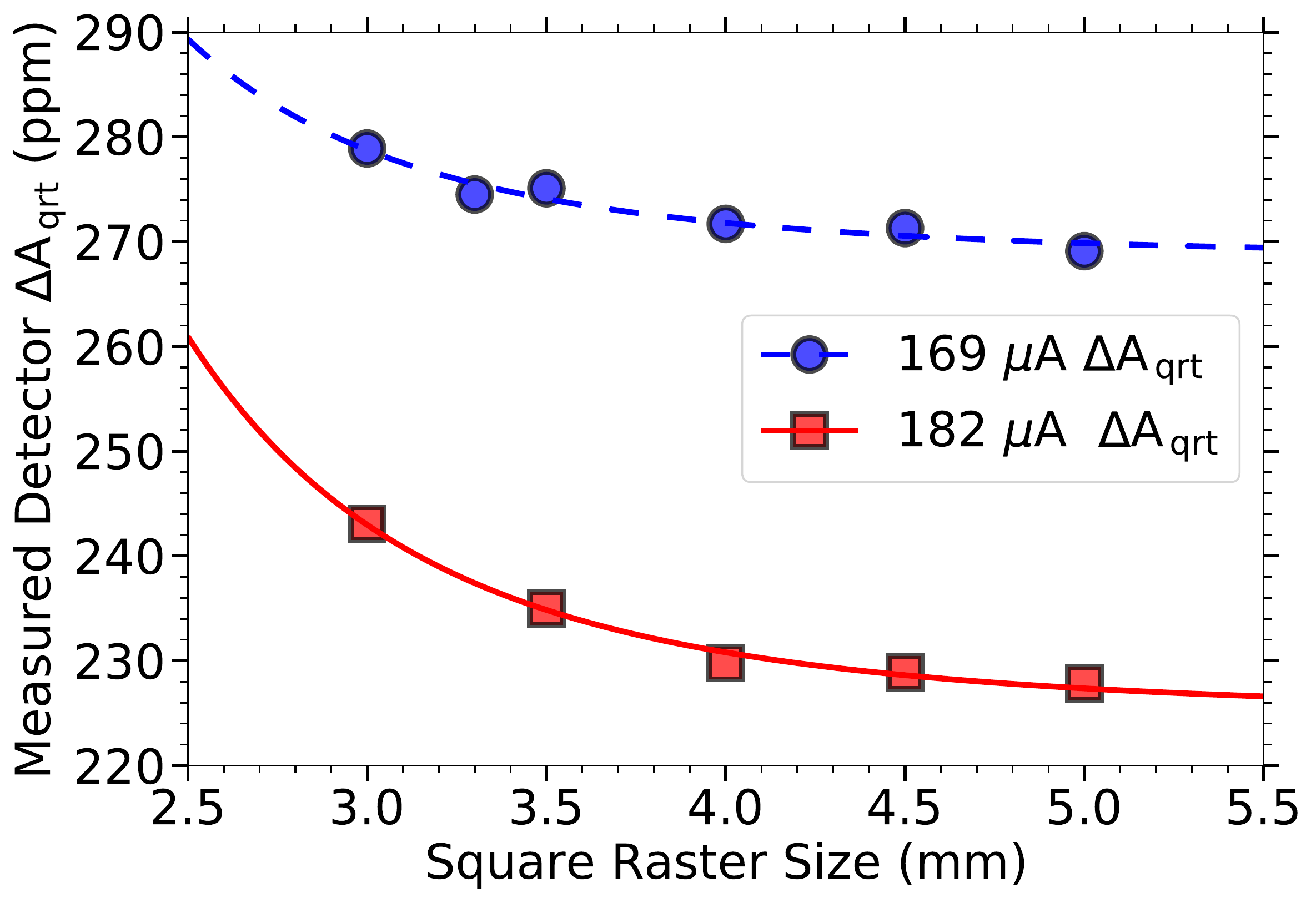}} 
\centerline{\includegraphics[width=0.7\textwidth,angle=0]{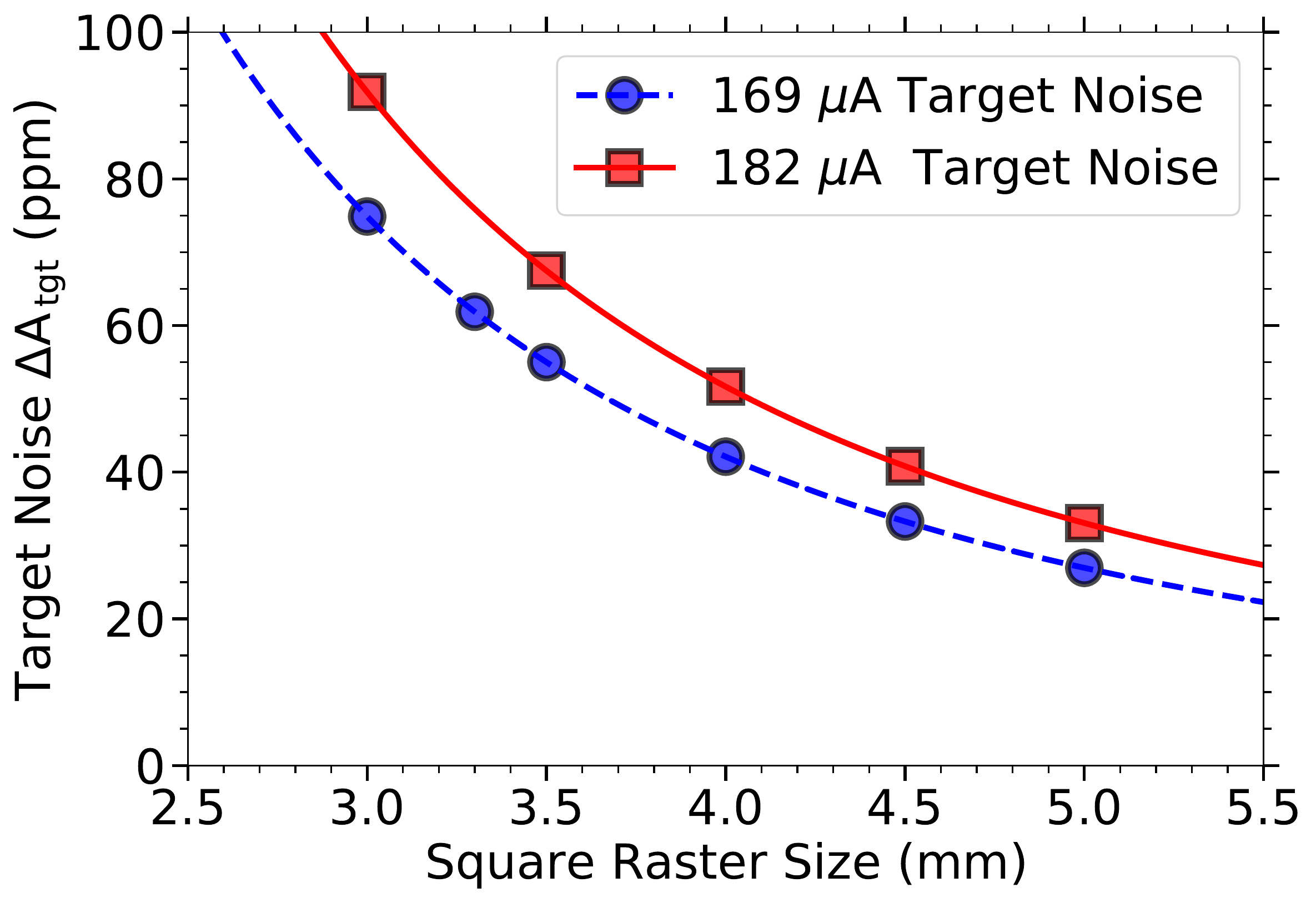}}
\caption[Target noise vs Raster Size] {\label{fig:RasterScan} 
Upper: The detector asymmetry width \dAqrt measured over helicity quartets at 169 $\mu$A (blue circles and  dashed line ) and 182 $\mu$A (red squares and solid line) as a function of raster size at the target with the LH$_2$ recirculation pump at 30 Hz. Fits to these data using Eq.~\ref{eq:raster_scan} are shown for each beam energy. Lower: The target noise term \dAtgt extracted in quadrature from the data in the upper figure at 169 $\mu$A (blue circles) and 182 $\mu$A (red squares) as a function of raster size at the target with the LH$_2$ recirculation pump at 30 Hz. Fits to these data are shown for each beam energy.  }
\end{figure}

In contrast to the beam current scans discussed in Sec.~\ref{sec:CurrentScan}, raster scans are not affected by changing counting statistics or BCM resolution. However, eventually at large enough raster areas second-order effects can arise due increased beam halo and corresponding beam scraping on collimators and flanges.  
Smaller raster areas can become dangerous since eventually the aluminum target cell windows could melt. For the scans presented in Fig.~\ref{fig:RasterScan},  raster dimensions between 3 and 5 mm were studied at 2 different beam currents. The measured helicity-quartet asymmetry width \dAqrt   at each raster area was assumed to consist of the quadrature sum of a fixed term  and  the target noise term \dAtgt which was assumed to be inversely proportional to the raster area:

\begin{equation}
\label{eq:raster_scan}
    \Delta {{A_{\, qrt}}} = \sqrt{a^2 +
    \left ( \frac{b}{Area} \right)^2 },
\end{equation}
where $Area$ represents the area 
of the nominally square raster on the face of the target. 

The fits shown in Fig.\ref{fig:RasterScan}  made use of Eq.~\ref{eq:raster_scan}. The fit parameters are $a=268.5$ ppm, $b=673.8$ ppm-mm$^2$ for the 169 $\mu$A  data, and $a=224.9$ ppm, $b=826.6$ ppm-mm$^2$ for the 182 $\mu$A data. The coefficient of determination ($R^2$) for each fit is 0.95 and 0.99, respectively. Extracting the target noise term \dAtgt $={b}/{Area}$ using Eq.~\ref{eq:raster_scan} at the nominal $4\times 4$ mm$^2$ raster area used for most of the experiment, we obtain 42.1 ppm at 169 $\mu$A and 51.7 ppm at 182 $\mu$A. Scaling the 169 $\mu$A target noise result to 182 $\mu$A using the the exponent $e=2.715$ determined in the previous section (Sec.\ref{sec:CurrentScan}) which established the dependence of \dAtgt  on beam current, the 42.1 ppm grows to 51.5 ppm, in good agreement with the measured result at 182 $\mu$A of 51.7 ppm. 

\subsection{ Pump speed scan }
\label{sec:Pump Scan}

\begin{figure}[!hhhtb]
\centerline{\includegraphics[width=0.6\textwidth,angle=0]{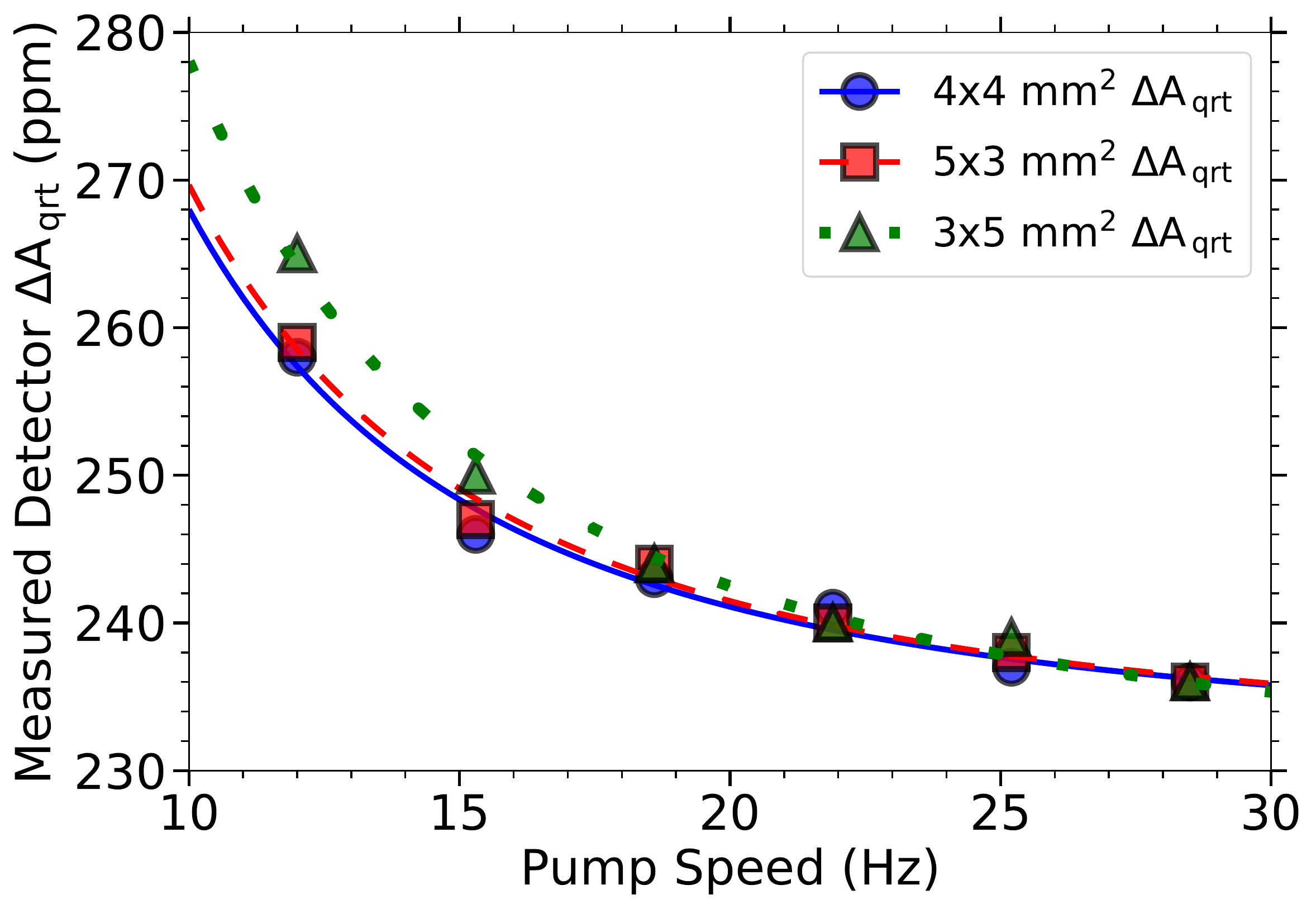}} 
\centerline{\includegraphics[width=0.6\textwidth,angle=0]{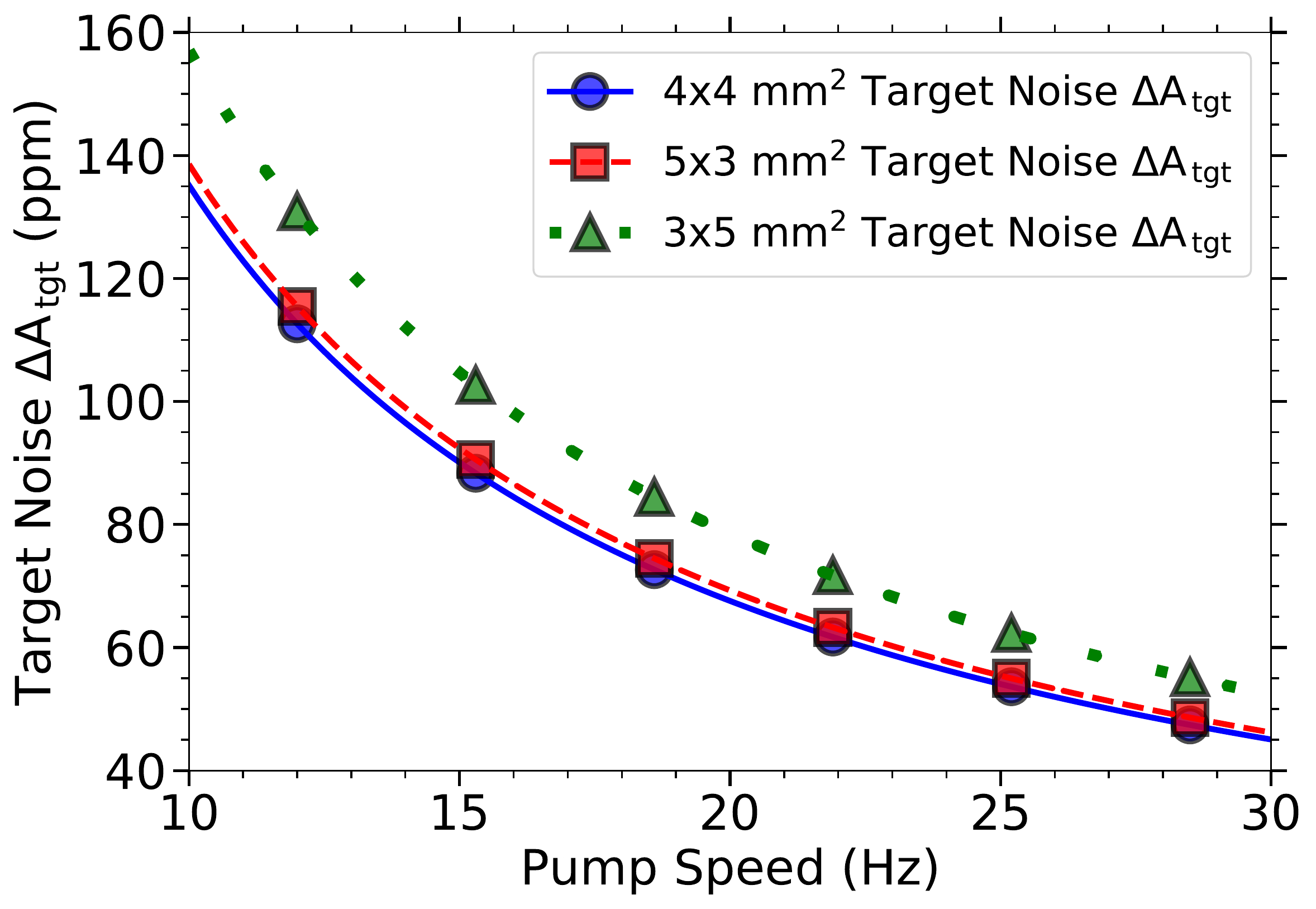}}
 
\caption[Target noise vs Pump Speed at 3 raster sizes] {\label{fig:PumpScanv1}  
Upper: The detector asymmetry width \dAqrt measured at 170 $\mu$A over helicity quartets with ($x,y$) raster dimensions of  $4\times 4$ mm$^2$ (blue circles), $5\times 3$ mm$^2$ (red squares), and $3\times 5$ mm$^2$ (green triangles) as a function of the  LH$_2$ recirculation pump speed. Fits to these data using Eq.~\ref{eq:pumpscanv1} are shown for each raster dimension as solid, dashed, and dotted lines in the corresponding color, respectively. Lower: The target noise term \dAtgt extracted in quadrature from the \dAqrt data in the upper figure with the same symbols and line types used in the upper figure.   }
\end{figure}

The cleanest way to measure the target noise contribution is to vary the LH$_2$ recirculation pump speed, because nothing else changes except the target noise term. As before, we characterize the measured quartet asymmetry widths \dAqrt as the sum in quadrature of a fixed term independent of the pump speed, and a target noise term term \dAtgt inversely proportional to the pump speed $f$.
\begin{equation}
\label{eq:pumpscanv1}
    {\mathrm{A_{\, qrt}}} = \sqrt{a^2 +
    \left ( \frac{b}{f} \right)^2 }.
\end{equation}
In Fig.~\ref{fig:PumpScanv1} we compare asymmetry width measurements made with three different raster configurations of similar area, but different horizontal and vertical ($x$ \& $y$) dimensions: $4\times 4$ mm$^2$, $5\times 3$ mm$^2$, and $3\times 5$ mm$^2$. The fits return ($a,b,R^2$) of (231.4, 1351.5, 0.98), (231.3, 1385.7, 0.99), and (229.4, 1570.3, 0.99) respectively. Note that the fixed  term, $a$,   returned from the fits is about the same for each raster configuration. The $4\times 4$ and $5\times 3$ mm$^2$ results shown in Fig.~\ref{fig:PumpScanv1} look very similar, indicating that increasing the raster $x$-dimension from 4 to 5 mm in the direction of the LH$_2$ flow across the beam axis didn't negatively impact the target boiling. Moreover, the decrease of the raster height in the vertical direction  from 4 to 3 mm didn't make much difference either. However, decreasing the raster $x$-dimension from 4 to 3 mm in the flow direction did have a detrimental effect on the target noise, even though the vertical raster dimension increased to 5 mm. This indicates that the canonical 4 mm raster $x$-dimension is about optimal for target noise in the \Qweak target, and changes in the vertical dimension about 4 mm are unimportant at the 1-mm-scale.  The extracted target noise at each of these three raster configurations is scaled to a common raster size of 16 mm$^2$, pump speed of 28.5 Hz, and beam current of 180 $\mu$A  
in Table~\ref{tab:NoiseScaling}. 

\subsubsection{Temperature dependence}
\label{Temperature dependence}

\begin{figure}[!hhhtb]
\centerline{\includegraphics[width=0.6\textwidth,angle=0]{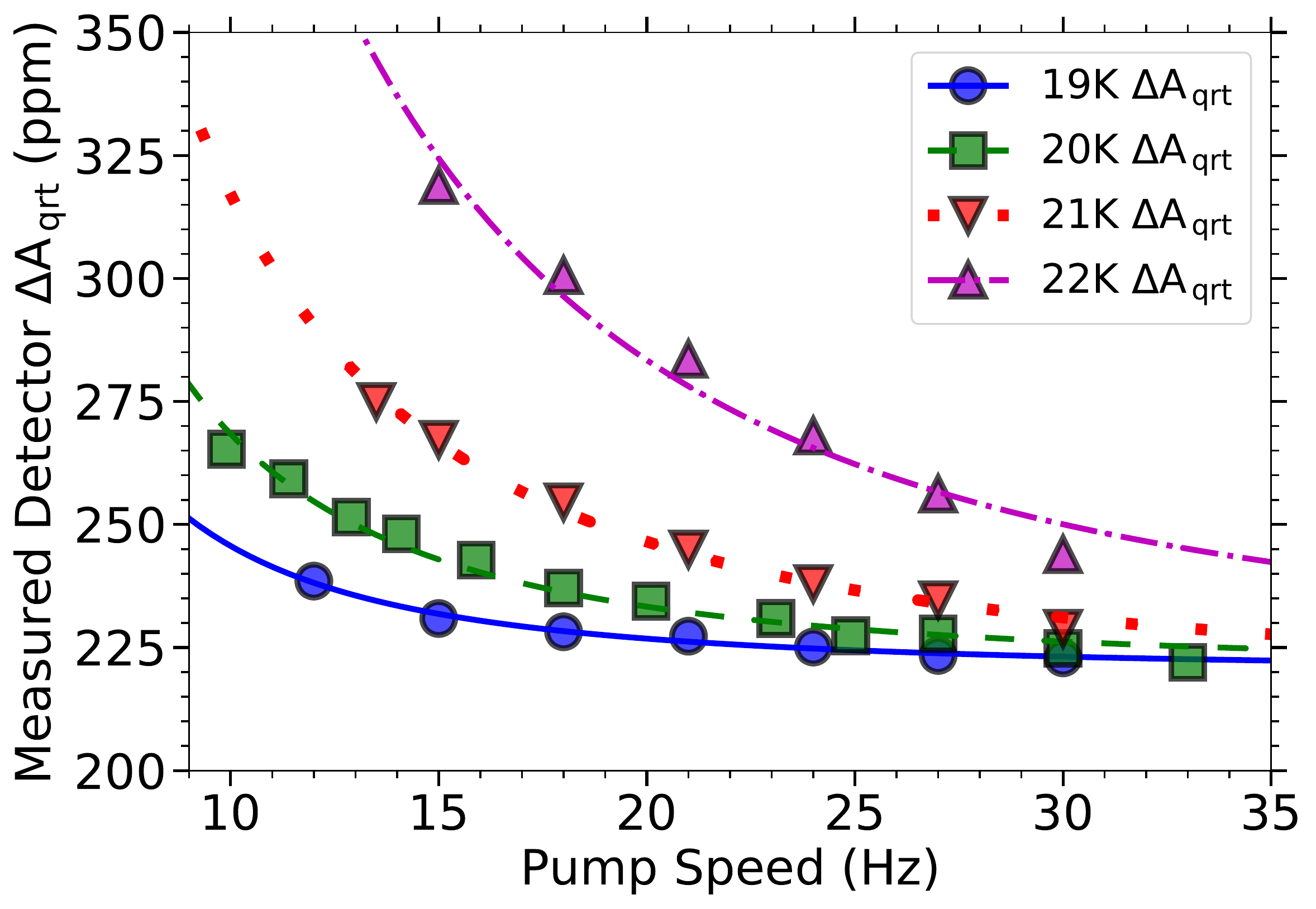}} 
\centerline{\includegraphics[width=0.6\textwidth,angle=0]{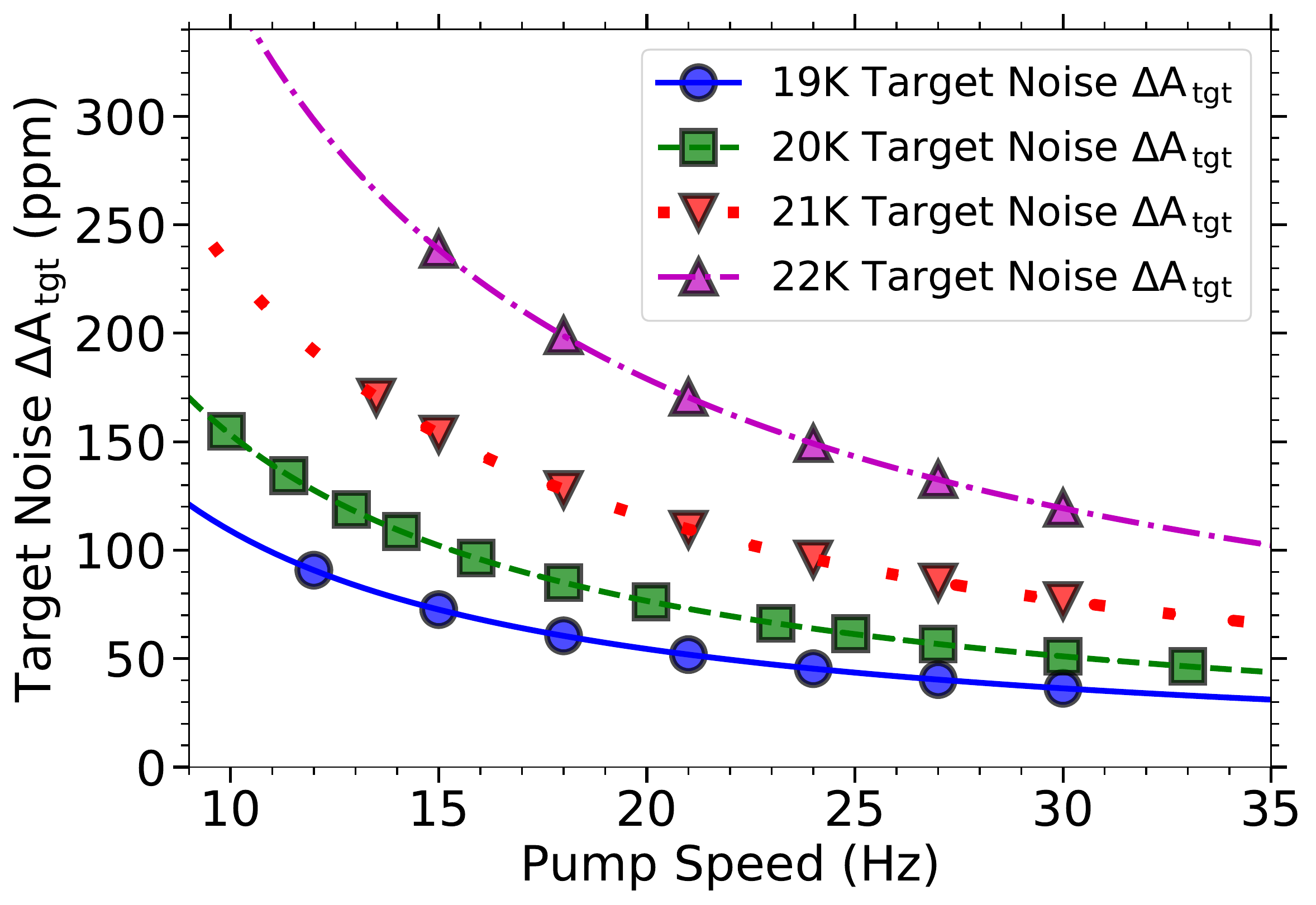}}
\caption[Target noise vs Pump Speed at different temperatures] {\label{fig:PumpScanTemp}  
Upper: The detector asymmetry width \dAqrt measured at 180 $\mu$A over helicity quartets  as a function of the LH$_2$ pump recirculation speed with the LH$2$ temperature at 19~K (blue circles), 20~K (green squares), 21~K (red downward-pointing triangles) and 22~K (magenta upward-pointing triangles). Fits to these data using Eq.~\ref{eq:pumpscanv1} are shown for each target operating temperature  as solid, dashed, dotted and dash-dotted lines in the corresponding colors. Lower: The target noise term \dAtgt extracted in quadrature from the data in the upper figure with the same symbols and line types used in the upper figure.}
\end{figure}

Here we explore how the target noise \dAtgt is affected by what operating temperature the \LH ~target is held at. Eq.~\ref{eq:mdot=QCdT} says that the cooling power is proportional to $\Delta T$, the difference between the coolant supply and return temperatures.
According to Table~\ref{tab:coolant}, the ``15 K'' supply temperature was 14.8~K, and the return temperature was of course close to the LH$_2$  operating temperature. So for the 15~K component of the cooling power, $\Delta T$ varies from about 4.2~K for a target maintained at a \LH ~temperature of 19~K, to a $\Delta T \approx 7.2$~K for a target held at 22~K, a factor of 1.7 improvement in cooling power. The impact on the 4~K cooling power is only a factor of 1.2. However it's clear that maintaining enough cooling power for these temperature studies is more challenging at 19~K than at 22~K.

Although the cooling power improves with higher target operating temperature, the target noise gets worse as the  temperature of the LH$_2$ target rises closer to its boiling point. At the typical LH$_2$ operating pressure of 220 kPa, the target is 3.2~K sub-cooled at an operating temperature of 20 K, but only 1.2~K sub-cooled at 22~K.

To explore how the target noise is affected by different LH$_2$ operating temperatures, pump scans were performed at 19, 20, 21, and 22~K. The measured asymmetry widths and extracted target noise results are  shown in Fig.~\ref{fig:PumpScanTemp}. The fits presented in this figure were performed using Eq.~\ref{eq:pumpscanv1}.

\begin{figure}[!hhhtb]
\centerline{\includegraphics[width=0.6\textwidth,angle=0]{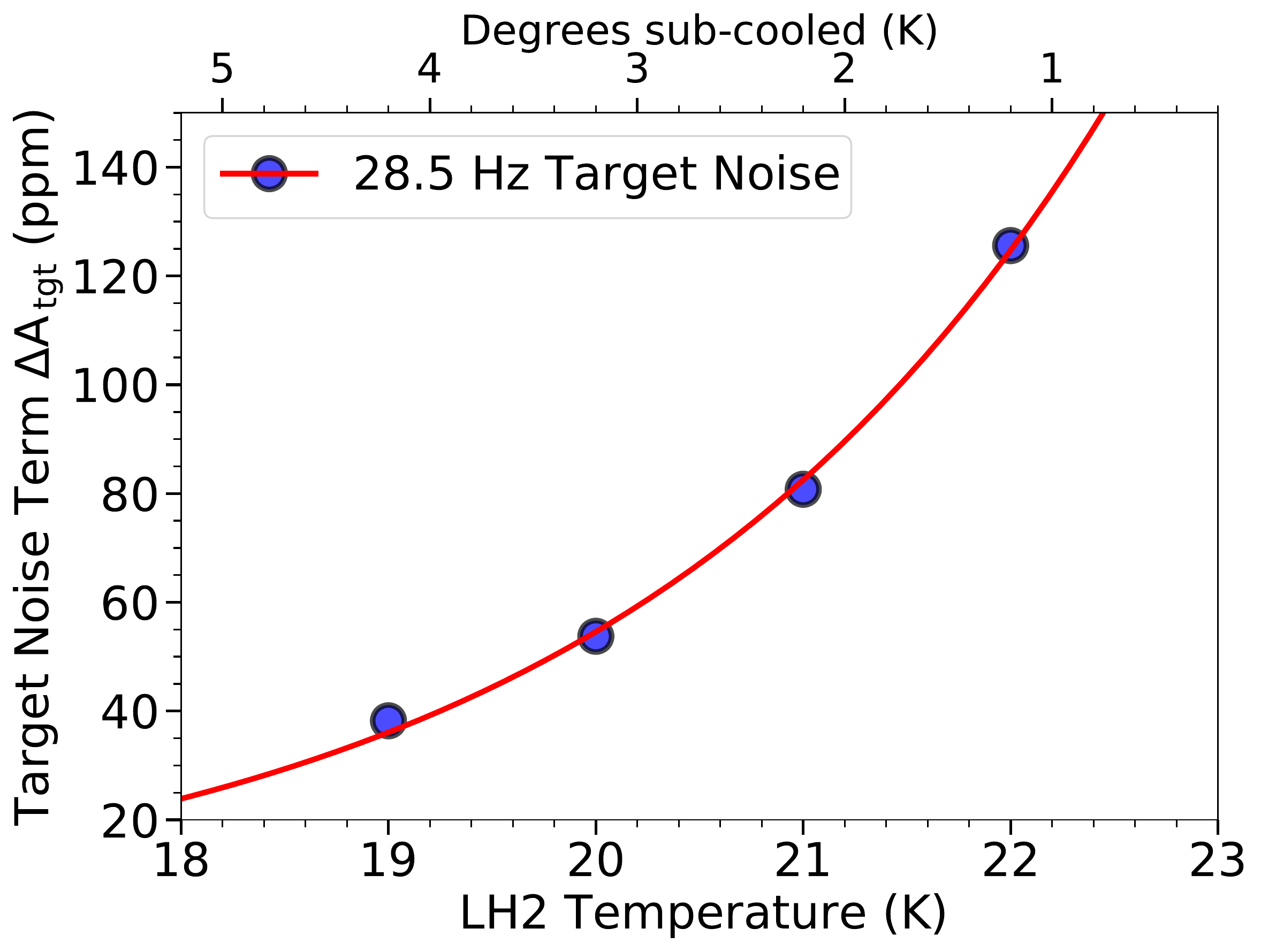}} 
\caption[Target noise vs LH2 Temperature] {\label{fig:BoilingvsLH2Temp}  
The target noise (blue circles) determined at a pump speed of 28.5 Hz and  179 $\mu$A   as a function of the LH$_2$ operating temperature (lower axis) or the amount of sub-cooling (upper axis).  A fit to these data using Eq.~\ref{eq:noisevsT} is shown by the red line. }
\end{figure}

In Fig.~\ref{fig:BoilingvsLH2Temp} the results for the target noise \dAtgt are shown as a function of the LH$_2$ operating temperature $T(LH_2)$ for the nominal pump speed of 28.5~Hz. The 2-parameter fit  to these four temperatures was performed using  
\begin{equation} \label{eq:noisevsT}
    \sigma_{tgt}=a*\exp{[b *T(LH_2)]}.
\end{equation}
The fit parameters were $a=0.0140$,  $b=0.4133$, and $R^2=0.998$.

The 23.2~K boiling point used to determine the amount of sub-cooling is calculated from the vapor-pressure curve at the approximate 220 kPa operating pressure of the target. Since for safety reasons the target's LH$_2$ re-circulation loop was always connected (through open valves) to H$_2$ storage tanks outside the experimental hall, the operating pressure could rise and fall a dozen kPa with the outside temperature according to the ideal gas law. As a result the boiling point and the amount of sub-cooling also varied by about $\pm 0.3$ K.

Fig.'s~\ref{fig:PumpScanTemp} and ~\ref{fig:BoilingvsLH2Temp} clearly show that target noise at the nominal 28.5 Hz pump speed could have been reduced from 54 ppm to 38 ppm by lowering the LH$_2$ operating temperature from 20~K to 19 K. This would still have been safely above the $\sim 14$~K  at which H$_2$ freezes. However as discussed above, the impact on the limited resources of the ESR would have made it difficult to run with the same luminosity at 19 K. If the same level of cooling power could have been sustained at the reduced $\Delta T$ associated with an operating temperature of 19 K, the amount of data  needed to achieve the same 0.0073 ppm statistical uncertainty $\Delta A$(stat) obtained at 20 K~\cite{QweakNature} would have been reduced only 3\% according to Eq.~\ref{eq:QuartetdA}. Accordingly, the compromise made for this experiment was to operate the target at 20~K and 180 $\mu$A.

On the other end of the scale, the target noise was much worse (126 ppm) at the higher operating temperature of 22~K than it was at 20~K (54 ppm). Considering that at 22~K there was a margin of only 1.2~K before the target would boil  
it's surprising the results were not worse. More surprising still is that the target could have been operated only 1.2~K sub-cooled with the full 179 $\mu$A of beam, even though the beam contributes about 2/3 of the nearly 3 kW total heat load seen by the LH$_2$. From Eq.~\ref{eq:QuartetdA} we see that the penalty for doing so would have been having to acquire an additional 25\% more data to reach the same 7.3 ppb statistical uncertainty that was achieved in the experiment at 20~K. 

\subsection{Summary of target boiling noise results}

All the target noise measurements discussed above are tabulated in Table~\ref{tab:NoiseScaling}. The average of all the 20~K results scaled to 179 $\mu$A, $4\times 4$ mm$^2$ raster, and 28.5 Hz pump speed (but excluding the $3\times 5$ result which has more noise for a known reason- see the beginning of Sec.~\ref{Temperature dependence}), is 53.1 ppm. The standard deviation is 2.5~K, which we adopt as the uncertainty in the target noise determinations from all the different techniques discussed in this section. The fact that several independent techniques employed to determine the target noise all give consistent results within this uncertainty provides great confidence in this result, and in the trends observed in the measurements.

\begin{table*}[ht]
\centering
\begin{tabular}{ c  c  c c c c c}
\toprule
 &  &  &  & & Fit & Scaled   \\
Scan & $I_{beam}$ & Raster & Pump & $T$(LH$_2$)& Noise &  Noise  \\
 Type & ($\mu$A)& (mm$^2$) & (Hz) &(K) & (ppm) & (ppm)  \\ \midrule
Current & 179 & $3.5\times 3.5$ & 28.5 &20& 73.1 & 55.1 \\ \midrule
Raster & 180 & $4\times 4$ & 28.5 & 20& 51.7 & 50.9 \\ 
Raster & 169 & $4\times 4$ & 28.5 & 20& 42.1 & 49.2 \\ \midrule
Pump & 169 & $4\times 4$ & 28.5 & 20& 47.4 &  55.4\\ 
Pump & 169 & $5\times 3$ & 28.5 & 20& 48.6 &  53.3\\ 
Pump & 169 & $3 \times 5$ & 28.5 & 20& 55.1 &  (60.4)\\ \midrule
Pump & 179 & $4\times 4$ & 28.5 & 19 & 38.2 &  -\\
Pump & 179 & $4\times 4$ & 28.5 & 20 & 53.7 & 54.5\\
Pump & 179 & $4\times 4$ & 28.5 & 21 & 80.8 &  -\\
Pump & 179 & $4\times 4$ & 28.5 & 22 & 125.6 &  -\\
\bottomrule
\end{tabular}
\caption[Noise Scaling]{Helicity-quartet target noise determinations using various methods. For each method, the beam current, raster size, and pump speed are tabulated. The last column indicates the target noise \dAtgt scaled to a common set of running conditions: 179 $\mu$A, $4\times 4$ mm$^2$ raster area, and 28.5 Hz pump speed.}
\label{tab:NoiseScaling}
\end{table*}

\subsection{Noise dependence on helicity reversal}
\label{sec:helicity}

The scattered electron yield measured in the experiment's detectors was examined in the frequency domain using a fast Fourier transform (FFT). 
Under typical conditions with the \LH ~target, the spectrum was relatively flat except for frequencies below about 50 Hz, where microphonics and sub-harmonics of the 60 Hz line frequency contribute. Spectra taken at lower beam currents or with solid targets were completely flat; hence they did not show any rise below 50 Hz. 

To mitigate these low frequency noise contributions in the experiment, the beam helicity reversal rate 
was increased from the 30 Hz typically used at JLab to 960 Hz  for the \Qweak experiment. In practice this means that the helicity state of $+$ or $-$ was selected every 1/960 s by switching the polarity of the  Pockels cell high voltage  in the polarized source. A settling time of 70 $\mu$s was lost each helicity reversal for the 2.5 kV Pockels cell voltage to stabilize, and another 40 $\mu$s delay for ADC gates in the data acquisition electronics. This means that the expected improvements in the asymmetry width  from faster helicity reversal rates are partially offset by the 110 $\mu$s lost every 1041.65 $\mu$s-long helicity state.

\begin{figure}[!hhhtb]
\centerline{\includegraphics[width=0.6\textwidth,angle=0]{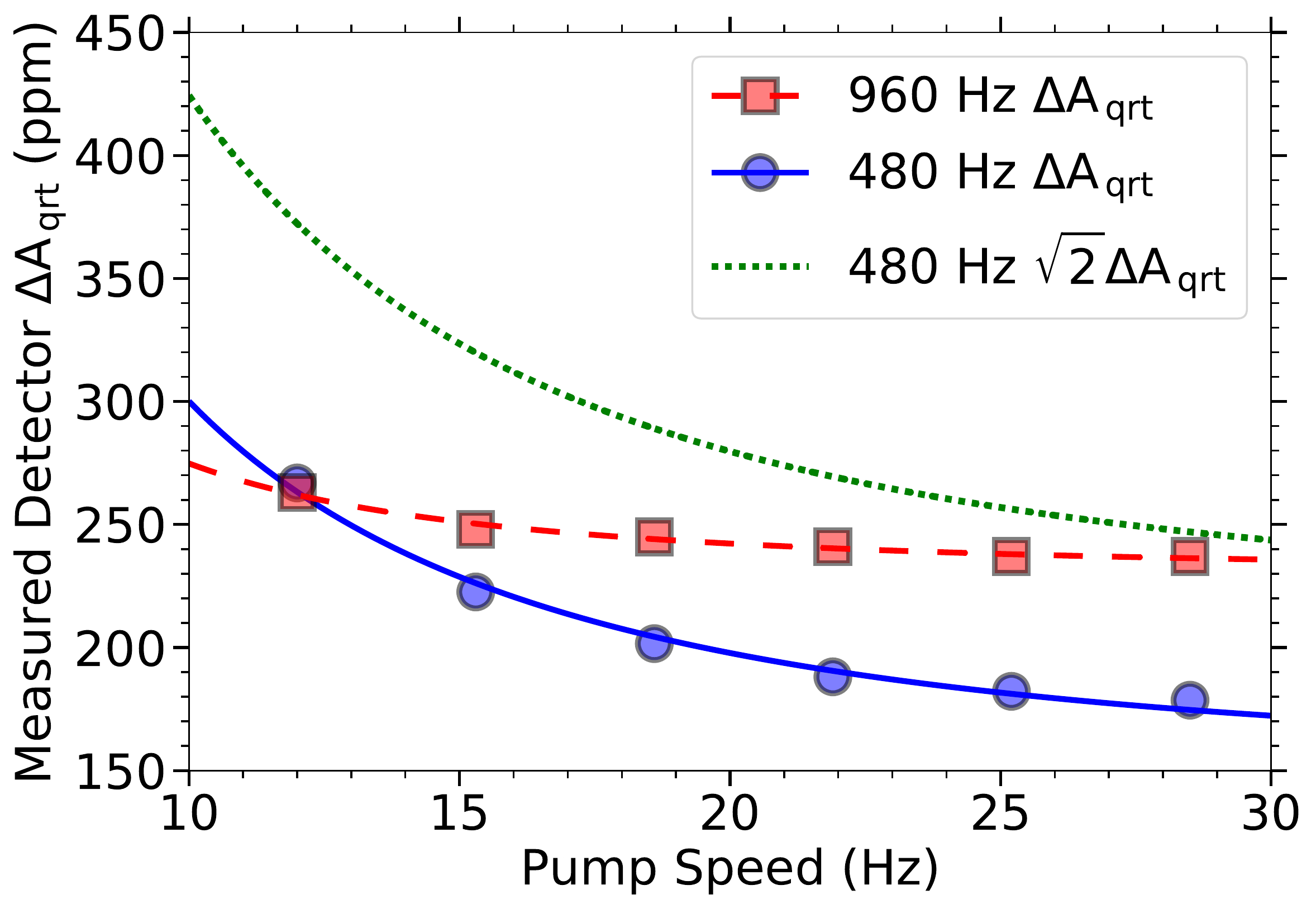}} 
\centerline{\includegraphics[width=0.6\textwidth,angle=0]{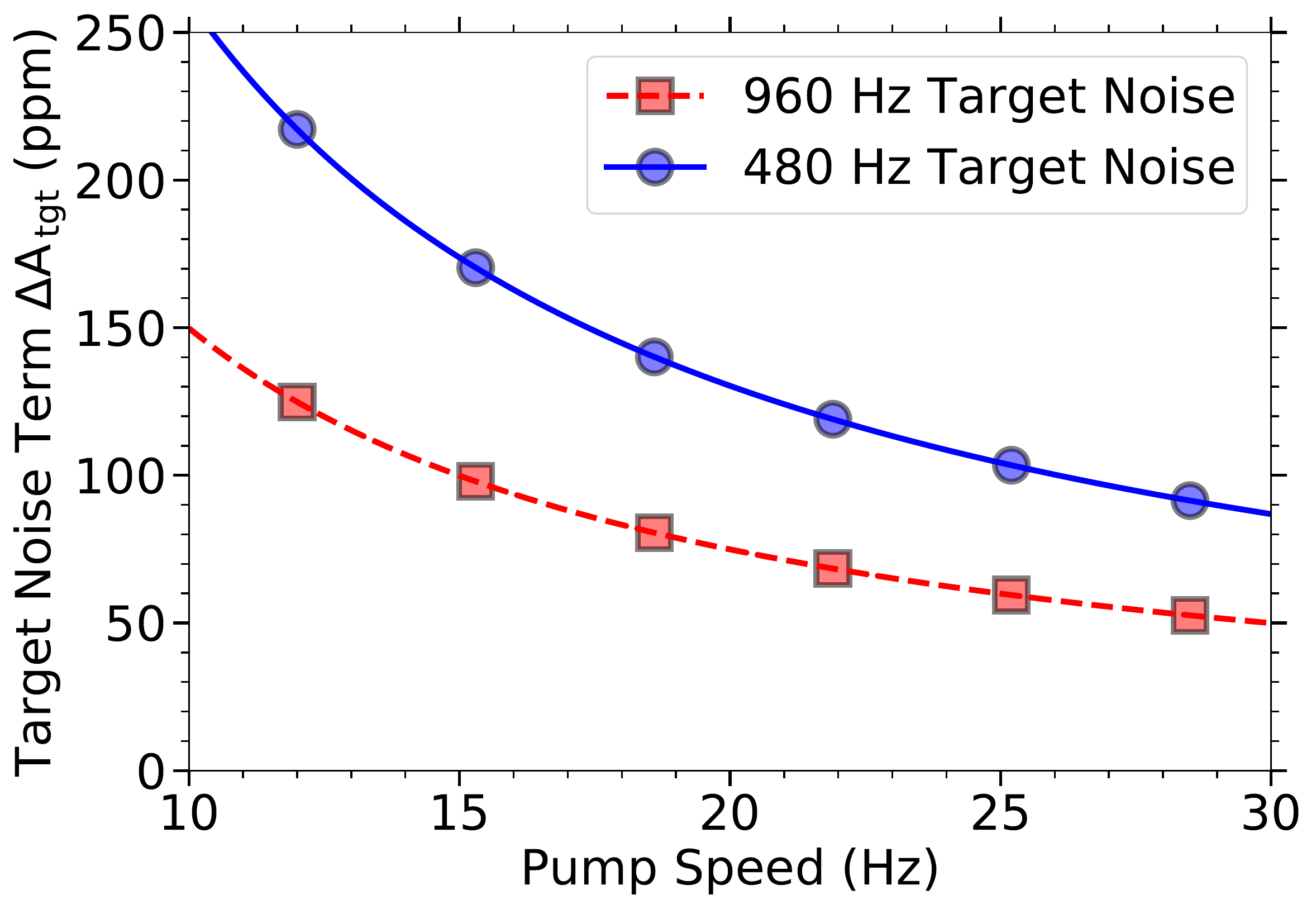}}

\caption[Target noise from pump scans at different helicity reversal frequencies] {\label{fig:Pump_Helicity_Scan} 
Upper: The detector asymmetry width measured over helicity quartets \dAqrt at 170 $\mu$A with a $4\times 4$ mm$^2$ raster area,  
as a function of the LH$_2$ pump recirculation speed with the helicity reversal frequency in the polarized source at the nominal 960 Hz  (red squares) and adjusted to 480 Hz (blue circles). 
Since the overall statistical width \dApv of the experiment's asymmetry depends on the quartet asymmetry width \dAqrt divided by $\sqrt{N_{qrt}}$,  the dotted green curve $\sqrt{2} \Delta {{A_{\, qrt}}(480)}$ is what should be compared to $\Delta {{A_{\, qrt}}(960)}$. Fits to these data are shown for each helicity reversal frequency as solid or dashed lines in the corresponding colors. Lower: The target noise \dAtgt extracted in quadrature from the \dAqrt data in the upper figure with the same symbols and line types used in the upper figure. 
}
\end{figure}

A test was performed during the \Qweak experiment to 
explore this further by acquiring a small amount of data with a helicity reversal rate $\nu=480$ Hz instead of the canonical 960 Hz. The results from the test are shown in Fig.~\ref{fig:Pump_Helicity_Scan}. The helicity-quartet asymmetry width  \dAqrt is much smaller at $\nu=480$ Hz  (178.6 ppm) than it is for the nominal $\nu=960$ Hz rate (237.0 ppm), a consequence of the better statistics at the slower helicity reversal rate. However, to gauge the impact on the statistical width  of the asymmetry \dApv the experiment aims to measure, one must account for the fact that there are twice as many helicity quartets at 960 Hz than at 480 Hz (see Eq.~\ref{eq:QuartetdA}). Accordingly, Fig.~\ref{fig:Pump_Helicity_Scan} also shows the $\nu=480$ Hz result multiplied by $\sqrt{2}$, where even at the canonical 28.5 Hz pump speed, \dApv  would be about 6.6\% larger  than the $\nu=960$ Hz result. The advantage of faster helicity reversal rates is made clear by this figure. It's also clear from the steeper slope of the 480 Hz \dAqrt results that target noise plays a much bigger role at lower helicity reversal rates. Not only is the $\nu=480$ target noise \dAtgt larger than it is at the higher helicity reversal rate, the $\nu=480$ statistical width in each quartet is smaller. The relative contribution of the target noise \dAtgt at $\nu=480$ is thus much larger, as highlighted by Fig.~\ref{fig:Pump_Helicity_Scan}. 

In the introduction of this article (see Sec.~\ref{sec:Performance_Requirements}), we pointed out that the reason target noise is so important for an experiment like \Qweak that sits at the precision frontier    is that it increases the time required to achieve a given precision. Equivalently, the target noise \dAtgt increases the experiment's statistical precision \dApv which is proportional to \dAqrt$/\sqrt{N_{\mathrm{qrt}}}$. So now at the end of this article we evaluate the impact of each of these two  helicity reversal rates on the time it takes for the experiment to achieve its precision goal. The results  are presented in Table~\ref{tab:NuScaling}. These results are drawn from the analyses shown in Fig.~\ref{fig:PumpScanv1} using the 28.5 Hz pump-speed data. The table clearly shows the impact that target noise has on the experiment's precision, in terms of the additional time required to achieve a given precision for each helicity-reversal rate $\nu$. At the canonical $\nu=960$ Hz the target noise penalty is only 5\%, but at the $\nu=480$ Hz rate it rises to 15\%, emphasizing the benefit of faster helicity reversal and less target noise in general on the precision an experiment like \Qweak can achieve.

\begin{table*}[ht]
\centering
\begin{tabular}{ c  c  c c c}
\toprule
Helicity &          &           & &  \\
Reversal & Measured & Extracted & Deduced &  Time\\
 Frequency & \dAqrt & \dAtgt &  ${A_{\, \rm{qrt}}^{\rm{no tgt}}}$ & Penalty\\
 $\nu$ (Hz) & (ppm) & (ppm) & (ppm) &  $\left( \frac{A\mathrm{_{\, qrt}}}{A\mathrm{_{\, qrt}^{no tgt}}} \right)^2$  \\
 \midrule
960 & 237.0 & 52.6 & 231.1 & 1.05    \\
480 & $\sqrt{2}\times178.6$ & 91.4 & 235.4 & 1.15    \\
\bottomrule
\end{tabular}
\caption[Nu Scaling]{Time penalties incurred  from the target noise analysis presented in Fig.~\ref{fig:PumpScanv1}. All entries correspond to a recirculation pump speed of 28.5 Hz. The first column denotes  the helicity-reversal frequencies $\nu$ used for the measured helicity-quartet asymmetry widths \dAqrt  in column 2. The \dAqrt for $\nu=480$ Hz is corrected for the fact that there are half as many quartets $N_{qrt}$ at 480 Hz as there are at 960 Hz. The target noise \dAtgt in the 3rd column is subtracted in quadrature from the \dAqrt in the second column to deduce what the measured \dAqrt would have been without any target noise. The last column takes the square of the ratio of the \dAqrt with and without the target noise term to obtain the time-penalty associated with each helicity-reversal frequency.}
\label{tab:NuScaling}
\end{table*}

\section{Summary}
A high-power liquid hydrogen target was built for the \Qweak experiment at Jefferson Lab to obtain the first measurement of the proton's weak charge, and to set limits on physics beyond the standard model of particle physics. The target was the highest power target used so far in an electron scattering experiment, and the first at Jefferson Lab (and anywhere else we are aware of) to employ CFD in its design. The total heat load of the target was about 3 kW, of which 2.1 kW came from beam heating in the ~\LH. It employed a custom-made centrifugal \LH ~recirculation pump, a novel hybrid heat-exchanger employing separate 4~K and 15~K supplies of helium coolant, a resistive wire heater, and a conical transverse-flow target cell with thin aluminum windows. It also featured a 2-axis target motion system that provided 24 different solid target options. 

Consistent results for the target boiling noise were obtained using a variety of independent techniques, by varying the incident beam current, the overall raster area, the width and height of the rectangular raster,  the recirculation pump speed, the \LH ~operating temperature, and the helicity reversal frequency. The target was well suited for the studies reported in this article, because the statistical noise in each helicity-quartet asymmetry width measurement was only about four times larger than the target boiling (noise) term. The average target noise was $53.1 \pm 2.5$ ppm for typical beam current, raster size, and \LH ~recirculation pump rotation of 179 $\mu$A, $4\times 4$ mm$^2$, and 28.5 Hz.  Ultimately the contribution of the target noise \dAtgt to the final asymmetry result \Apv and uncertainty \dApv obtained in the experiment was negligible. 

\section{Acknowledgments}
\label{sec:Acknowledgments}

This work was supported by DOE contract No. DE-AC05-06OR23177, under which Jefferson Science Associates, LLC operates Thomas Jefferson National Accelerator Facility. Construction and operating funding for the target was provided through  the US Department of Energy (DOE). 
We gratefully acknowledge help from JLab designers R.~Anumagalla, B.~Metzger, P.~Medeiros, S. Furches,  and G.~Vattelana. The authors thank the Dept. of Physics at the Univ. of New Hampshire and  Prof. F. William Hersman in particular for support with the Fluent license during the design of the target with CFD.
We are also indebted to JLab engineers P.~Brindza, D.~Young, J. Henry, E. Daly, K. Dixon, R.~Ganni, and P. Knudson for useful discussions, the technical staff of Hall C and the JLab target group, and software expert S.~Witherspoon for her support helping us control our many devices. We are also grateful to P. Degtiarenko for simulations that helped us reduce the radiation from the target, as well as D. Hamlette and the other members of the radiation control group at JLab who helped make operation of the target possible. Finally, we thank J. Mei, J. Mammei, and all the members of the \Qweak collaboration who operated the target during the experiment.

\pagebreak

\bibliographystyle{elsarticle-num-names}
\bibliography{ref.bib}
\end{document}